\documentclass[fleqn,usenatbib]{mnras}

\usepackage{newtxtext,newtxmath}
\usepackage[T1]{fontenc}
\usepackage{silence}
\usepackage{graphicx}
\usepackage{xcolor}
\usepackage{amsmath}
\usepackage{hyperref}

\newcommand{\kms}{{\rm km\,s}^{-1}}
\newcommand{\Msun}{{\rm M}_{\odot}}

\newcommand{\ld}{{\rm light\mbox{-}days}}
\newcommand{\cloudy}{\textsc{Cloudy}}
\newcommand{\memecho}{\textsc{MEMEcho}}
\newcommand{\Hx}{{H_{\rm LP}}}

\title[Non-Axisymmetric Pillars in AGN Discs]{Non-Axisymmetric Pillar Structures due to Embedded Stars in AGN Accretion Discs}

\author[Huang et al.]{
Jiamu Huang$^{1}$\thanks{E-mail: jiamu\_huang@ucsb.edu},
Douglas N.\ C.\ Lin$^{2,3}$,
Keith Horne$^{4}$
\\
$^{1}$Department of Physics, University of California, Santa Barbara, CA 93106, USA\\
$^{2}$Department of Astronomy and Astrophysics, University of California, Santa Cruz, CA 95064, USA\\
$^{3}$Institute for Advanced Studies, Tsinghua University, Beijing 100086, China\\
$^{4}$SUPA School of Physics and Astronomy, University of St Andrews, St Andrews, KY16 9SS, Scotland, UK
}

\pubyear{2026}

\begin{document}
\label{firstpage}
\pagerange{\pageref{firstpage}--\pageref{lastpage}}
\maketitle

\begin{abstract}
Gravitationally unstable outer regions of AGN discs may host
in situ formation of massive stars whose heating produces
localized vertical ``pillars'' on the disc surface, in the same region that
emits the broad lines. Here we investigate the reverberation-mapping signatures that
such pillars would produce. We compute
velocity--delay maps $\Psi(\lambda, \tau)$, time-evolving emission-line
spectra, and continuum reverberation lag spectra for these structures. We model each pillar as a 2D
Gaussian perturbation to the disc height under lamp-post
irradiation and weight the line response with \cloudy{}
photoionization grids covering the major BLR lines. We show H$\alpha$, Mg\,{\sc ii}, and C\,{\sc iv} as representative
examples. A single pillar produces a
localized hotspot in $\Psi$ and a clear ``S-wave'' pattern in
the time-evolving spectrum that persists, though diluted, for
$N \sim 100$ pillars. 
Emission-line fluxes are suppressed in shadows behind the pillars,
but when normalized by H$\beta$ (or by the local
continuum) the line-to-H$\beta$ ratios move in opposite directions:
Mg\,{\sc ii}/H$\beta$ rises while C\,{\sc iv}/H$\beta$ drops. This
contrast distinguishes pillars from purely kinematic features that
would modulate all lines by similar factors. Across scans of the pillar shape and
lamp-post height, the pillar models produce an optical continuum
reverberation lag spectrum of the observed magnitude.
The once-per-orbit modulation grows with $\sin i$ from
foreshortening, and at 
sufficiently high inclination the pillar's body
geometrically self-occults its lamp-facing inner slope, adding a
sharper near/far contrast. 
These signatures provide
observational diagnostics for detecting in situ star formation in
AGN discs through velocity-resolved reverberation mapping campaigns.
\end{abstract}



\section{Introduction}
\label{sec:intro}

Reverberation mapping exploits the time delay between variations in the AGN
continuum and the corresponding response of broad emission lines to constrain
the size, geometry, and kinematics of the broad-line region
\citep[BLR;][]{Blandford1982, Peterson1993}. In recent years, the development
of velocity--delay mapping \citep{Horne2004} has made it possible to resolve the
BLR kinematics in detail, revealing signatures consistent with Keplerian
rotation around supermassive black holes
\citep[SMBHs;][]{Bentz2010vdm, Horne2021}. The disc-like BLR geometry
underlying these signatures is also supported by direct interferometric
resolution of 3C\,273 \citep{GRAVITY2020} and by a range of theoretical
models including failed-wind dust-inflated discs \citep{BaskinLaor2018}
and dust-driven outflows
\citep{NaddafCzerny2022}. Forward-modelling
codes have been developed to fit reverberation data with parameterized
disc-BLR geometries \citep{Pancoast2011, Li2013}, with recent
applications extending to high-mass AGN that constrain the virial
factor and support the disc-BLR picture \citep{Wang2026dyn}. Complementary to mean-delay diagnostics, simulations of discy BLRs
reverberating in response to stochastic continuum variability
\citep{Ogborn2026} show that observed line-shape parameters carry
additional information about BLR geometry. These parameters include
the centroid velocity, whose stochastic variations are observed in
many monitored AGN \citep{Barth2015}; the velocity dispersion, whose
coordinated variations with continuum luminosity are referred to as
``breathing'' \citep[or ``anti-breathing'';][]{KoristaGoad2004,
CackettHorne2006, Park2012, GuoGu2014, Goad2014, Wang2020breathing};
and the Pearson skewness coefficient, which quantifies asymmetry of
the line profile. Such diagnostics provide a route to distinguish
between competing structural models from time-resolved spectroscopic
campaigns such as the SDSS-V Black Hole Mapper Reverberation Mapping
project \citep{Almeida2023, Fries2024, Stone2025}.

Independently of these observational developments, models of the
outer, gravitationally unstable regions of AGN discs predict in situ
formation of massive stars whose heating raises localized vertical
structures on the disc surface (see Section\,\ref{sec:background}). Such
structures would be non-axisymmetric and would orbit at the local
Keplerian speed. The question we address is what signatures they leave
in reverberation data.

\citet{Starkey2023} demonstrated that the standard thin-disc
model cannot simultaneously reproduce the inter-band continuum lags and the
spectral energy distribution (SED) observed for NGC\,5548 in the Space Telescope and Optical Reverberation Mapping (STORM) campaign \citep{DeRosa2015}.
To resolve this tension, they introduced steep vertical structures
(rims and ripples) on the disc surface, modelled as axisymmetric sinusoidal perturbations
$H(r) = H_1(r)[1 + A(r)\cos(kr)]$, where $H_1(r)$ is the
unperturbed disc surface height, $A(r)$ is the ripple amplitude
relative to $H_1(r)$, and $k$ is the radial wavenumber of the ripple pattern, that
shadow the intervening disc valleys and concentrate lamp-post irradiation
onto a small fraction of the disc surface \citep{Starkey2023}. The
inward-facing crests of these structures intercept the lamp-post radiation and
reach temperatures of $\sim 6000$\,K, while the bulk of the disc remains at the
viscous temperature $T \propto r^{-3/4}$.

The same STORM data also contain direct evidence for orbiting,
non-axisymmetric structure. After subtracting a smooth model of the
continuum and emission-line variations, \citet{Horne2021} found that the
residuals of the C\,{\sc iv} profile, and less clearly those of
Ly$\alpha$, show a helical ``barber-pole'' pattern of diagonal stripes
moving from red to blue across the line, with a peak-to-trough amplitude
of $\pm 8$ per cent of the continuum flux density. A localized structure
orbiting in the line-emitting disc produces such a stripe, because its
line-of-sight velocity, and hence the wavelength at which it perturbs the
profile, varies sinusoidally with orbital phase. The same S-wave
signature is familiar from Doppler tomography of cataclysmic-variable
discs \citep{MarshHorne1988} and has been seen in the debris disc of a
white dwarf \citep{Manser2019}. The stripes cross about half of the
C\,{\sc iv} profile during the 180-day campaign, implying a period of
$\sim 2$\,yr, which for the black-hole mass and inclination inferred from
the velocity--delay maps corresponds to a Keplerian orbit at
$R/c \approx 4$ light-days with $V\sin i \approx 7000$\,km\,s$^{-1}$,
compatible with the velocity amplitude of the stripes. \citet{Horne2021}
tentatively attributed the pattern to shadows cast on the emission-line
region by vertical structure in the inner disc. The stripes therefore
require azimuthal structure moving on a Keplerian orbit
\citep{Horne2021}.

These two observations provide direct tests for any model that
adds vertical structure to the disc surface. In this paper we start from
the embedded-star model of Section\,\ref{sec:background}, which predicts
localized vertical structures on the disc, and compute the reverberation
signatures such structures produce, so that the model can be confronted
with both. To do so we  extend the axisymmetric framework of \citet{Starkey2023} to
non-axisymmetric structures, namely individual ``pillars'' that are localized in both
radius and azimuth. We compute their signatures in emission-line
velocity--delay maps. The key advantage of emission lines over continuum
observations is that lines carry velocity information through the Doppler shift
associated with Keplerian orbital motion, allowing us to map pillar positions
in the $(v, \tau)$ plane. Because pillars cast shadows that reduce the ionizing
flux on the disc surface behind them, they produce differential responses in
high- and low-ionization emission lines, and these responses evolve as the
pillars orbit on Keplerian timescales.

Localized orbiting structure on an otherwise smooth disc has been
invoked before to explain the profile variability of double-peaked
emitters, as a single transient hot spot in Arp\,102B \citep{Newman1997},
as bright spots on the approaching side of the disc of 3C\,390.3
\citep{Jovanovic2010}, and as ensembles of stochastically generated spots
that rotate, shear and decay \citep{FlohicEracleous2008}. Strongly
inhomogeneous discs with many independently fluctuating temperature zones
have likewise been proposed to reconcile microlensing sizes with quasar
variability \citep{DexterAgol2011}. In all of these the spot is a local
emissivity or temperature enhancement of the disc itself. The pillars
considered here are instead opaque vertical structures under lamp-post
irradiation, so each one brightens on its lamp-facing side and also casts
a shadow that suppresses the ionizing flux over the disc behind it, giving
a line-dependent response that a bright spot alone cannot produce.

Structure of this kind is invisible to axisymmetric temperature
reconstructions such as that of \citet{StoneShen2023}, who fitted
multi-epoch SDSS Reverberation Mapping (SDSS-RM; see e.g.,
\citealt{Shen2015, Almeida2023, Shen2024, Fries2024, Stone2025})
spectra of $\sim 100$ quasars with a disc whose temperature varies
with radius and time but is the same at every azimuth, and found slow
radial patterns favouring internal disc processes over lamp-post
reprocessing. A pillar shadow co-rotates with its pillar, so the
shadowed fraction of each annulus is constant in time and the
azimuthal average that such a fit measures does not change as the
pillars orbit. Detecting pillars therefore requires the azimuthal
information that only the Doppler shift of the emission lines
provides, which is why this paper concentrates on velocity-resolved
reverberation.

We organize this paper as follows. In Section\,\ref{sec:background}, we qualitatively 
recapitulate a stellar evolution and pollution in AGN discs (SEPAD) model,
based on which we infer hot spots in the proximity of 
a population of massive stars.  Detailed construction of the SEPAD model
will be presented elsewhere.  In Section\,\ref{sec:model} we describe the
disc geometry and pillar parameterization. Section\,\ref{sec:cloudy} presents the
\cloudy{} photoionization modelling that maps ionizing flux to line emissivity.
In Section\,\ref{sec:vdm} we construct velocity--delay maps and describe how
they evolve as pillars orbit.
We present results in Section\,\ref{sec:results}, turn to the continuum
reverberation signatures in Section\,\ref{sec:continuum_lags}, and discuss the observational
implications in Section\,\ref{sec:discussion}.

\section{Massive stars embedded in AGN discs}
\label{sec:background}

Active galactic nuclei (AGNs) are powered by disc accretion onto supermassive
black holes \citep[SMBHs;][]{lyndenbell1969}
with masses  $M_\bullet \sim 10^{6}$--$10^{9}\,M_\odot$ and accretion rate ${\dot M}_\bullet 
\sim (M_\bullet /10^8 M_\odot) M_\odot {\rm yr}^{-1}$\citep{shankar2009}.
Application of the conventional $\alpha$-prescription for turbulent viscosity
\citep{shakura1973} indicates that at radii $R \gtrsim R_{\rm GI} \simeq 
10^{2-3} R_\bullet$ (where $R_\bullet=G M_\bullet /c^2$ is SMBH's 
gravitational radius), these discs are subject to gravitational 
instability (GI) \citep{goodman2003}, prone to fragmentation and 
spontaneous star formation \citep{chenGI2023}.  
\citet{Hopkins2024,Gerling-Dunsmore2026} 
have raised the possibility that GI may be suppressed by sufficiently 
intense magnetic fields. However, with inefficient angular momentum
transport induced by magneto-hydrodynamic turbulence, the accretion 
flow through the disc onto the supermassive black holes would require 
a large disc aspect-ratio (Eq.\,4 of \citealt{chenpop2024})
which is inconsistent with that inferred from reverberation
mapping analysis \citep{Starkey2023}. Moreover, recent radiation magneto-hydrodynamic simulations indicate that radiation-pressure-dominated outer regions of AGN disks at high accretion rates 
can still undergo gravitational collapse 
owing to their much higher compressibility, 
even in the presence of strong magnetic turbulence (Chen et al., in preparation).

Stars formed in these dense environments 
continue to accrete until they acquire masses $M_\star \gtrsim $ a few
$10^2 M_\odot$ with luminosity $L_\star \rightarrow L_{\rm \star, Edd}$,
their Eddington limit. Thereafter, radiative feedback leads to an 
accretion-wind equilibrium \citep{chenfeedback2024}.
If the freshly accreted disc gas can reach the stars' nuclear burning cores,
the replenishment of H-laden fuel  indefinitely sustains their structure on
the main sequence.  These stars have been referred to as immortal stars \citep{cantiello2021ApJ}.  However, elemental mixing
through an insulating radiative layer between the stellar surface and nuclear
burning core \citep{spruit2002} is inadequate to provide adequate resupply
of H fuel to compensate the H-to-He conversion rate \citep{fryer2025}.  
Consequently, the stars' mass-to-light ratio decreases with the He abundance 
in their core $Y_\star$ as well as their mass \citep{owocki2012}.  An 
accretion-wind equilibrium (with $L_\star \sim L_{\rm \star, Edd}$) is
maintained with decreases in $M_\star$ and $L_\star$. These stars undergo
metamorphic transition to post-main-sequence evolution 
\citep{alidib2023, xuetal2026} and eventually core-collapse supernova
\citep{fryer2025}.  

In contrast to the immortal stars, metamorphic stars retain their $\alpha-$
element and Fe-peak yields to pollute their natal accretion disc. Super-solar metallicities
have been established from AGN's emission line ratios \citep{nagao2006}. 
Evidence for in situ heavy-element pollution in AGN discs has been
inferred from 1) the higher metallicity of the broad lines in comparison
with the narrow lines; 2) the redshift independence of heavy element abundance;
and 3) sub-solar Fe-to-$\alpha-$element ratio \citep{huang2023}.  

The embedded stars provide auxiliary power to maintain a state of marginal
gravitational (in)stability \citep{goodman2003}.  At $R \gtrsim R_{\rm GI}$
the radiative flux from the disc surface $Q^- _\gamma$ is supplied by the heating
flux of the stellar population $Q^+ _\star \sim s_\star \langle L_\star \rangle$ 
where $s_\star$ is the surface density of the stars and $\langle L_\star 
\rangle$ is the average luminosity of the stars \citep{chenpop2024}.  Immortal stars have
very similar masses so that $\langle L_\star \rangle = L_\star$.  Metamorphic
stars have well defined mass function \citep{alidib2023, xuetal2026} and 
$\langle L_\star \rangle$ is weighted by the massive stars.  

Thermal equilibrium, $Q^+ _\star = Q^- _\gamma$, determines the
magnitude of $s_\star \sim 10^3 m_8^{5/6} r_{\rm pc}^{-3/2}$ pc$^{-2}$
where $m_8=M_\bullet/10^8 M_\odot$ and $r_{\rm pc} = R/{\rm pc}$
\citep{chenpop2024}.  Within a given logarithmic radius fraction $f_{\rm R}= 
\Delta R/R$ there are $\Delta N \simeq f_{\rm R} R 2 \pi R s_\star \simeq 
10^3 f_{\rm R} m_8^{5/6} r_{\rm pc}^{1/2}$.  Although star formation is
suppressed in the gravitational stable region at $R \lesssim R_{\rm GI}$,
stars formed at $R \gtrsim R_{\rm GI}$ may migrate inwards, albeit their
$s_\star$ and $\Delta N$ are smaller than those inferred for the star-forming
region with $R \gtrsim R_{\rm GI}$.

In conventional theory \citep{lyndenbell1974, pringle1981}, accretion discs
are heated by axi-symmetric viscous dissipation.  In the SEPAD model,
heating by the individual massive stars leads to local hot spots. 

Assessment of the enhanced disc thickness near the hot spot requires
three-dimensional radiative hydrodynamics simulations which are beyond the scope of
this paper.
Under the assumption that heat diffusion
over the disc mid-plane is efficient to induce average
axi-symmetric temperature ($T_{\rm c}$), pressure
($P_{\rm c}$) and sound speed ($c_{\rm s}$) there,
a marginally (un)stable hydrostatic state can be
maintained with a disc thickness $H = hR = c_{\rm s} /
\Omega$ where the aspect ratio $h \simeq 0.02
\,m_8 ^{-1/6} r_{\rm pc}^{1/2}$ and $\rho_{\rm c}
= M_\bullet/2 \pi R^3$ at radius $R$ with Keplerian
angular frequency $\Omega$ \citep{goodman2003, chenpop2024}.
In a radiation pressure ($P_{\rm rad} = a T_{\rm c}^4/3$)
dominated region, $P_{\rm c} \simeq 2 Q^+ /3 c = 2 s_\star
\langle L_\star \rangle /3c$, $P_{\rm c}/ \rho_{\rm c}
= c_s^2 = \Omega^2 H^2$, and $H = (2 s_\star \langle L_\star
\rangle / 3 c \rho_{\rm c} )^{1/2}/\Omega$.

In the proximity of an average embedded star (with an average
luminosity $\langle L_\star \rangle$), the temperature scale height in the
radial direction is much larger than that along the column perpendicular to
the disc so that heat diffusion leads to a flux
($F_{\rm hotspot} \simeq \langle L_\star \rangle/ f H_{\rm hotspot}^2$) with
$4 \pi > f \sim {\mathcal O} (1)$ and radiation pressure
($P_{\rm hotspot} \simeq 2 F_{\rm hotspot} /3 c =2 \langle L_\star \rangle/
3 c f H_{\rm hotspot}^2$) over a scale height
$H_{\rm hotspot} \simeq (P_{\rm hotspot}/\rho_{\rm c})^{1/2} /\Omega \simeq
(2\langle L_\star \rangle/ 3 c \rho_{\rm c} f H_{\rm hotspot}^2)^{1/2}/\Omega$.
From these relations, we find
\begin{equation}
{H_{\rm hotspot} / H} \simeq (f s_\star H^2)^{-1/4}
\simeq 1.3 f^{-1/4} m_8^{-1/8} r_{\rm pc} ^{-3/8}.
\label{eq:hotspot}
\end{equation}
The embedded stars have well defined mass spectrum \citep{xuetal2026} and
$H_{\rm hotspot}$ around more massive stars is expected to be larger than the
above estimate.
A more accurate follow-up treatment of radiation transfer
will be carried out with 3D radiative hydrodynamic simulation
elsewhere.

The total area filling
factor of the hot spots is $\tau_{\rm hot \ spot} \simeq \pi s_\star H_{\rm hotspot}^2
\sim 2 f^{-1/2} m_8 ^{1/4} r_{\rm pc}^{3/4}$
in the self-regulated star forming 
region.  Due to star-star interaction, some stars may migrate into the
gravitationally stable region of the disc at $R \lesssim R_{\rm GI}$
where $\tau_{\rm hot \ spot}$ is expected to be smaller and
$H_{\rm hotspot}/H$ larger than those in the above estimate.   

Above these hot spots, the disc
thickness swells to form columns with radius $\sim H_{\rm hotspot}$. 
Keplerian shear may extend the disc thickness perturbation by much more than $H_{\rm hotspot}$ in the azimuthal direction.
The structure of these pillars will be simulated
and analysed with 3D radiative hydrodynamic codes elsewhere.  
In this paper, we primarily consider their observable signatures.


\begin{figure}
    \centering
    \includegraphics[width=\columnwidth]{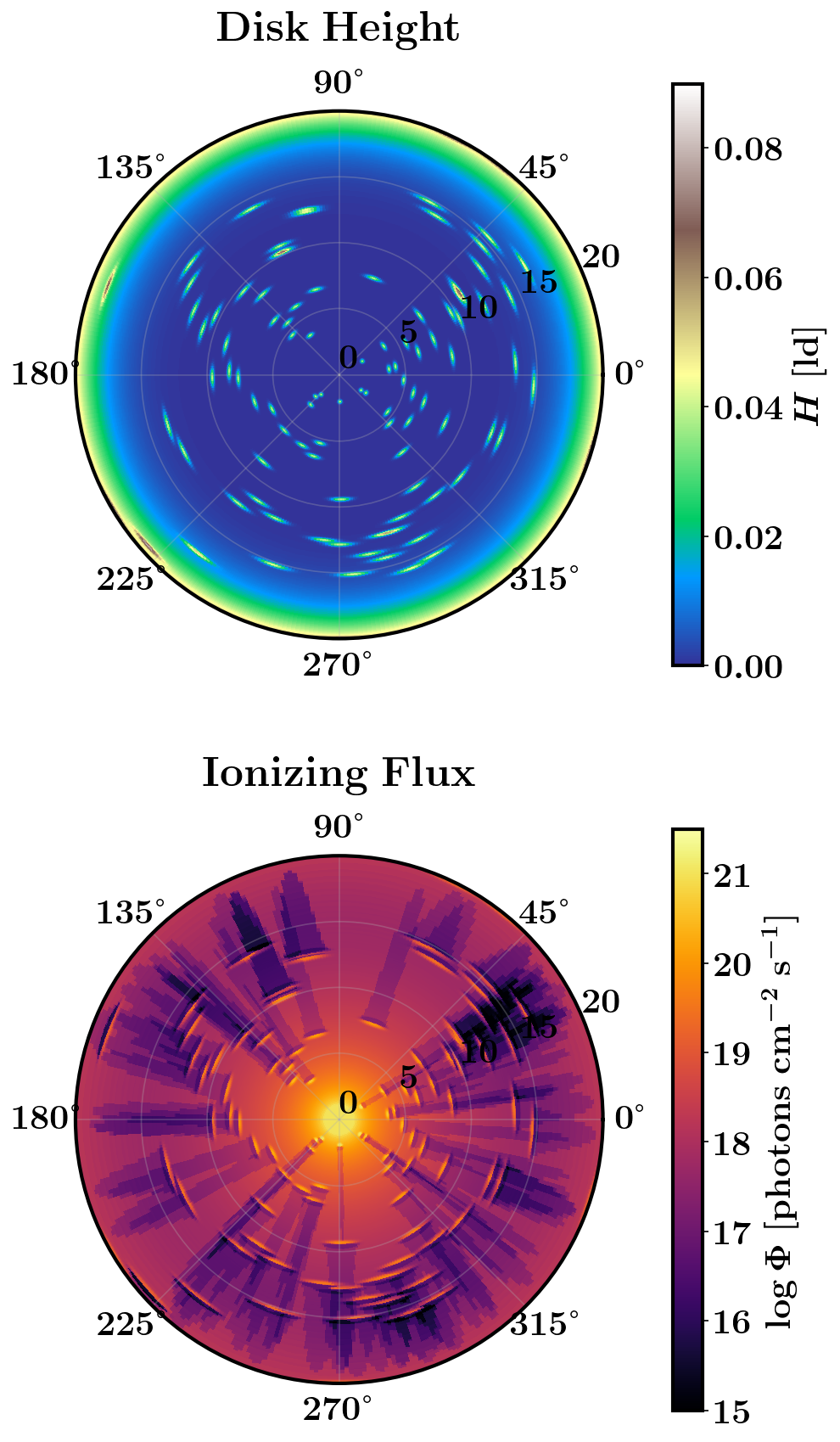}
    \caption{Face-on polar maps for the $N = 100$ pillar configuration.
    \textbf{\textit{Top:}}~disc surface height $H(r, \phi)$.
    \textbf{\textit{Bottom:}}~ionizing photon flux $\log\Phi_{\rm H}(r, \phi)$.
    The disc extends from $r_{\rm in} = 0.008$\,$\ld$ to
    $r_{\rm out} = 20$\,$\ld$ with bowl-height parameters
    $h_0 = 0.05$\,$\ld$, $r_0 = 20$\,$\ld$, $\beta = 10$, lamp height
    $\Hx = 0.05$\,$\ld$, and $T_{\rm irr}/T_{\rm v} = 10$ at $r_0$.
    The 100 pillars are drawn from a Gaussian radial distribution
    with mean $r_{\rm mean} = 10$\,$\ld$, standard deviation $\sigma_{\rm dist} = 5$\,$\ld$, and
    $r_{\rm min} = 2$\,$\ld$, uniform in azimuth, with per-pillar
    height $h_p = 0.04$\,$\ld$, radial width $\sigma_r = 0.1$\,$\ld$,
    and azimuthal width $\sigma_\phi = 0.05$\,rad. Multiple shadow
    lanes create a complex pattern of illuminated and shadowed
    regions.}
    \label{fig:ionflux}
\end{figure}

\section{Disc Model and Pillar Geometry}
\label{sec:model}

\subsection{Bowl-shaped disc}
\label{sec:disc_model}

We adopt the bowl-shaped disc geometry introduced by \citet{Starkey2023}, in
which the bare (unperturbed) disc surface height follows a power-law profile
\begin{equation}
H_0(r) = h_0 \left( \frac{r}{r_0} \right)^\beta.
\label{eq:h_profile}
\end{equation}
Here $r_0$ is a reference radius (chosen to coincide with the outer disc
edge $r_{\rm out}$); $h_0 \equiv H_0(r_0)$ is the disc surface height at
that reference radius, which sets the overall amplitude of the bowl flare;
and $\beta$ controls the steepness of the rise from the inner disc to
the rim. Subscript ``$0$'' on $H_0$ denotes the bare bowl, before any
pillar perturbations are added (Eq.\,\ref{eq:pillar}). We adopt the
fiducial values $h_0 = 0.05$\,$\ld$, $r_0 = 20$\,$\ld$, and
$\beta = 10$ throughout this paper. The large value of $\beta$
produces a strongly concave (bowl-shaped) geometry in which the inner
disc remains essentially flat and is exposed to the central lamp-post
source, while the outer disc rises steeply at the rim
($r \gtrsim 0.7\,r_0$). We set the inner and outer radii to
$r_{\rm in} = 2\,r_{\rm g} \approx 0.008\,\ld$, the innermost stable orbit of a
prograde Kerr black hole, and $r_{\rm out} = 20$\,$\ld$, respectively,
consistent with the BLR extent inferred for NGC\,5548
\citep{Horne2021}.  It is also compatible with the
``failed wind'' model for the broad emission line region in AGN discs
\citep{czerny2015, owen2025MNRAS}.
Only the upper disc surface, the one facing the observer, is
modelled. The disc and its pillars are assumed to be mirror-symmetric
about the midplane. For any inclination $i<90^\circ$ the lower surface is
hidden behind the optically thick disc and does not contribute to the
observed response, so Figs.\,\ref{fig:ionflux} and \ref{fig:geometry}
show the upper surface only.

For the temperature structure, we combine viscous heating with lamp-post
irradiation. The effective temperature at each disc surface element is given by
\begin{equation}
T_{\rm e}^4(r, \phi) = T_{\rm v}^4(r) + T_{\rm irr}^4(r, \phi),
\label{eq:temp}
\end{equation}
where the viscous temperature follows $T_{\rm v}(r) = T_{\rm v,0} (r_0/r)^{3/4}\,[(r - r_{\rm in})/r]^{1/2}$
with $T_{\rm v,0} \approx 450$\,K at $r_0$, set so that
$T_{\rm v}(1\,{\rm light\mbox{-}day}) \approx 4000$\,K; the last factor imposes zero torque at the
inner edge, and $T_{\rm irr}$ is the irradiation
temperature from a point-source lamp-post at height $\Hx = 0.05$\,$\ld$ on the
symmetry axis. We adopt a ratio $T_{\rm irr}/T_{\rm v} = 10$ at $r_0$ for
illuminated regions, reflecting the dominance of lamp-post heating over viscous
dissipation. In shadowed regions behind pillars, we set the irradiation
component to zero so that only viscous heating remains.

We discretize the disc surface on a grid of $n_r \times n_\phi = 600 \times 360$
cells, with logarithmic spacing in radius and uniform spacing in azimuth.

\subsection{Pillar parameterization}
\label{sec:pillar_param}

We model each pillar as a two-dimensional Gaussian perturbation to the disc
surface height:
\begin{equation}
\Delta H(r, \phi) = h_p \exp\left[
-\frac{(r - r_p)^2}{2 \sigma_r^2}
-\frac{r_p^2 (\phi - \phi_p)^2}{2 (r_p \sigma_\phi)^2}
\right],
\label{eq:pillar}
\end{equation}
where $r_p$ and $\phi_p$ denote the pillar's radial and azimuthal position,
$h_p$ is the peak height above the ambient disc surface, and $\sigma_r$ and
$\sigma_\phi$ control the radial and azimuthal extent. The total disc surface
height is the sum of the smooth background and all pillar contributions,
$H(r, \phi) = H_0(r) + \sum_{i=1}^N \Delta H_i(r, \phi)$. Each pillar is
therefore characterized by five parameters
$(r_p, \phi_p, h_p, \sigma_r, \sigma_\phi)$, with the radial and azimuthal
widths chosen to be small compared to the orbital radius so that each pillar
occupies a localized patch on the disc surface.

\subsection{Shadow computation}
\label{sec:shadows}

The lamp-post geometry naturally produces shadows behind each pillar. For each
disc surface element at position $(r, \phi)$, we trace a ray from the lamp-post
at $(0, 0, \Hx)$ and check whether the ray is intercepted by any pillar at
smaller radius. We encode this prescription in a binary shadow mask $S(r, \phi)$, where
$S = 1$ for illuminated regions and $S = 0$ for shadowed regions. When $S = 0$,
we set the irradiation temperature to zero so that only the viscous temperature
contributes.

The lamp ray that grazes the pillar top at $(r_p, \phi_p,
H_0(r_p)+h_p)$ defines the boundary of the radial shadow lane. The
shadow ends where this grazing ray meets the disc surface
$z = H_0(r)$, i.e. at the radius $r_{\rm shadow}$ that satisfies
\begin{equation}
H_0(r_{\rm shadow}) - \Hx
   = \frac{r_{\rm shadow}}{r_p}\,
     \bigl[H_0(r_p) + h_p - \Hx\bigr].
\label{eq:shadow_extent}
\end{equation}
For a flat disc ($H_0 \equiv 0$), the solution is
\begin{equation*}
r_{\rm shadow} = r_p\,\frac{\Hx}{\Hx - H_0(r_p) - h_p}\quad
,~\text{for } h_p + H_0(r_p) < \Hx,
\end{equation*}
and the shadow has finite extent. When the pillar top is higher than
the lamp ($h_p + H_0(r_p) > \Hx$) and the disc is flat, the grazing
ray rises monotonically and the shadow extends to infinity. 
For the partially transparent pillars of our fiducial model
($f_{\rm trans} = 0.3$; see Appendix\,\ref{sec:transparent_pillars}) the mask floors at
$S = f_{\rm trans}$ rather than zero, lowering $\Phi_{\rm H}$ in the lane by
0.5--1\,dex.
We solve
Eq.\,(\ref{eq:shadow_extent}) numerically by ray-tracing on the
$(r, \phi)$ grid; the result is the elongated shadow lanes visible
behind each pillar in Fig.\,\ref{fig:ionflux}. The dependence of the
shadow length and of the foreshortening/occultation modulation on
$h_p$ is explored in Appendix\,\ref{sec:height_sweep}, where we
present a pillar-height sweep at fixed inclination.

\subsection{Pillar distributions}
\label{sec:pillar_dist}

We adopt parameters motivated by NGC\,5548, for which \citet{Horne2021}
inferred $M_{\rm BH} \approx 7 \times 10^7$\,$\Msun$, a disc-like BLR
extending from 2 to 20\,$\ld$ with inclination $i \approx 45^\circ$, and
 the barber-pole pattern in the C\,{\sc iv} residuals
(see Section\,\ref{sec:intro}) with an estimated
orbital period of $\sim 2$\,years, corresponding to a Keplerian radius
$R/c \approx 4$\,$\ld$.

\begin{figure}
    \centering
    \includegraphics[width=\columnwidth,trim=0 0 0 220,clip]{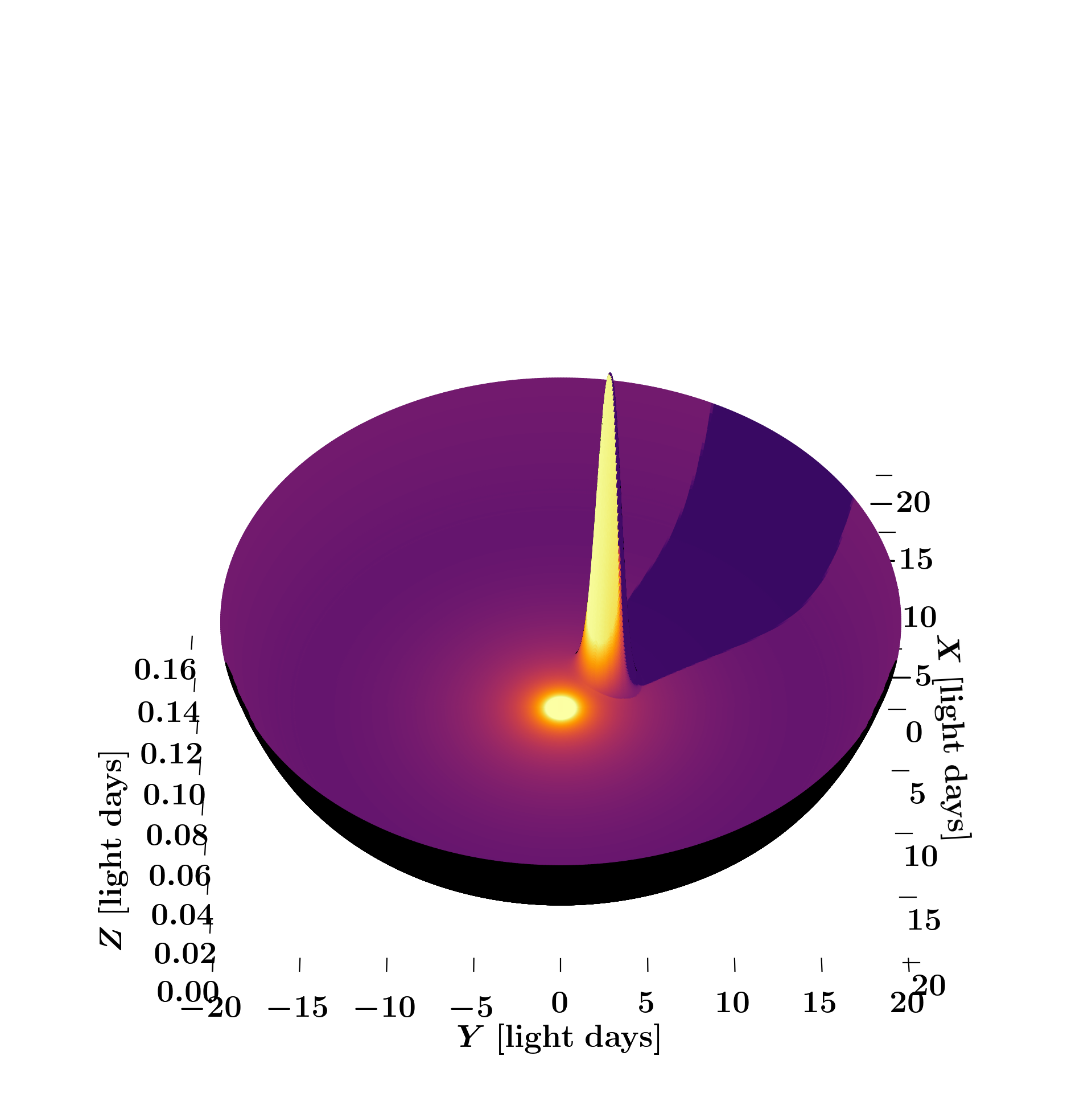} \\
    \includegraphics[width=\columnwidth,trim=0 0 0 180,clip]{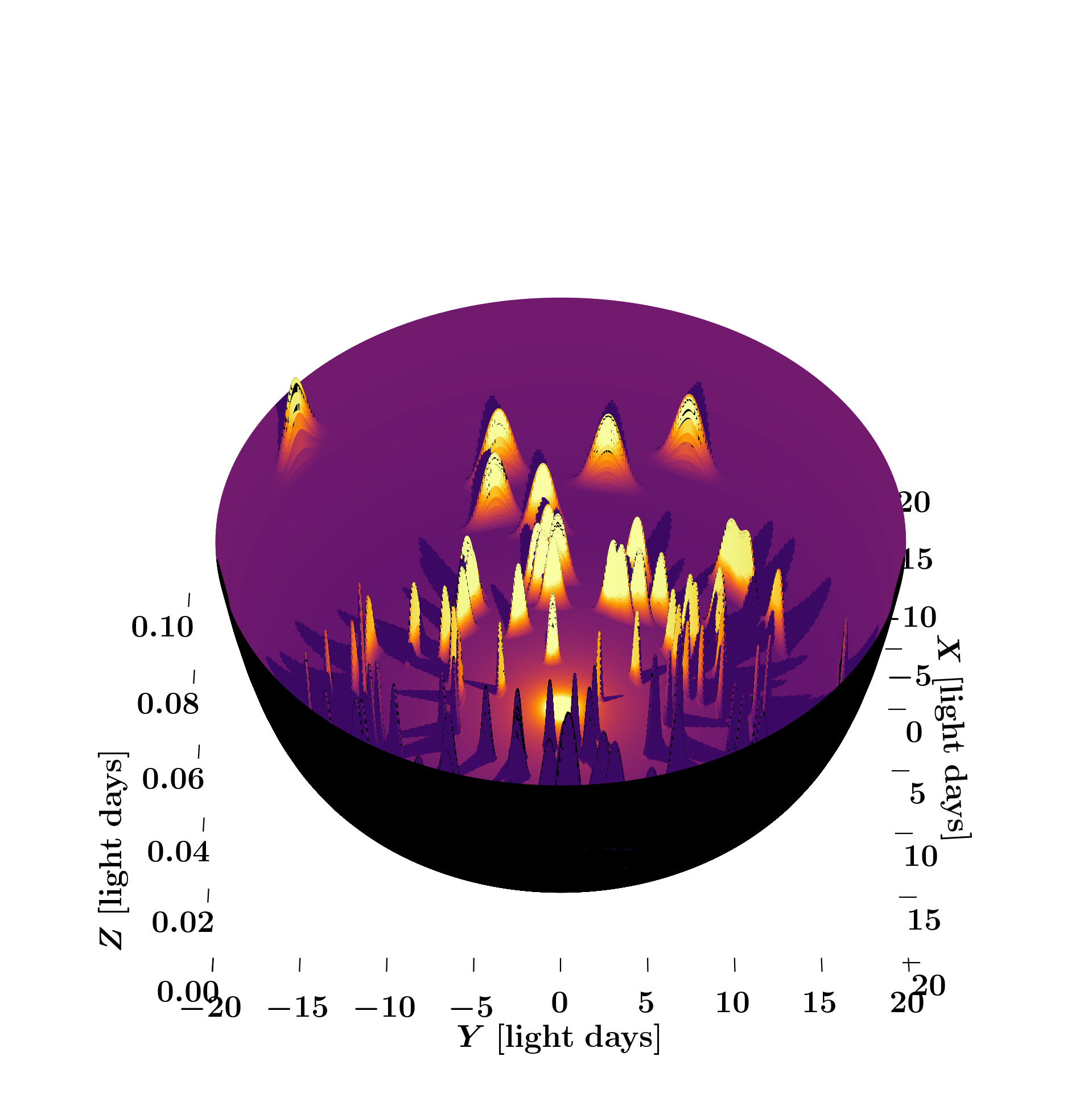} \\
    \includegraphics[width=0.75\columnwidth]{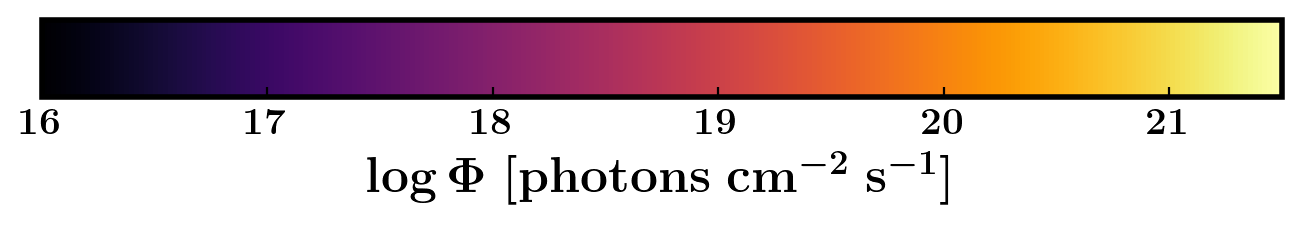}
    \caption{Three-dimensional disc geometry coloured by ionizing flux,
    viewed at inclination $i = 45^\circ$. The vertical axis is
    stretched by a factor of $\sim 100$ relative to the radial axes so
    the very shallow bowl ($h_0 = 0.05$\,\ld\ at $r_{\rm out}
    = 20$\,\ld) is visible; convex-underside facets, which the lamp
    cannot illuminate, are masked black.
    The bowl-shaped disc extends from $r_{\rm in} = 0.008$\,$\ld$ to
    $r_{\rm out} = 20$\,$\ld$ with height profile
    $H_0(r) = h_0 (r/r_0)^\beta$ ($h_0 = 0.05$\,$\ld$, $r_0 = 20$\,$\ld$,
    $\beta = 10$). The lamp-post (yellow star) sits at height
    $\Hx = 0.05$\,$\ld$ on the symmetry axis.
    \textbf{\textit{Top:}} Single pillar at $r_p = 4$\,$\ld$,
    $\phi_p = 3\pi/4$, with height $h_p = 0.15$\,$\ld$ and widths
    $\sigma_r = 0.5$\,$\ld$, $\sigma_\phi = 0.2$\,rad.
    \textbf{\textit{Bottom:}} $N = 100$ pillars drawn from a Gaussian
    radial distribution with mean $r_{\rm mean} = 10$\,$\ld$,
    standard deviation $\sigma_{\rm dist} = 5$\,$\ld$, and $r_{\rm min} = 2$\,$\ld$.
    }
    \label{fig:geometry}
\end{figure}

We consider two configurations, illustrated in Fig.\,\ref{fig:geometry}.
The first is a single test pillar ($N = 1$)
placed at $r_p = 4$\,$\ld$, $\phi_p = 3\pi/4$, with height
$h_p = 0.15$\,$\ld$ and widths $\sigma_r = 0.5$\,$\ld$,
$\sigma_\phi = 0.2$\,rad. This radius matches the barber-pole feature
identified by \citet{Horne2021} in NGC\,5548 and serves as a diagnostic
for understanding the physics of individual pillars.
Its size is chosen for demonstration and is much larger than the hot spot
of a single embedded star (Eq.\,\ref{eq:hotspot}), which the $N = 100$ pillars
below are scaled to. Structures this large may be produced by other
mechanisms, such as a slow tidal disruption event that inflates the disc
locally, which we will study elsewhere.

The second configuration is a realistic population of $N = 100$ pillars,
motivated by the expected number of embedded massive stars in self-gravitating
disc regions. As an idealization, we draw the radial positions from a Gaussian
distribution
\begin{equation}
p(r) \propto \exp\left[-\frac{(r - r_{\rm mean})^2}{2 \sigma_{\rm dist}^2}\right],
\label{eq:r_dist}
\end{equation}
with mean $r_{\rm mean} = 10$\,$\ld$ and standard deviation $\sigma_{\rm dist} = 5$\,$\ld$, clipped at a
minimum radius $r_{\rm min} = 2$\,$\ld$, comparable to the inner boundary
($R_{\rm GI}$) of the gravitationally unstable regions of the disc where 
star formation
may take place \citep{chenpop2024}. Inside $r_{\rm GI}$ the disc is
gravitationally stable and star formation no longer takes place. 
The exact domain of the gravitational instability is subject 
to uncertainties in the turbulent viscosity prescription 
\citep{kingpringle2007}.
Although the embedded stars may undergo 
stochastic migration away from their birth-place \citep{wu2024},
the surface density of embedded massive stars is likely to drop 
sharply interior to $R_{\rm GI}$, motivating the cutoff.

This Gaussian profile is adopted as a simple proxy that concentrates
pillars around the reverberation-mapped BLR radius of NGC\,5548, and is
not intended to reproduce the embedded-star surface density
$s_\star \propto r^{-3/2}$ derived in Section\,\ref{sec:background}. A more
detailed match to $s_\star(r)$ would be appropriate when modelling a
specific source. We distribute the azimuthal positions
uniformly in $[0, 2\pi)$ and assign each pillar a height
$h_p = 0.04$\,$\ld$, a radial width $\sigma_r = 0.1$\,$\ld$, and an
azimuthal width $\sigma_\phi = 0.05$\,rad. The compact widths
($\sigma_r \ll r_p$) reflect the localized nature of individual stellar
heating sources within the disc.
The hot-spot height of Eq.\,(\ref{eq:hotspot}) grows with radius
($\propto r^{9/8}$). For the parameters adopted here it equals
$h_p = 0.04\,\ld$ at $r \simeq 3$--$5\,\ld$ and reaches 0.08--0.15\,\ld{}
at $r \simeq 10\,\ld$. We adopt a single height for simplicity.
By Eq.\,(\ref{eq:hotspot}), hot spots rise above the lamp
($H_{\rm hotspot} > \Hx$) beyond $r \simeq 4$--$6\,\ld$, where their shadows
reach the disc edge (Section\,\ref{sec:shadows}).
In our configurations, the bowl rim ($h_0 = 0.05\,\ld$) stays below the
grazing ray of the single pillar ($h_p = 0.15\,\ld > \Hx$), so its lane
reaches the disc edge, while the lower $N = 100$ pillars
($h_p = 0.04\,\ld < \Hx$) cast finite lanes ending near $r \approx 10$, 17
and 19\,\ld\ for $r_p = 2$, 4 and 10\,\ld.

\section{Photoionization Modelling with \cloudy{}}
\label{sec:cloudy}
\label{sec:ionflux}
\label{sec:cloudy_grid}

The pillars and their shadows change the ionizing flux that reaches
each part of the disc surface, and the broad emission lines respond to
that flux in a line-dependent way. To predict these signatures we need
the emissivity of each line as a function of the local hydrogen-ionizing
photon flux, which we obtain from the photoionization code \cloudy{}
\citep[version 23.01;][]{Ferland2017, Gunasekera2023}.
We first describe how the ionizing flux map is computed from the disc,
pillar and lamp geometry, and then the grid of \cloudy{} models and its
reduction to a one-dimensional emissivity table.

The emission-line emissivity at each disc surface element depends on
the local hydrogen-ionizing photon flux $\Phi_{\rm H}$
(photons\,cm$^{-2}$\,s$^{-1}$, used throughout), which we compute directly from the
disc--pillar--lamp geometry as the sum of a lamp-post contribution and
a small viscous-blackbody contribution,
\begin{equation}
\Phi_{\rm H}(r, \phi) = \Phi_{\rm lamp}(r, \phi) + \Phi_{\rm visc}(r).
\label{eq:phi_total}
\end{equation}
The viscous component, computed from a Wien-tail integral of the local
disc blackbody at $T_{\rm v}(r)$, is comparable to $\Phi_{\rm lamp}$
only at the innermost radii; it falls below the \cloudy{} grid floor
beyond $r \sim 0.5$\,\ld{} and is many orders of magnitude below
$\Phi_{\rm lamp}$ throughout the pillar region
(see Appendix\,\ref{sec:appendix_flux}).
We retain it in
Eq.\,(\ref{eq:phi_total}) for completeness but in practice
$\Phi_{\rm H} \approx \Phi_{\rm lamp}$ on the unshadowed disc, and
$\Phi_{\rm H} \to 0$ inside shadows casted by opaque pillars.

We model the central source as an isotropic point lamp on the disc
symmetry axis, at height $\Hx$ above the midplane, emitting $Q$
hydrogen-ionizing photons per second. At a disc surface element with
cylindrical coordinates $(r, \phi)$ and surface height
$H(r, \phi) = H_0(r) + \sum_i \Delta H_i(r,\phi)$
(Eq.\,\ref{eq:h_profile} and\,\ref{eq:pillar}), the three-dimensional
lamp-to-element distance is
\begin{equation}
d(r, \phi) = \sqrt{r^2 + \bigl[\Hx - H(r, \phi)\bigr]^2},
\label{eq:lamp_distance}
\end{equation}
and the cosine of the incidence angle on the locally tilted disc
surface is
\begin{equation}
\cos\theta_{\rm inc}(r, \phi)
   = \hat{\boldsymbol n}(r, \phi) \cdot \hat{\boldsymbol\ell}(r, \phi),
\label{eq:cos_theta}
\end{equation}
where $\hat{\boldsymbol\ell}$ is the unit vector from the surface
element toward the lamp and $\hat{\boldsymbol n}$ is the outward
surface normal computed from the local height field $H(r, \phi)$.
Because pillars introduce steep azimuthal as well as radial gradients
in $H$, the surface normal carries both a radial and an azimuthal
component near a pillar, and the lamp-facing pillar wall sees a much
larger $\cos\theta_{\rm inc}$ than the surrounding flat disc. The
lamp-post contribution to the ionizing flux is then
\begin{equation}
\Phi_{\rm lamp}(r, \phi) = S(r, \phi)\,
\frac{Q\,\cos\theta_{\rm inc}(r, \phi)}{4\pi\,d^2(r, \phi)},
\label{eq:phi_lamp}
\end{equation}
where $S(r, \phi) \in [0, 1]$ is the geometric shadow mask
(Section\,\ref{sec:shadows}): $S = 0$ deep inside the geometric shadow
of a fully opaque pillar, $S = 1$ on fully illuminated parts of the
disc surface, and $S \to f_{\rm trans}$ in the inner shadow of a
partially-transparent pillar wall, where $f_{\rm trans}$ represents
an effective transmission of both the ionizing continuum and the line
emission through the pillar (Appendix\,\ref{sec:transparent_pillars}).

The lamp-post normalization $Q$ is fixed by requiring the unshadowed
flat-disc flux at the inner edge to match a target value $\Phi_0$:
\begin{equation}
Q = 4\pi\,d_{\rm in}^{2}\,\Phi_0\,/\,\cos\theta_{\rm inc, in},
\label{eq:Q_calib}
\end{equation}
where, since the disc is geometrically thin at the inner edge
($H(r_{\rm in}, \phi) \ll H_{\rm LP}$ ), Eq.\,\eqref{eq:lamp_distance} reduces to
$d_{\rm in} = (r_{\rm in}^2 + \Hx^2)^{1/2}$ and Eq.\,\eqref{eq:cos_theta} to
$\cos\theta_{\rm inc,in} = \Hx / d_{\rm in}$.
We adopt
$\log\Phi_0 = 24$, which ensures that the resulting $\log\Phi_{\rm H}$ radial profile
spans the full \cloudy{} grid range $[15, 21]$
(Fig.\,\ref{fig:ionflux}). 
The bowl-shaped disc tilts the surface
toward the lamp at large radii and so partly compensates for the
$1/d^2$ falloff. The extended \cloudy{} grid extends down to
$\log\Phi_{\rm H} = 15$, which sets the floor of our ionizing-flux map:
cells in deep shadow where $\log\Phi_{\rm H} < 15$ are evaluated at
this floor value.

In practice, illuminated cells at moderate radii reach
$\log\Phi_{\rm H} \approx 19$--$20$, while the radial shadow lanes
behind pillars drop by 0.5--1 dex (Fig.\,\ref{fig:ionflux}). $\log\Phi_{\rm H}$
is clipped to the \cloudy{} grid range $[15, 21]$ before being looked
up in the pre-computed emissivity tables described next
to obtain the local emissivity of each emission line.

To predict how each emission line responds to the local ionizing flux, we
pre-compute a two-dimensional grid of photoionization models using
\cloudy{}. The grid varies 
two physical parameters: the hydrogen-ionizing photon flux
$\Phi_{\rm H}$ (photons\,cm$^{-2}$\,s$^{-1}$) and the hydrogen number density $n_{\rm H}$
(cm$^{-3}$). The ionizing flux spans $\log\Phi_{\rm H} = 15$--21 in
steps of 0.5\,dex, the gas density spans
$\log n_{\rm H} = 9$--12 in steps of 0.5\,dex.
Each combination defines a single \cloudy{} model. 
All models are computed at solar metallicity. In future work we
will relax this assumption, since the pillar gas sits close to the
embedded stars that enrich the disc (see Section\,\ref{sec:background}) and may
therefore be more metal-rich than the underlying disc.

For every grid point we adopt the ionizing continuum of
\citet{nagao2006}, a broken power-law SED with a strong UV thermal
bump matching the quasar template of \citet{scott2004}.
We adopt this generic quasar SED rather than an object-specific one,
such as the NGC\,5548 SED distributed with \cloudy{}, so that the predicted
signatures are not specific to a single AGN.
We stop each calculation when the hydrogen column
density reaches $N_{\rm H} = 10^{25}$\,cm$^{-2}$ or the temperature
drops below 3000\,K, whichever occurs first.
Columns of $10^{24}$--$10^{25}$\,cm$^{-2}$ are adopted in photoionization
studies of the low-ionization lines
\citep{FerlandPersson1989, Panda2020, Sarkar2021}, and the emission-line
spectrum constrains the column only from below, since gas beyond the
ionization front contributes little to the lines. In the grid, C\,{\sc iv} and
C\,{\sc iii}] are complete within $N_{\rm H}\sim10^{22}$\,cm$^{-2}$, ahead of
the ionization front, whereas H$\beta$ and Mg\,{\sc ii} gain their last
20 per cent between $10^{23}$ and $10^{25}$\,cm$^{-2}$, so a cut at
$10^{23}$\,cm$^{-2}$ would bias the line ratios by about 0.1\,dex. What sets
the emitting layer is the density structure of the star-heated spot rather than
the width of the pillar, and we will compute it by post-processing
hydrodynamic simulations of an embedded star in future work. We include a turbulent
velocity of 100\,$\kms$, following \citet{Sarkar2021}, which broadens the
line absorption profiles, increasing escape probabilities and affecting the ionization structure of the
cloud. Each model is run in single-iteration mode, saving the emergent
line fluxes for a predefined list of broad-line region transitions
spanning a wide range of ionization potential and excitation
mechanism. In this paper we focus on three representative lines that
sample the dynamic range of the response: H$\alpha$~$\lambda 6563$
(low-ionization recombination), Mg\,{\sc ii}~$\lambda 2798$
(intermediate-ionization, sensitive to the temperature of the
partially-ionized zone), and C\,{\sc iv}~$\lambda 1549$
(high-ionization, collisionally-excited). The \cloudy{} grid itself
contains many additional BLR lines (e.g.\ Ly$\alpha$~$\lambda1216$,
N\,{\sc v}~$\lambda1240$, Si\,{\sc iv}+O\,{\sc iv}]~$\lambda1400$,
He\,{\sc ii}~$\lambda1640$, C\,{\sc iii}]~$\lambda1909$,
H$\beta$~$\lambda4861$) which the same machinery can be applied to.

Because the BLR density is not well constrained and likely varies widely across
and above the disc surface, 
we marginalize over the density axis by averaging the line
emissivities over all seven $\log n_{\rm H}$ values at each
 $\log\Phi_{\rm H}$ grid point. This reduces the  grid to  a one-dimensional function: emissivity versus
$\log\Phi_{\rm H}$ alone. We interpolate this function with a cubic
spline. At each disc surface element, the local $\log\Phi_{\rm H}$
computed from the disc geometry (Eqs.\,\ref{eq:phi_total}--\ref{eq:Q_calib}) is mapped
onto this interpolation curve to obtain the local line emissivity.

\begin{figure}
    \centering
    \includegraphics[width=0.47\textwidth]{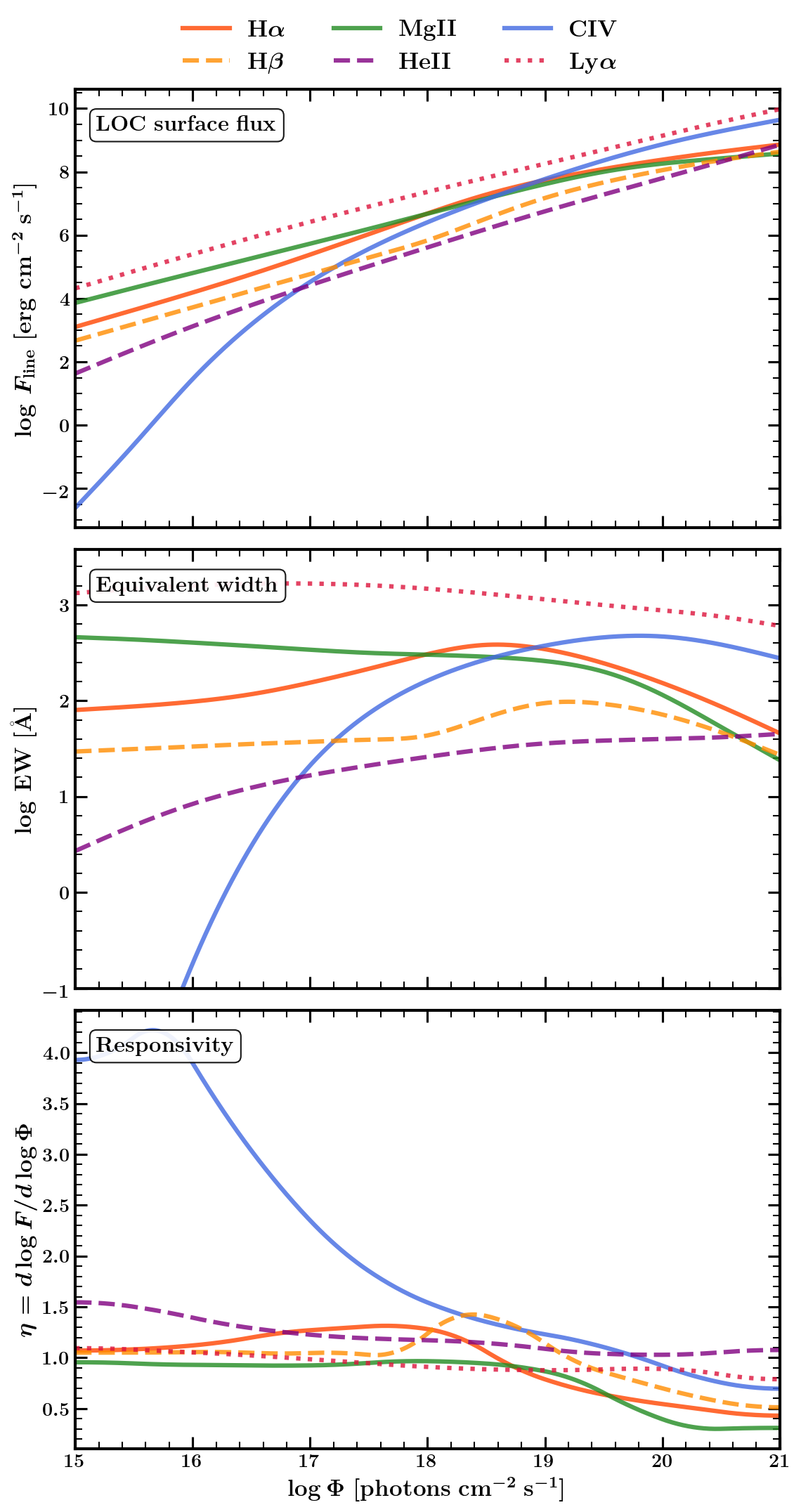}
    \caption{LOC-averaged emission-line properties as functions of ionizing photon flux $\log\Phi_{\rm H}$, evaluated at solar metallicity ($Z = Z_{\odot}$) over the extended \cloudy{} grid $\log\Phi_{\rm H} = 15$--$21$. \textbf{\textit{Top:}} surface flux $F_{\rm line}$ for the six example BLR lines. All lines brighten monotonically with $\Phi_{\rm H}$; C\,\textsc{iv} rises most steeply at low $\Phi_{\rm H}$.
    \textbf{\textit{Middle:}} equivalent width ${\rm EW} = F_{\rm line} \times 1215\,\mbox{\AA} / \nu F_{\nu}^{\rm inc}(1215\,\mbox{\AA})$, referenced to the incident continuum at $1215$\,\AA{} \citep{KoristaGoad2000}. The H$\alpha$ and H$\beta$ EWs peak near $\log\Phi_{\rm H} \sim 18.5$--$19$ and decline at higher fluxes, while the Mg\,\textsc{ii} EW declines monotonically, steepening beyond $\log\Phi_{\rm H} \sim 19$ (Baldwin effect). The C\,\textsc{iv} EW increases with flux and peaks near $\log\Phi_{\rm H} \sim 20$.
    \textbf{\textit{Bottom:}} local responsivity $\eta = d\log F_{\rm line}/d\log\Phi_{\rm H}$, the logarithmic slope of each $F_{\rm line}(\Phi_{\rm H})$ curve. C\,\textsc{iv} responds strongly superlinearly ($\eta \approx 4$) at low $\Phi_{\rm H}$ and approaches sublinear response at the highest fluxes, while the low-ionization lines respond nearly linearly at low $\Phi_{\rm H}$ and become sublinear ($\eta \approx 0.3$--$0.6$) at high $\Phi_{\rm H}$. Because the incident continuum scales linearly with $\Phi_{\rm H}$, the EW declines wherever $\eta < 1$, which produces the Baldwin trends in the middle panel.}
    \label{fig:loc_emissivity}
\end{figure}

This procedure produces a Locally Optimally-emitting Cloud (LOC)
surface flux $F_{\rm line}(\Phi_{\rm H})$ and the corresponding equivalent
width ${\rm EW}(\Phi_{\rm H}) = F_{\rm line} \times 1215\,\mbox{\AA} /
\nu F_{\nu}^{\rm inc}(1215\,\mbox{\AA})$, referenced to the incident
continuum at $1215$\,\AA{} for full source coverage
\citep{KoristaGoad2000}, each given as a
function of the local ionizing photon flux. These two quantities are
shown in Fig.\,\ref{fig:loc_emissivity} for the main BLR lines over
the extended \cloudy{} grid $\log\Phi_{\rm H} = 15$--$21$. The shapes of
the curves explain why pillar shadows affect each line differently:
removing flux at a given radius shifts that cell along the
$F_{\rm line}(\Phi_{\rm H})$ curve, and lines whose curves are steeply
varying (or change sign of slope) at the relevant $\Phi_{\rm H}$ respond
strongly, while lines with flat curves do not.

\section{Velocity--Delay Map Construction}
\label{sec:vdm}

Following \citet{Blandford1982} and \citet{Horne2004}, we relate the
emission-line flux at wavelength $\lambda$ and time $t$ to the
continuum variations $X(t)$ through the velocity--delay map
$\Psi(\lambda, \tau)$:
\begin{equation}
F(\lambda, t) - \bar{F}(\lambda) =
\int_0^\infty \Psi(\lambda, \tau)\,
\bigl[X(t - \tau) - \bar{X}\bigr]\, d\tau,
\label{eq:echo}
\end{equation}
where $\bar{F}(\lambda)$ and $\bar{X}$ are the time averages of the
line and continuum light curves, respectively.
We do not adopt a continuum light curve in this work. The maps shown are
$\Psi$ itself, the line response to a delta-function impulse in the lamp
luminosity, constructed below from the disc geometry and the \cloudy{}
response.
The velocity--delay map encodes the geometry and kinematics of the emitting
gas: the delay $\tau$ gives the excess light-travel time of the reprocessed
path (continuum source $\to$ emitting element $\to$ observer) relative to
the direct path from the continuum source to the observer, while the
wavelength axis maps the line-of-sight Doppler shift from orbital motion. For NGC\,5548,
\citet{Horne2021} recovered velocity--delay maps for multiple emission lines,
revealing signatures consistent with Keplerian disc-like kinematics.

For each disc surface element at radius $r$ and azimuth $\phi$, we compute the
light-travel time delay from the lamp-post via the disc surface to the observer
as $\tau(r, \phi) = c^{-1}[\sqrt{r^2 + \Delta h^2} + \Delta h \cos i
- r \cos\phi \sin i]$, where $\Delta h = \Hx - H(r, \phi)$, $i$ is the disc
inclination, and $c$ is the speed of light. We compute the line-of-sight
velocity from Keplerian orbital motion as
$v_{\rm los}(r, \phi) = v_{\rm K}(r) \sin i \sin\phi$, where
$v_{\rm K}(r) = \sqrt{GM_{\rm BH}/r}$ is the local Keplerian velocity and
$M_{\rm BH}$ is the mass of the central SMBH, so that the observed wavelength
becomes $\lambda_{\rm obs} = \lambda_0 (1 + v_{\rm los}/c)$.

We then construct the velocity--delay map by summing the per-cell
linear response of the line surface flux to a small fractional
perturbation in the lamp luminosity over all disc surface elements:
\begin{equation}
\Psi(\lambda, \tau) = \sum_{i,j} w_{ij} \,
F_{\rm line}(\log\Phi_{ij}) \,
\eta_{\rm line}(\log\Phi_{ij}) \,
\delta(\tau - \tau_{ij}) \, \delta(\lambda - \lambda_{{\rm obs},ij}),
\label{eq:psi_sum}
\end{equation}
where $w_{ij} = dA_{ij} \cos\theta_{ij}$ is the projected surface area
of the $(i,j)$ cell as seen by the observer (accounting for foreshortening),
$dA_{ij}$ is the physical area of the cell, and
$\cos\theta_{ij}$ is the dot product of the surface normal with the observer
direction. The \cloudy{}-derived weighting splits into two factors:
$F_{\rm line}(\log\Phi_{ij})$ is the LOC-averaged line surface
flux at the local ionizing flux (Section\,\ref{sec:cloudy_grid}), and
\begin{equation}
\eta_{\rm line}(\log\Phi_{\rm H}) \equiv \frac{{\rm d}\log F_{\rm line}}{{\rm d}\log\Phi_{\rm H}}
\label{eq:eta}
\end{equation}
is the local responsivity, the logarithmic slope of the
$F_{\rm line}(\Phi_{\rm H})$ curve, computed by analytic
differentiation of the cubic-spline interpolation of
$\log F_{\rm line}$ vs $\log\Phi_{\rm H}$
(Fig.\,\ref{fig:loc_emissivity}, bottom panel). The product
$F_{\rm line}\,\eta_{\rm line} = \partial F_{\rm line}/\partial\log\Phi_{\rm H}$
is the per-cell linear response coefficient, so that with this
weighting Eq.\,(\ref{eq:psi_sum}) is the partial-derivative response
map defined in \citet[][Eq.\,3]{Horne2004}. A global
$1/\bar{L}_{\rm ion}$ factor common to all cells is absorbed into the
overall normalization. The alternative weighting
$w \propto F_{\rm line}$ alone gives the steady-state line
surface brightness distribution $\bar{F}_l(v, \tau)$ of
\citet[][Eq\,6]{Horne2004}, which we discuss in
Appendix\,\ref{sec:appendix_emiss_weighting}.

In practice, we replace the delta functions with finite bin widths,
binning the velocity--delay map onto a grid of $n_\lambda \times n_\tau
= 200 \times 100$ pixels that spans $\pm 3 v_{\rm K}(r_0)$ in
velocity and $0 < \tau < 50$\,days in delay. We smooth each element's
contribution over neighboring bins using a Gaussian kernel with a
width of 1.5 bins.

For a pillar at position $(r_p, \phi_p)$, its azimuthal extent
$\pm 2\sigma_\phi$ maps to a range of line-of-sight velocities, producing
an arc-like feature in the velocity--delay map rather than a point.

Integrating the velocity--delay map over wavelength gives the one-dimensional
transfer function
\begin{equation}
\Psi(\tau) = \int \Psi(\lambda, \tau) \, d\lambda,
\label{eq:1d_transfer}
\end{equation}
while integrating over delay gives the emission-line response profile
$\Delta F(\lambda) = \int \Psi(\lambda, \tau) \, d\tau$. 
The mean delay spectrum is
\begin{equation}
\langle\tau\rangle(\lambda) = \frac{\int \tau \, \Psi(\lambda, \tau) \, d\tau}
{\int \Psi(\lambda, \tau) \, d\tau}.
\label{eq:mean_delay}
\end{equation}
For a smooth disc, $\langle\tau\rangle(\lambda)$ traces the characteristic
Keplerian envelope ($\tau \propto v^{-1/2}$), with maximum delay at line centre ($v = 0$) and
decreasing delay in the wings ($|v| > 0$). 


\subsection{Keplerian rotation of pillars}
\label{sec:rotation}
\label{sec:time}

Each pillar orbits at its local Keplerian velocity
$v_{\rm K}(r_p) = \sqrt{GM_{\rm BH}/r_p}$ with angular velocity
$\omega(r_p) = v_{\rm K}(r_p)/r_p$, so that the orbital period is
$T_{\rm orb} = 2\pi r_p / v_{\rm K}(r_p) = 2\pi r_p^{3/2} / \sqrt{GM_{\rm BH}}$.
For NGC\,5548 parameters ($M_{\rm BH} = 7 \times 10^7$\,$\Msun$), a pillar at
$r_p = 4$\,$\ld$ has $T_{\rm orb} \approx 2$\,years and
$v_{\rm K} \approx 9000$\,$\kms$, consistent with the barber-pole period estimated by \citet{Horne2021}.
As the disc rotates, each pillar's azimuthal position advances as
$\phi_p(t) = \phi_{p,0} + \omega(r_p) \, t$. In the $N = 100$ case, pillars
at different radii rotate at different rates (differential Keplerian
rotation), so the pillar distribution shears over time.

\subsection{Spectral time evolution}
\label{sec:spec_time}

The continuum is held steady throughout, so the time
dependence of $F(\lambda, t)$ comes from the motion of the pillars and not from
continuum variability.
At each time step $t$, we update all pillar positions, recompute the shadow
pattern and the ionizing flux map, and reconstruct the velocity--delay map. We
then collapse the map over delay to obtain the time-evolving emission-line
spectrum:
\begin{equation}
F(\lambda, t) = \int_0^{\tau_{\rm max}} \Psi(\lambda, \tau; t) \, d\tau.
\label{eq:spectrum_t}
\end{equation}
This produces a two-dimensional $(\lambda, t)$ map that shows how the line
profile varies as pillars orbit. To isolate the time-varying component,
we form the residual
$\Delta F(\lambda, t) = F(\lambda, t) - \bar{F}(\lambda)$, where
$\bar{F}(\lambda) \equiv T^{-1}\!\int_0^T F(\lambda, t')\, dt'$ is the
time average over a baseline of duration $T$.
In practice we sample $F(\lambda, t)$ at 100 epochs spaced 12\,d apart,
covering $T = 1200$\,d, about 1.5 orbits at $r_p = 4$\,\ld.
We show these residual maps in
Section\,\ref{sec:barberpole}.


\section{Results}
\label{sec:results}

For the 100-pillar configuration, the resulting ionizing flux maps are shown in Fig.\,\ref{fig:ionflux}. 
With 100 pillars, the individual shadow lanes overlap to produce a complex
patchwork of partially-shadowed regions. Approximately 60 per cent of the
disc surface in the pillar belt ($5 < r < 15\,\ld$) is shadowed to below half its unshadowed flux, but the cumulative shadowing remains spatially
incoherent: rather than a single dark sector, the illumination is
broken into many narrow radial shadow lanes. 

\subsection{Emission-line maps}
\label{sec:emission_maps}

\begin{figure}
    \centering
    \includegraphics[width=\columnwidth]{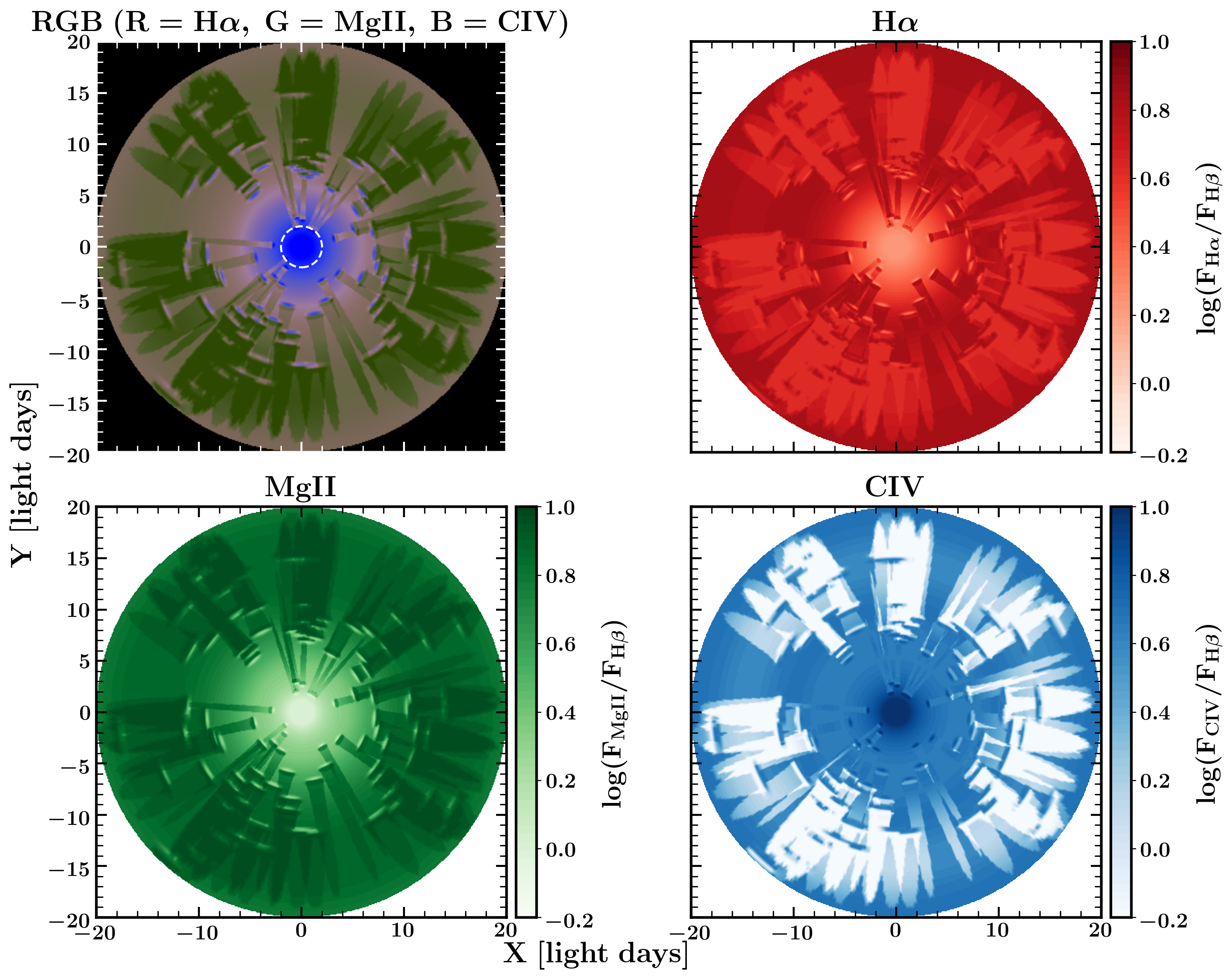}
    \caption{RGB composite of disc-surface emission-line map for 100
    partially-transparent pillars ($f_{\rm trans} = 0.3$, see
    Appendix\,\ref{sec:transparent_pillars} for a detailed discussion on
    the transmission of pillars). \textbf{\textit{Top-left:}}
    composite with R = H$\alpha$, G = Mg\,{\sc ii}, B = C\,{\sc iv},
    formed from the $F_{\rm line}/F_{\rm H\beta}$ ratios. The white
    dashed circle marks the minimum allowed pillar radius
    $r_{\rm min} = 2\,\ld$. \textbf{\textit{Top-right:}} H$\alpha$.
    \textbf{\textit{Bottom-left:}} Mg\,{\sc ii}.
    \textbf{\textit{Bottom-right:}} C\,{\sc iv}. Each shown as
    $\log(F_{\rm line}/F_{\rm H\beta})$ on the same colour scale.}
    \label{fig:rgb}
\end{figure}

In Fig.\,\ref{fig:rgb} we show the spatial distribution of emission-line
intensity across the disc surface for the 100-pillar case, rendered as an RGB
composite with R = H$\alpha$, G = Mg\,{\sc ii}, and B = C\,{\sc iv}. Each
pixel is assigned an emissivity from the  LOC integral of the \cloudy{} grid \citep[][see also
\citealt{Krolik1981} for the original two-phase BLR
formulation]{Ferland2017,Gunasekera2023}, evaluated at the local ionizing
flux $\Phi_{\rm H}(r,\phi)$ from Section\,\ref{sec:ionflux}. The local emissivity
is taken from the LOC-averaged \cloudy{} tables, in which the density
axis has been marginalized following the 
LOC prescription \citep{Baldwin1995,KoristaGoad2000,KoristaGoad2001,KoristaGoad2004};
no explicit radial dependence in $n_{\rm H}$ is imposed.

The \cloudy{} models reveal significant differences in the shadow
responses of the three lines (Fig.\,\ref{fig:loc_emissivity}).
H$\alpha$ and Mg\,{\sc ii} are predominantly powered by recombination following
photoexcitation from the diffuse radiation field: when the
ionizing flux drops in a shadow region (low $\log\Phi_{\rm H}$), the
recombination rate per ion increases because a larger fraction of
hydrogen remains neutral and the ionization front recedes, producing
a denser partially-ionized zone where recombination emission is
efficient. 
The incident continuum scales linearly with $\Phi_{\rm H}$, whereas
$F_{\rm line}$ for H$\alpha$ and Mg\,{\sc ii} grows sub-linearly at
high fluxes ($\eta < 1$; bottom panel of
Fig.\,\ref{fig:loc_emissivity}). The combination yields a Baldwin-like
trend in EW. In the inner disc, where the unshadowed flux reaches
$\log\Phi_{\rm H} \sim 21$, the partially transparent pillars of our fiducial
model ($f_{\rm trans} = 0.3$) lower the local ionizing flux by 0.5--1\,dex in
their shadows, and the EWs of Mg\,{\sc ii} and H$\alpha$ rise by
0.3--0.7\,dex (middle panel of Fig.\,\ref{fig:loc_emissivity}).
C\,{\sc iv}, on the
other hand, is a collisionally-excited high-ionization line that requires
temperatures $T \gtrsim 10^4$\,K and a substantial C$^{3+}$ ion
fraction. A modest reduction in
ionizing flux lowers the C$^{3+}$ abundance only slightly, so the
C\,{\sc iv} EW changes by $\lesssim 0.25$\,dex across the
$\log\Phi_{\rm H} \sim 18$--$21$ range. In deeper shadows
($\log\Phi_{\rm H} \lesssim 18$), however, the three lines diverge:
the C\,{\sc iv} EW collapses with the C$^{3+}$ fraction,
Mg\,{\sc ii} continues to rise gently, and H$\alpha$ turns over and
declines slowly from its peak (Fig.\,\ref{fig:loc_emissivity}).

The inner $r \lesssim 3\,\ld$ appears blue, where the high local $\Phi_{\rm H}$
drives C\,{\sc iv}/H$\beta$ above the other two ratios. The bulk of the
disc at $r \gtrsim 5\,\ld$ is olive-green: at moderate $\Phi_{\rm H}$,
Mg\,{\sc ii}/H$\beta$ and H$\alpha$/H$\beta$ are both substantial while
C\,{\sc iv}/H$\beta$ is suppressed. Shadow lanes radiating outward from
each pillar appear as darker green streaks, where the suppressed $\Phi_{\rm H}$
both dims the local emission and shifts the ratio toward an even
stronger Mg\,{\sc ii} dominance as C\,{\sc iv} collapses. The three
single-line panels show this directly: C\,{\sc iv}/H$\beta$ drops
sharply inside the shadows, while H$\alpha$/H$\beta$ and
Mg\,{\sc ii}/H$\beta$ stay comparatively flat. 

This differential response is the key observational diagnostic of
pillar shadows. Relative to the local continuum, shadows raise the
equivalent widths of H$\alpha$ and Mg\,{\sc ii}, while the C\,{\sc iv}
equivalent width changes little for modest flux reductions and collapses
only in deep shadows. In the velocity--delay maps and time-evolving
spectra this appears as opposite shifts of Mg\,{\sc ii}/H$\beta$ and
C\,{\sc iv}/H$\beta$ as the pillars orbit.

\subsection{Velocity--delay maps}
\label{sec:vdm_results}

\begin{figure}
    \centering
    \includegraphics[width=\columnwidth]{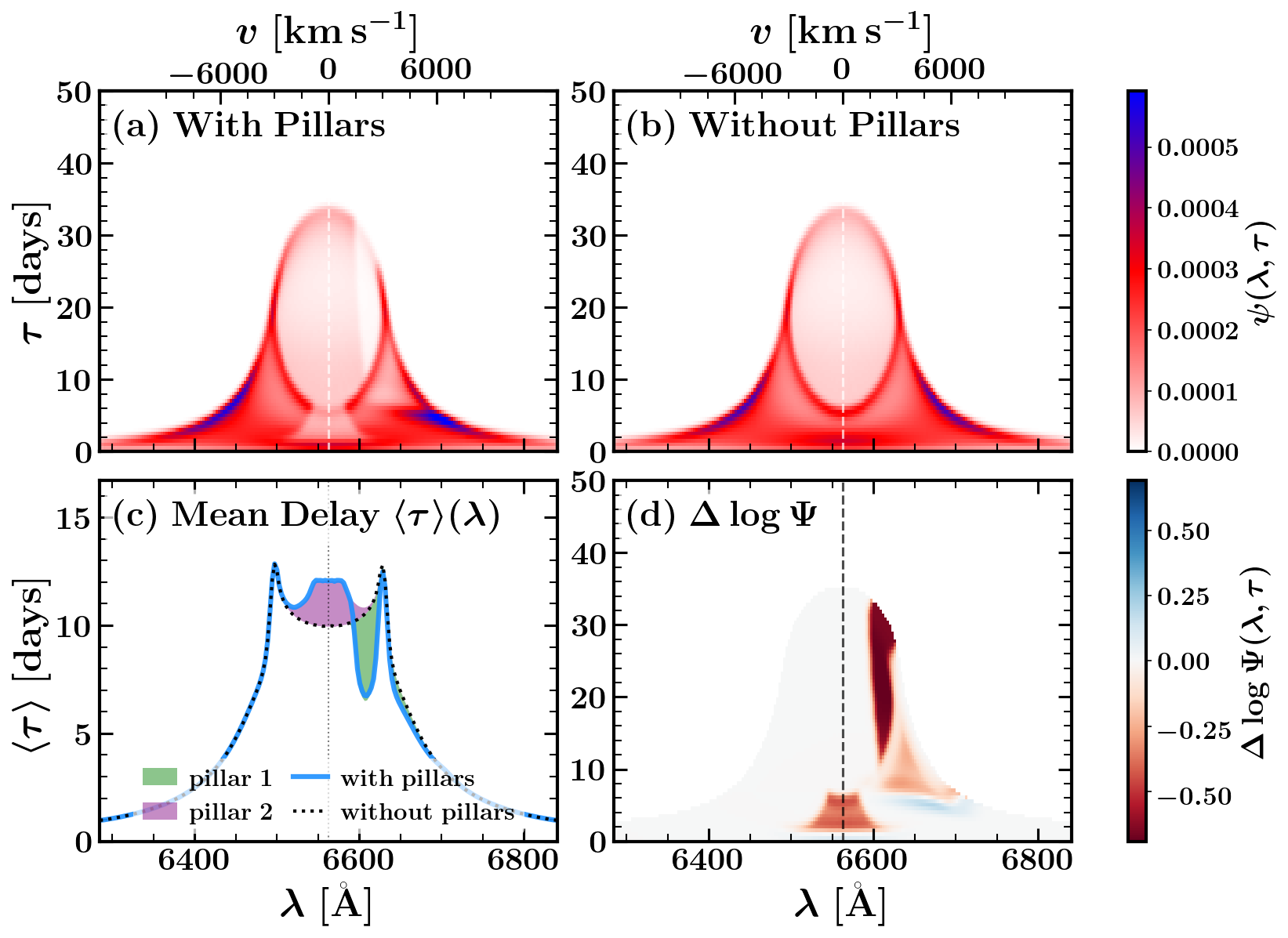}
    \caption{Velocity--delay map for the two-pillar model (H$\alpha$),
    with both pillars at $r_p = 4$\,$\ld$ and azimuths
    $\phi_p = 3\pi/4$ and $\phi_p = 0$. The normalized responsivity maps
    in (a) and (b) are
    $\psi(\lambda, \tau) = \Psi(\lambda, \tau)/\!\int\!\!\int
    \Psi\,d\lambda\,d\tau$.
    \textbf{\textit{(a):}} Map with pillars, showing the Keplerian
    envelope plus localized hotspots from the two pillars.
    \textbf{\textit{(b):}} Bowl-shaped disc (no pillars).
    \textbf{\textit{(c):}} Mean delay spectrum
    $\langle\tau\rangle(\lambda)$ with pillars (blue solid)
    and without (black dotted). The
    shaded regions show the contribution of each
    pillar, filling the gap between the no-pillar curve and the mean delay of the
    disc with that pillar alone (green for the pillar at $\phi_p = 3\pi/4$, purple
    for the pillar at $\phi_p = 0$).
    \textbf{\textit{(d):}} Log-ratio map
    $\Delta\log\Psi = \log_{10}(\Psi_{\rm with}/\Psi_{\rm without})$,
    using the absolute (un-normalized) response functions and computed
    only where the bare-bowl baseline $\Psi_{\rm without}$ is non-zero.
    Positive values (blue) mark cells where the pillars boost the
    response; negative (red) where they suppress it.}
    \label{fig:vdm_1}
\end{figure}

To show how a pillar's position maps onto the velocity--delay plane, we use
two pillars at the same radius, $r_p = 4\,\ld$, each with the same height and
widths as the single pillar of Section\,\ref{sec:pillar_dist}. One sits at
$\phi_p = 0$ (i.e., on the near side), where the line-of-sight velocity
vanishes and the delay is shortest, and the other at $\phi_p = 3\pi/4$, where
the projected Keplerian velocity is large.
We present the velocity--delay map for this two-pillar case
in Fig.\,\ref{fig:vdm_1}, computed for H$\alpha$. The two pillars map to two distinct
hotspots in the $(v, \tau)$ plane. 
The smooth-disc component
produces the characteristic Keplerian signature first derived by
\citet{Blandford1982}.
As discussed by \citet{Welsh1991},
each annulus on the inclined Keplerian disc corresponds to an ellipse  
spanning the delay range $\tau=(r/c)(1\pm\sin{i})$ and 
projected Kepler velocity range $v=\pm v_{\rm K}(r)\,\sin{i}$.
Most prominent in Fig.\,\ref{fig:vdm_1} is the ellipse spanning 6 to 34 days, which corresponds to the outer annulus at $r_{\rm out}=20$\,\ld\ and determines the maximum delay
$\tau_{\rm max} \approx (r_{\rm out}/c)(1+\sin i) \simeq 34$\,d at line centre.

Annuli inside $r \approx 2\,\ld$ fill the shortest delays, and their
projected velocities extend beyond the plotted velocity range.
The virial envelope with 
$(v\,\sin{i})^2\,(\tau\,c)\approx G\,M_{\rm BH}$
is a caustic pattern tangent to the ellipses from intermediate annuli,
connecting the inner and outer ellipses.

The pillar at $\phi_p = 3\pi/4$ appears as a compact ``hotspot'' at
$(v,\tau)
\approx (+4700\,{\rm km\,s^{-1}},\, 6\,{\rm d})$, on the red side of the line
because this pillar is receding, while the
pillar at $\phi_p = 0$ produces a second hotspot at
$v \approx 0$ and a much shorter delay,
$\tau \approx 1.2$\,d, because it lies on the near side (the line-of-sight velocity
vanishes for that azimuth and only the light-travel offset
contributes). Each hotspot has extent
$\Delta v \approx 2\sigma_\phi v_{\rm K}(r_p)\sin i\,|\cos\phi_p|$ in velocity and
$\Delta\tau \sim \sigma_r(1-\sin i)/c$ in delay, i.e. 2000--2700
km\,s$^{-1}$ wide and a fraction of a day deep. Their locations are
set by the projected Keplerian velocity
$v = +v_{\rm K}(r_p)\sin\phi_p\sin i$,
with positive velocities receding, and light-travel delay
$\tau = (r_p/c)(1-\cos\phi_p\sin i)$, giving a direct geometric
mapping from $(r_p,\phi_p)$ to $(v,\tau)$ \citep{Horne2004}.
Each map is a snapshot with the pillars held fixed. This is a good
approximation because a pillar at $r_p = 4\,\ld$ advances by less than
$3^\circ$ in azimuth during the $\sim 6$\,d delay of its own response, much
less than its azimuthal width $2\sigma_\phi \approx 23^\circ$.
Panel (c) shows the same structure in the mean delay spectrum. The
pillars add localized perturbations to $\langle\tau\rangle(\lambda)$ at the
velocities of the two hotspots, displacing it from the smooth Keplerian
envelope of the bare bowl.

Panel (d) of Fig.\,\ref{fig:vdm_1} isolates the pillar contribution
and exhibits a positive excess on
the lamp-facing face of each pillar and a deficit along its shadow lane, which
lies at larger radii and hence at longer delays and lower velocities. The amplitude
of the residual reaches $+58$ per cent on the face of the
pillar at $\phi_p = 3\pi/4$ (about $+5000\,{\rm km\,s^{-1}}$ and 5\,d) and
$-80$ per cent along its shadow lane (about $+2000\,{\rm km\,s^{-1}}$ and
23\,d), well above the photon-noise
floor of high-fidelity reverberation experiments
\citep[$\sim 1$--5 per cent; see][]{Horne2021,Williams2021}.

\begin{figure*}
    \centering
    \includegraphics[width=\textwidth]{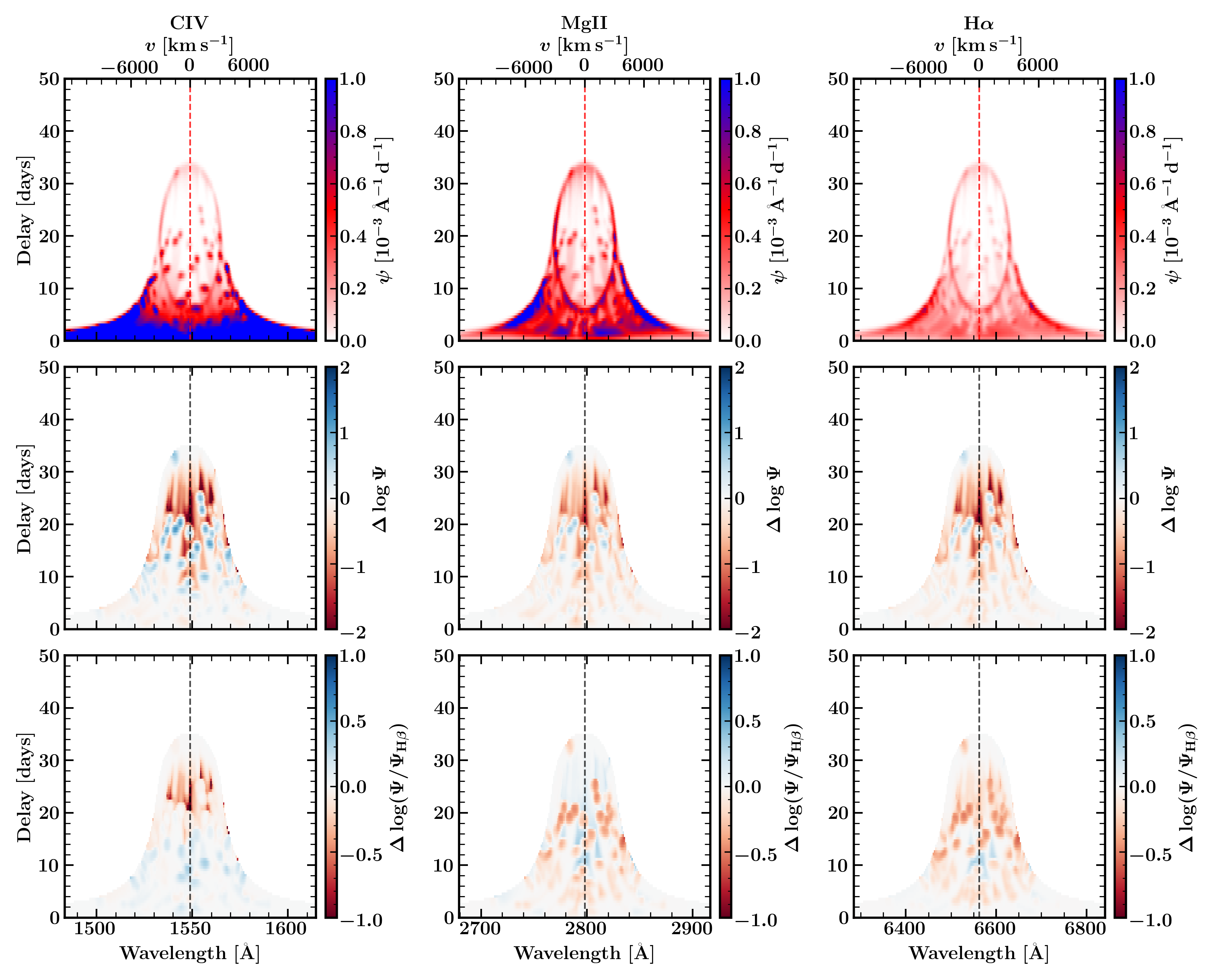}
    \caption{Velocity--delay map for the 100-pillar configuration.
    Each column shows one emission line (left to right, in order of
    increasing rest wavelength: C\,{\sc iv}, Mg\,{\sc ii}, H$\alpha$);
    each row shows one specific quantity built from the same set of
    \cloudy{}-weighted response maps $\Psi(\lambda, \tau)$, evaluated
    once with the pillars present and once for the bare bowl baseline.
    Throughout, $\Delta\log X \equiv \log_{10} X_{\rm with} -
    \log_{10} X_{\rm without}$ is the dex shift in the quantity $X$
    produced by adding the pillars.
    \textbf{\textit{Top row:}} normalized responsivity
    $\psi(\lambda, \tau) = \Psi(\lambda, \tau)/\!\int\!\!\int \Psi\,d\lambda\,d\tau$
    of the with-pillars response, normalized per line so each panel
    integrates to unity.
    \textbf{\textit{Middle row:}}
    $\Delta\log\Psi(\lambda, \tau) =
    \log_{10}\bigl[\Psi_{\rm with}(\lambda, \tau)\,/\,
    \Psi_{\rm without}(\lambda, \tau)\bigr]$,
    the dex shift in the line's own response under the pillar
    perturbation. Red means the line response drops in that
    $(\lambda, \tau)$ cell (shadow suppression); blue means it rises
    (illuminated pillar face).
    \textbf{\textit{Bottom row:}}
    $\Delta\log\!\bigl(\Psi\,/\,\Psi_{\rm H\beta}\bigr)$
    --- the dex shift in the line-to-H$\beta$ ratio. The H$\beta$ map
    $\Psi_{\rm H\beta}(\lambda, \tau)$ is computed on the same
    $(v, \tau)$ grid as the line maps; positive (blue) values mean the
    line strengthens relative to recombination in that cell
    (Mg\,{\sc ii} in shadows), while negative (red) values mean it
    collapses faster than H$\beta$ does (C\,{\sc iv} in shadows).
    This row isolates the 
    line-dependent
    differential shadow response that is
    a diagnostic signature of the pillar geometry.
    }
    \label{fig:movie_frame}
\end{figure*}

Figure\,\ref{fig:movie_frame} extends the velocity--delay analysis to
three lines spanning a wide range of ionization potential ---
C\,{\sc iv}, Mg\,{\sc ii}, and H$\alpha$ --- in the 100-pillar
configuration, with each column showing one line and each row a
different quantity built from the same \cloudy{}-weighted responses
$\Psi(\lambda, \tau)$ evaluated with and without the pillars. The
upper panels confirm that all three responses share the same
Keplerian envelope expected from a thin disc, with
pillar-induced fine structure concentrated where the pillars
themselves sit ($\tau \sim 10$--$30\,$d and
$|v| \lesssim 4000\,{\rm km\,s^{-1}}$); C\,{\sc iv}, whose emissivity
rises steeply with $\log\Phi_{\rm H}$, is the most centrally
concentrated and reaches the highest velocities, in line with its
preference for small radii.
The hotspots in Fig.\,\ref{fig:movie_frame} are much smaller. The
$N = 100$ pillars are four times narrower in azimuth ($\sigma_\phi = 0.05$
against 0.2\,rad) and mostly lie at larger radii, where the Keplerian speed is
lower, so by the same relation each spans $\lesssim 400\,{\rm km\,s^{-1}}$, a
few pixels of the map, and appears as a compact spot.

In absolute terms, however, the three lines respond to the pillars
in essentially the same way. Where a pillar's illuminated face faces
the lamp post, the local ionizing flux is enhanced and the
\cloudy{}-weighted emissivity rises in every line; where the pillar
casts a shadow on the disc behind it, the ionizing flux drops by
0.5--1\,dex and the line response drops by up to 1--2\,dex. This is what
the middle row of Fig.\,\ref{fig:movie_frame} makes explicit: the
dex shift in each line's own response is dominated by the shadow
geometry itself, and the patterns in the C\,{\sc iv}, Mg\,{\sc ii},
and H$\alpha$ panels look almost identical, with strong negative
dex shifts in shadowed cells (red) and positive enhancements on
illuminated faces (blue).

The lines diverge once they are referenced to H$\beta$. As a Balmer
recombination line, H$\beta$ tracks the ionizing photon budget
approximately linearly and largely cancels out the bulk illumination
response when one forms a ratio. The bottom row of
Fig.\,\ref{fig:movie_frame} plots
$\Delta\log\bigl(\Psi_{\rm line}\,/\,\Psi_{\rm H\beta}\bigr)$ ---
the dex shift in the line-to-H$\beta$ ratio between the with-pillars
and bare-bowl maps --- and this is where the 
line-dependent photo-ionization physics
of the pillars shows up. On the illuminated face,
C\,{\sc iv}/H$\beta$ rises (blue, $\lesssim 0.2$\,dex) because the locally enhanced
ionizing flux pushes the emitting gas onto the steep,
high-$\log\Phi_{\rm H}$ part of the C\,{\sc iv} emissivity surface,
where C\,{\sc iv} grows faster than H$\beta$; in the shadow behind
the pillar the same ratio drops (red) because the collapse of
$\Phi_{\rm H}$ erodes C\,{\sc iv} more sharply than it does
H$\beta$. Mg\,{\sc ii} shows the opposite pattern: the illuminated
face is suppressed in Mg\,{\sc ii}/H$\beta$ (red) because at high
ionization parameter the Mg$^{+}$ population is depleted by further
ionization to Mg\,{\sc iii}, while in the shadow the lower ionization
parameter favours Mg$^{+}$ and the ratio brightens (blue). H$\alpha$ behaves like Mg\,{\sc ii} on the illuminated faces,
falling by 0.2--0.5\,dex relative to H$\beta$, and changes by less than
0.2\,dex in the shadows. The take-away message is that absolute responses cannot
distinguish lines of different ionization stages --- they merely
trace the shadow geometry --- whereas the H$\beta$-normalized ratios
encode the local ionization parameter, and pillars therefore drive
anti-correlated shifts in C\,{\sc iv}/H$\beta$ and
Mg\,{\sc ii}/H$\beta$ in the velocity--delay regions occupied by
the orbiting bumps.

\subsection{Barber-pole patterns in time-evolving spectra}
\label{sec:barberpole}

The most distinctive signature of non-axisymmetric pillars is the
time-evolving emission-line profile. As a pillar orbits at the Keplerian
angular speed
$\Omega_K(r_p) = (G M_{\rm BH}/r_p^3)^{1/2}$, its line-of-sight velocity
$v_{\rm los}(t) = +v_{\rm K}(r_p)\sin[\phi_p+\Omega_K(r_p)t]\sin i$ varies
sinusoidally, tracing a diagonal stripe in the $(\lambda, t)$ plane whose
slope $d\lambda/dt = +\lambda_0\,v_{\rm K}\Omega_K\cos\phi(t)\sin i / c$
locally measures the projected acceleration. When the pillar's shadow
simultaneously modulates the emission from surrounding gas, the combined
effect produces the characteristic ``barber-pole'' pattern, in which a
single localized perturbation maps onto a sinusoidal stripe whose period,
amplitude, and slope are jointly diagnostic of $r_p$, $\phi_p$, and
$M_{\rm BH}$.

\begin{figure*}
    \centering
    \includegraphics[width=\textwidth]{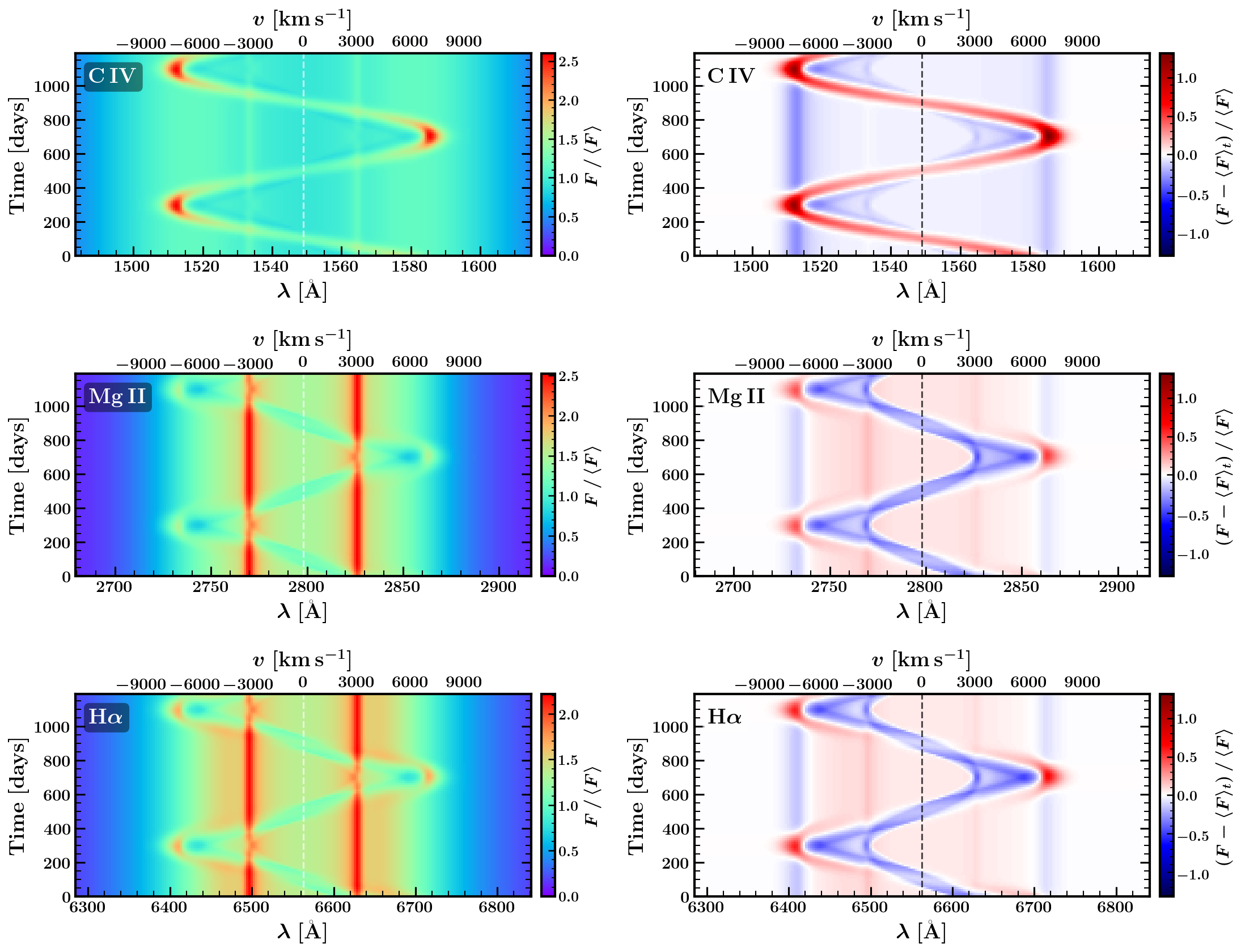}
    \caption{Time-evolving emission-line spectra for a single pillar.
    \textbf{\textit{Left column:}} normalized flux
    $F(\lambda, t)/\langle F\rangle$, where
    $\langle F\rangle \equiv \langle F(\lambda, t)\rangle_{\lambda, t}$
    is the disc-integrated mean (a single scalar per line) computed
    over both the wavelength axis and the time axis.
    \textbf{\textit{Right column:}} residual
    $(F(\lambda, t) - \langle F(\lambda)\rangle_t)/\langle F\rangle$,
    where $\langle F(\lambda)\rangle_t$ is the time-averaged spectrum
    at each wavelength. Rows from top to bottom: C\,{\sc iv},
    Mg\,{\sc ii}, H$\alpha$ (in order of increasing rest wavelength,
    indicated in each panel). Diagonal stripes in the residual trace
    the pillar's Keplerian orbit. C\,{\sc iv} is enhanced along the orbit, while
    Mg\,{\sc ii} and H$\alpha$ are suppressed except at the largest $|v|$.
    The residual panels share one colour scale. The orbital period at $r_p = 4$\,$\ld$ for
    $M_{\rm BH} = 7 \times 10^7$\,$\Msun$ is
    $T_{\rm orb} \approx 800$\,d.}
    \label{fig:velocity_time_1}
\end{figure*}

\begin{figure*}
    \centering
    \includegraphics[width=\textwidth]{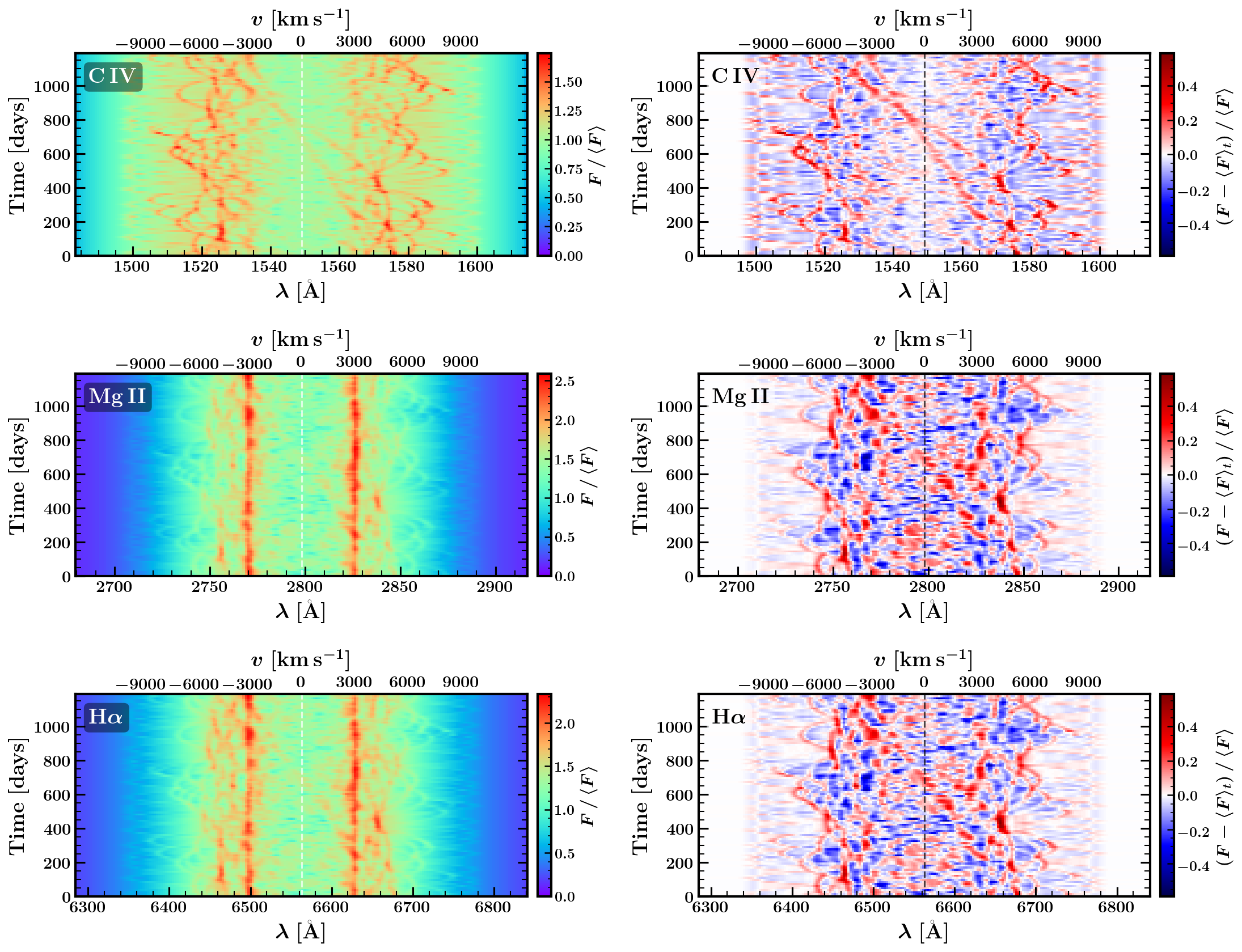}
    \caption{Same as Fig.\,\ref{fig:velocity_time_1} but for the
    100-pillar configuration. The superposition of many pillars with
    different orbital periods creates a complex modulation pattern:
    coherent red/blue stripes from the dominant pillar population
    near $r_{\rm mean} = 10$\,$\ld$ are visible in all three residual columns.}
    \label{fig:velocity_time_N100}
\end{figure*}

We show the $(\lambda, t)$ maps for the single-pillar case in
Fig.\,\ref{fig:velocity_time_1}, with the normalized flux
$F(\lambda, t)/\langle F\rangle$ in the left column and the residual
$(F - \langle F\rangle_t)/\langle F\rangle$ in the right column,
arranged from top to bottom for C\,{\sc iv}, Mg\,{\sc ii}, and
H$\alpha$. Throughout this section
(Figs.\,\ref{fig:velocity_time_1} and \ref{fig:velocity_time_N100}) we
adopt  $i = 45^\circ$, as in the rest of the paper.
Near edge-on inclinations would expose the foreshortening modulation and the
geometric self-occultation of the pillar's lamp-facing inner slope more strongly
(Appendix\,\ref{sec:inclination_appendix}, Fig.\,\ref{fig:inclination_civ}), but
at such inclinations the broad-line region is expected to be hidden by the
dusty torus. The residual makes the orbital trail most visible: the
residuals of all three lines show diagonal stripes,
alternating between enhanced (red) and suppressed (blue) emission as
the pillar moves between the near and far sides of the disc. The peak
velocity excursion $|v_{\rm K}(r_p)\sin i| \approx 6700\,{\rm
km\,s^{-1}}$ at $r_p = 4\,\ld$ matches the inferred half-width of the
line wings in the residual, providing an internal consistency check on
the Keplerian interpretation.  Crucially, the
relative amplitudes of the H$\alpha$/Mg\,{\sc ii}/C\,{\sc iv}
residuals are not arbitrary but track the differential
ionization response built into the LOC integration; this colour
pattern is in principle separable from purely kinematic features such
as those produced by orbiting BLR clouds
\citep[e.g.,][]{Alexander1997} or disc-wind streamlines
\citep{Murray1995,Chiang1996}, which would modulate all three lines
by similar factors.

The physical origin of this near-side/far-side asymmetry is the shadow
visibility effect. When the pillar is on the near side of the disc
($\phi \approx 0$, $\cos\phi > 0$), the observer looks at the shadow side
and the extended shadow lane behind the pillar is visible, so the reduced
ionizing flux enhances low-ionization line emission. When the pillar is on
the far side ($\phi \approx \pi$, $\cos\phi < 0$), the observer sees the
illuminated face and the shadow is hidden behind the pillar. This
foreshortening effect produces a half-period asymmetry in the residual
time series: enhanced low-ionization emission for $\phi \in (-\pi/2,
\pi/2)$ (near-side phases) and suppressed emission on the opposite half
of the orbit. The same effect produces a phase-dependent modulation of
the line ratio Mg\,{\sc ii}/C\,{\sc iv} that is in principle observable
even in low-resolution spectra, although its detection requires careful
control of the underlying continuum reverberation signal.

The amplitude of this once-per-orbit modulation depends sensitively on
disc inclination: at face-on geometry ($i = 0^\circ$) the observer's
line of sight aligns with the disc normal everywhere, so near and far
sides of the pillar appear identical and the modulation vanishes; as
$i$ increases the foreshortening dot product $\hat{\bf n}\cdot
\hat{\bf o}$ on the pillar's lamp-facing inner slope picks up a
$\sin i \cdot \cos\phi_p$ term that grows with $\sin i$, until at
near edge-on inclination the pillar's own body geometrically occults
its inner face on the near side and the contrast becomes near-binary.
At Type 1 inclinations such as the $i = 45^\circ$ adopted here, the
barber-pole trail persists, with a velocity amplitude smaller in proportion
to $\sin i$ and a weaker near-side/far-side asymmetry.
This trend is illustrated in Appendix\,\ref{sec:inclination_appendix}
(Fig.\,\ref{fig:inclination_civ}), where the C\,{\sc iv} velocity-time
map for a single pillar is shown for $i = 30^\circ, 45^\circ$, and
$80^\circ$.

The 100-pillar case is shown in Fig.\,\ref{fig:velocity_time_N100},
again with the normalized flux on the left and the residual on
the right. Unlike the clean single-pillar case, the superposition of
100 pillars at different radii produces a more complex modulation
pattern. The orbital periods scale as
$T_{\rm orb} \propto r_p^{3/2}$ and range from $\sim$ 3\,yr at
$r_p = 5\,\ld$ to $\sim$ 24\,yr at $r_p = 20\,\ld$ for our adopted
$M_{\rm BH} = 7\times 10^7\,\Msun$. The striping fragments into a salt-and-pepper
texture in which short-period inner pillars beat against
longer-period outer pillars, and partial phase-cancellation between
pillars on the near and far sides washes out the cleanest stripes.
Nevertheless, the collective modulation  remains clear in all three lines, with residuals of up to $\sim 0.5\,\langle F\rangle$.   We note that similar collective modulations are
predicted in scenarios involving self-gravitating clumps or disc-warp
precession \citep[e.g.,][]{Pringle1996,Sarkar2021}, but the
line-dependent shadow signature distinguishes the pillar mechanism
from these structurally-degenerate alternatives.

\section{Continuum Reverberation Signatures of Pillars}
\label{sec:continuum_lags}

The preceding sections have demonstrated how pillars imprint a
distinctive signature on the broad-line velocity--delay maps
(Section\,\ref{sec:vdm_results}) and on the time-evolving emission-line
spectra (Section\,\ref{sec:barberpole}), with the differential
ionization response of high- and low-ionization lines producing
line-dependent ``barber-pole'' modulations as the pillars orbit. We
now turn to a complementary continuum diagnostic and ask: if pillars
are present, how do they perturb the disc surface temperature profile
$T_{\rm eff}(r,\phi)$ and, in turn, the wavelength-dependent continuum
reverberation lag spectrum $\langle\tau\rangle(\lambda)$ measured in
inter-band lag campaigns \citep{Starkey2023,DeRosa2015,Cackett2021}? The
continuum signal depends only on the two-dimensional temperature
distribution and the geometric light-travel structure of the disc, and
so provides a self-contained probe of pillar geometry that does not
require photoionization equilibrium to be assumed. In this context the
pillars are a non-axisymmetric generalization of the rim and ripples that
\citet{Starkey2023} placed on the bowl-shaped disc to explain the long
continuum lags, and we keep their bowl as the baseline so that the two cases
can be compared directly. As we show below,
pillars systematically steepen the azimuthally-averaged $T(r)$ profile
in the outer disc, driving it from the canonical $T \propto r^{-3/4}$
slope toward $T \propto r^{-7/8}$. This is in the same direction as the
observational result of \citet{Weaver2022}, who measured a quasar
variability spectral index $\alpha = 0.71 \pm 0.02$ that is in tension
with a strict $T \propto r^{-3/4}$ disc and consistent with a steeper
outer-disc temperature gradient.

\subsection{Effective surface temperature}
\label{sec:T_eff}

The temperature at each disc surface element $(r,\phi)$ has two
contributions: viscous internal heating from accretion, and external
irradiation from the lamp post (with absorption by intervening
pillars). We sum them in quadrature \citep{pringle1981},
\begin{equation}
T_{\rm eff}^4(r,\phi) = T_{\rm visc}^4(r) + T_{\rm irr}^4(r,\phi),
\label{eq:teff_pillar}
\end{equation}
which is appropriate when the surface re-emits as a blackbody and the
two heating sources deposit their energy at distinct depths within the
optically-thick layer.

The viscous profile follows the \citet{shakura1973} thin-disc
solution with a no-torque inner boundary condition,
\begin{equation}
T_{\rm visc}(r) = T_{\rm visc}(r_0)\,
\left(\frac{r}{r_0}\right)^{-3/4}\,
\left(\frac{r - r_{\rm ISCO}}{r}\right)^{1/2}.
\label{eq:tvisc_pillar}
\end{equation}
Here $T_{\rm visc}(r_0)$ denotes the normalization
in the limit $r \gg r_{\rm ISCO}$, i.e. the value the profile would
take at $r = r_0$ if the no-torque factor were unity (its actual
value at $r_0$ is reduced by the small factor
$[(r_0 - r_{\rm ISCO})/r_0]^{1/2}$, which is $\approx$ 0.9998 for our
parameters). At $r \gg r_{\rm ISCO}$ the no-torque factor approaches
unity and the profile asymptotes to $T_{\rm visc} \propto r^{-3/4}$,
while at $r \to r_{\rm ISCO}$ the bracketed factor drives
$T_{\rm visc} \to 0$. We adopt $r_{\rm ISCO} = 2 r_{\rm g}$ corresponding
to a prograde Kerr black hole and use $r_0 = r_{\rm out} = 20\,\ld$
with $T_{\rm visc}(r_0) \approx 450$\,K, set so that
$T_{\rm visc}(1\,{\rm light\mbox{-}day}) \simeq 4\times10^3$\,K.

The irradiation term is computed by ray-tracing from the lamp post at
$(0,0,h_{\rm LP})$ to each surface element at $(r\cos\phi,\,r\sin\phi,\,
H(r,\phi))$. The absorbed flux is
\begin{equation}
F_{\rm irr}(r,\phi) =
\frac{L_{\rm X}}{4\pi\,d_{\rm LP}^2(r,\phi)}\;
\mu(r,\phi)\;\mathcal{S}(r,\phi),
\label{eq:firr_pillar}
\end{equation}
where $d_{\rm LP}$ is the lamp-post-to-element distance, $\mu$ is the
cosine of the angle between the local surface normal and the
lamp-to-element direction, and $\mathcal{S}\in[0,1]$ is the binary
shadow mask described in Section\,\ref{sec:shadows}; $T_{\rm irr}$ then
follows from the Stefan--Boltzmann law $\sigma_{\rm SB}T_{\rm irr}^4 =
F_{\rm irr}$. Pillar bumps therefore enter $T_{\rm eff}$ through three
distinct routes: (i)~by raising $H(r,\phi)$ at the pillar location,
modifying $\mu$ and producing a local enhancement on the lamp-facing
pillar surface; (ii)~by occluding the lamp-post pencil beam, suppressing
$T_{\rm irr}$ along the radial shadow lane; and (iii)~by altering
$d_{\rm LP}$ at the pillar top. The shadow effect dominates the
azimuthal averages discussed below.

\begin{figure*}
    \centering
    \includegraphics[width=\textwidth]{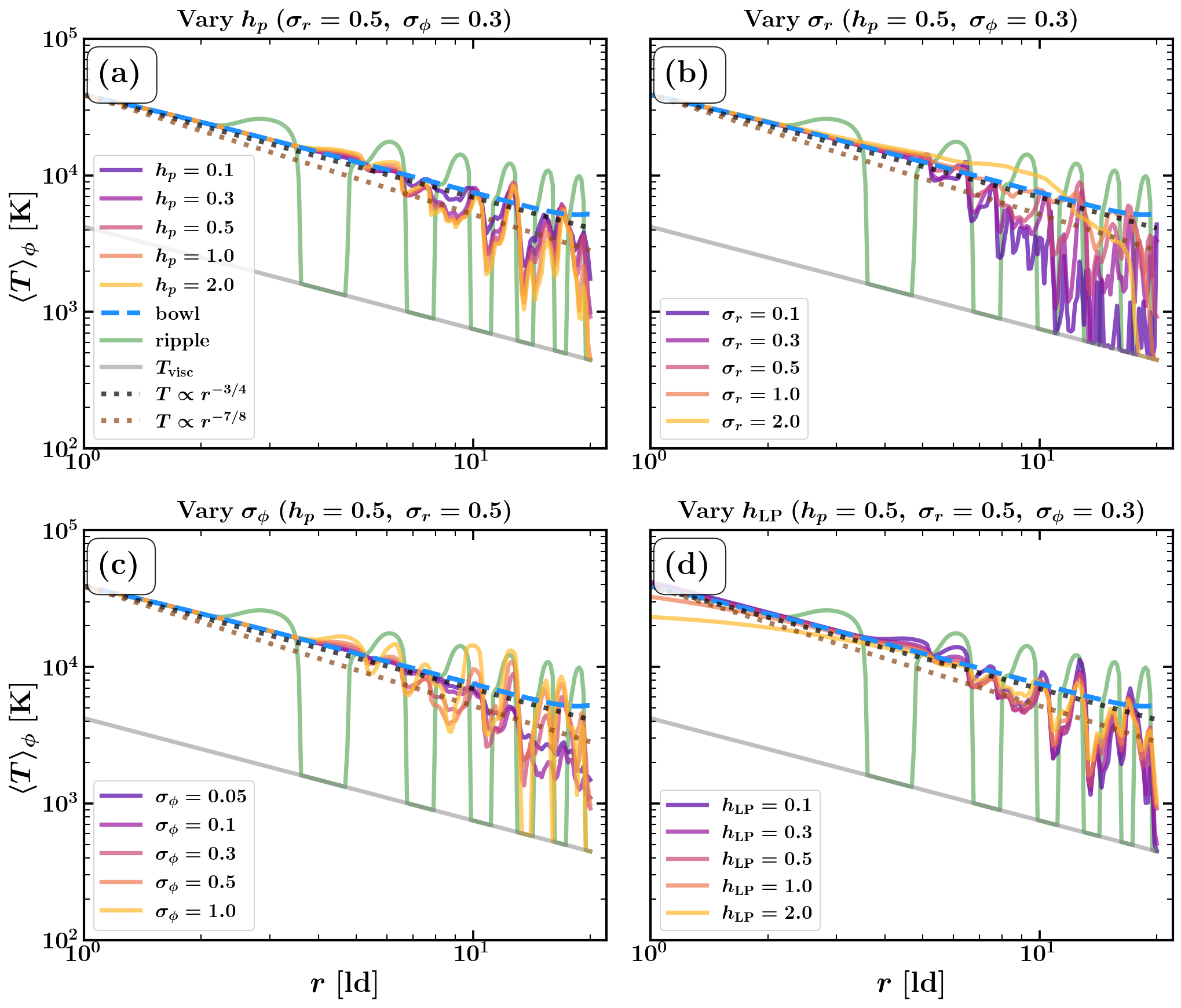}
    \caption{Azimuthally-averaged surface temperature profile
    $\langle T_{\rm eff}\rangle_\phi(r)$ for the bowl baseline (blue
    dashed) and four parameter scans, each showing five values
    of one parameter:
    \textbf{\textit{(a)}} pillar height $h_p$;
    \textbf{\textit{(b)}} radial width $\sigma_r$;
    \textbf{\textit{(c)}} azimuthal width $\sigma_\phi$;
    \textbf{\textit{(d)}} lamp-post height $h_{\rm LP}$.
    The green solid line is the axisymmetric rippled-disc reference of
    \citet{Starkey2023}, with peak ripple amplitude
    $h_{\rm ripple} = 0.5\,\ld$ matched to the pillar height
    parameter $h_p$ and the same inner edge $r_{\rm in,ripple} =
    2\,\ld$ at which our pillars start. The grey solid line is
    $T_{\rm visc}$ from Eq.\,(\ref{eq:tvisc_pillar}). Reference power-laws
    $T \propto r^{-3/4}$ (black dotted) and $T \propto r^{-7/8}$
    (brown dotted) are anchored to the total bowl temperature at
    $r = 1\,{\rm light\mbox{-}day}$, where viscous heating is sub-dominant. The plotted
    range is the pillar belt $r \in [1, 22]\,\ld$.}
    \label{fig:T_param_scan}
\end{figure*}

Fig.\,\ref{fig:T_param_scan} shows $\langle T_{\rm eff}\rangle_\phi(r)$
for our four parameter scans. The bowl baseline (no pillars) closely
tracks $T \propto r^{-3/4}$ in the outer disc (the canonical
\citet{shakura1973} slope) and gives a clean reference against which to
gauge the pillar perturbations. Adding 100 pillars introduces a
fluctuating component superposed on the smooth profile: shadow lanes
drag the azimuthal average below the bowl curve, while the illuminated
pillar tops can locally exceed the bowl temperature by a factor of a
few. Crucially, the average over the pillar belt steepens systematically
toward the $T \propto r^{-7/8}$ reference power law, particularly for
the configurations with larger $h_p$, $\sigma_r$, or $\sigma_\phi$. This
emergent steepening is in the same direction as the
$\alpha = 0.71 \pm 0.02$ spectral index inferred from the inter-band
variability of quasars by \citet{Weaver2022}, which excludes the
canonical $L_\nu \propto \nu^{1/3}$ disc spectrum corresponding to a
strict $T \propto r^{-3/4}$ profile and instead favours a steeper
outer-disc temperature gradient. By contrast, increasing $h_{\rm LP}$
raises the inner-disc plateau (the lamp now illuminates inner radii at
a less grazing angle) but barely affects the outer pillar belt. The
fluctuation amplitude $\Delta T / \langle T \rangle \sim 10$--50 per cent at
$r = 5$--$15\,\ld$ is the relevant input for the continuum lag
computation discussed next.

For comparison, Fig.\,\ref{fig:T_param_scan} also shows the
azimuthally-averaged temperature of the axisymmetric rippled-disc
model of \citet{Starkey2023} (green solid line). We follow the
\citet{Starkey2023} bowl for prescription but, because the bowl
adopted here is essentially flat across the pillar
belt, we add the ripple as an additive perturbation rather
than the multiplicative form used in the original code. The disc
surface height is
\begin{equation}
H(r) = H_0(r) +
\begin{cases}
0, & r < r_{\rm in,ripple}, \\
h_{\rm ripple}\,\bigl(r/r_0\bigr)^{w_{\rm pow}}\,
\Bigl[\dfrac{\sin\phi(r) + 1}{2}\Bigr]^{c}, & r \ge r_{\rm in,ripple},
\end{cases}
\label{eq:ripple_h}
\end{equation}
with the ripple phase
chosen so that $\sin\phi(r_{\rm in,ripple}) = -1$ and the ripple
amplitude vanishes at the inner edge, joining smoothly with the bare
bowl. We adopt the parameter set
$\{h_{\rm ripple}, w_{\rm pow}, w_{\rm num}, w_{\rm npow}, c,
r_{\rm in,ripple}\} =
\{0.5\,\ld, 0, 2\,\mathrm{rad}\,\ld^{-1}, 0, 2, 2\,\ld\}$,
matching the central pillar amplitude $h_p = 0.5\,\ld$ and the
pillar inner radius. The wavelength $2\pi/w_{\rm num} \simeq \pi\,\ld$
gives roughly six bumps across the pillar belt, comparable to the
typical pillar spacing in the 100-pillar ensemble.

With these parameters the rippled disc tracks the bowl in
$\langle T_{\rm eff}\rangle_\phi$ closely at all radii. This is the
expected outcome of azimuthal averaging: an axisymmetric rippled
height profile contributes the same $H(r)$ to every $\phi$, whereas
an ensemble of localized pillars modulates $H(r,\phi)$ in $\phi$ as
well, generating the wider scatter around the mean visible in the
colour curves. The two pictures therefore predict almost the same
$\langle T_{\rm eff}\rangle_\phi(r)$ for matched amplitude and
wavelength, but they make different predictions for the
line-dependent 
emission and time variability discussed in
Section\,\ref{sec:results} (which the rippled model cannot
produce).

\subsection{Continuum lag spectrum}
\label{sec:lag_spec}

Following \citet{Blandford1982} and the lamp-post reprocessing
formalism of \citet{Starkey2023}, the response of the disc surface
brightness at wavelength $\lambda$ to a small perturbation in the
ionizing luminosity is
\begin{equation}
\Psi(\lambda,\tau) =
\int_{\rm disc}\!
\frac{\partial B_\nu\!\bigl[T_{\rm eff}(r,\phi),\lambda\bigr]}
     {\partial T_{\rm irr}}\,
\delta\!\bigl[\tau - \tau_{\rm geom}(r,\phi)\bigr]\,
dA(r,\phi),
\label{eq:psi_pillar}
\end{equation}
where $B_\nu$ is the Planck function, the partial derivative is taken
at fixed $T_{\rm visc}$, and $\tau_{\rm geom}(r, \phi)$ is the
geometric light-travel delay from the lamp post to the surface
element $(r, \phi)$ and onward to a distant observer at inclination
$i = 45^\circ$ (Section\,\ref{sec:disc_model}), including the standard
near-side/far-side projection for an inclined disc. The mean delay
at wavelength $\lambda$ is
\begin{equation}
\langle\tau\rangle(\lambda)
= \frac{\int \tau\,\Psi(\lambda,\tau)\,d\tau}
       {\int \Psi(\lambda,\tau)\,d\tau}.
\label{eq:tau_mean_pillar}
\end{equation}
Numerically we evaluate Eq.\,(\ref{eq:psi_pillar}) on a polar grid
$(r_i,\phi_j)$ with $n_r = 600$ and $n_\phi = 360$, using a delay grid
of $n_\tau = 500$ uniformly spaced between $0$ and $50$\,d.

\begin{figure*}
    \centering
    \includegraphics[width=\textwidth]{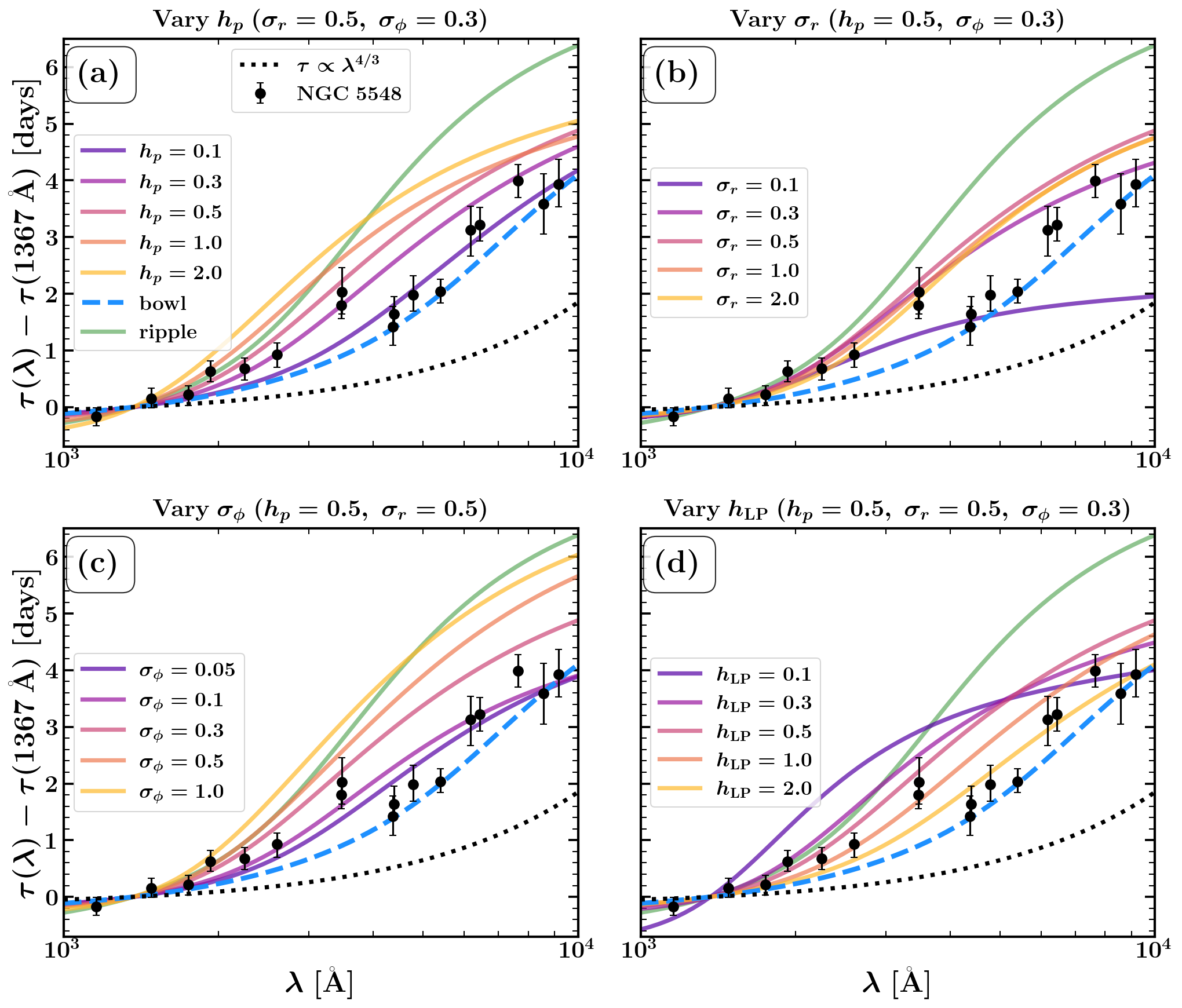}
    \caption{Relative continuum lag spectrum
    $\langle\tau\rangle(\lambda) - \langle\tau\rangle(1367\,\mbox{\AA})$
    (Eq.\,\ref{eq:tau_mean_pillar}) for the same four parameter scans
    as Fig.\,\ref{fig:T_param_scan} (panels labelled
    \textbf{\textit{(a)}}--\textbf{\textit{(d)}} for $h_p$, $\sigma_r$,
    $\sigma_\phi$, $h_{\rm LP}$ respectively).
    The bowl baseline (blue dashed)
    gives the characteristic $\propto\lambda^{4/3}$ ramp, and pillars
    systematically lengthen the lag at optical wavelengths by adding
    emitting area at larger effective reprocessing radii. The black
    dotted line is the standard thin-disc prediction for NGC\,5548 and the
    points are the AGN STORM lags, both from \citet{Fausnaugh2016}. The
    axisymmetric rippled-disc reference of \citet{Starkey2023} (green
    solid line) is shown for comparison, with the same amplitude and
    inner radius as in Fig.\,\ref{fig:T_param_scan}. The pillar curves
    bracket the rippled-disc lag at optical wavelengths, demonstrating
    that the localized, non-axisymmetric pillar geometry can reproduce
    a lag excess similar to that which the axisymmetric ripples in
    \citet{Starkey2023} produced. The deviation grows with both radial
    and azimuthal pillar size and is of comparable amplitude to the
    excess lag observed in the STORM campaign
    on NGC\,5548 \citep{Fausnaugh2016}.}
    \label{fig:tau_rel_param_scan}
\end{figure*}

In continuum reverberation campaigns, inter-band lags are measured by
cross-correlating each band against a chosen driving light curve
(typically an X-ray or far-UV variable continuum). For comparing
disc models the wavelength-dependent shape of $\langle
\tau\rangle(\lambda)$ is the diagnostic of interest, and a uniform
offset that depends on the (poorly-constrained) lamp-to-disc and
disc-to-observer geometry adds nothing to the comparison. We
therefore plot the difference relative to the
1367\,\AA\ reference band of the NGC\,5548 lags \citep{Fausnaugh2016},
$\langle\tau\rangle(\lambda) - \langle\tau\rangle(1367\,\mbox{\AA})$,
for the four parameter scans in
Fig.\,\ref{fig:tau_rel_param_scan}. With $\beta = 10$ our bowl
baseline is essentially flat for $r \lesssim 0.7\,r_0$ and contains
no inner-disc height structure, so its lag spectrum has the
shape of the
standard \citet{shakura1973} thin-disc prediction
$\tau \propto \lambda^{4/3}$, modified only at the longest
wavelengths by the single outer-edge flare, although with about 2.5 times
its amplitude because the irradiated disc adopted here is hotter. As is well established, though not yet clearly understood,
this standard prediction underestimates the inter-band continuum
lags inferred from intensive monitoring campaigns of NGC\,5548 \citep{Fausnaugh2016} and
similar Seyferts by a factor of $\sim 3$
\citep[e.g.,][]{Edelson2019,Lewin2025}.
\citet{Starkey2023} showed that this discrepancy can be resolved by
introducing a steep rim or rippled height structure near
$r \sim 5\,\ld$ (close to the dust sublimation radius), whose
inward-facing crests reach $T \sim 6000$\,K and contribute hot
reprocessed emission that lengthens the optical lag without changing
the $\propto \lambda^{4/3}$ overall shape.

Adding pillars (coloured curves) shifts the lag curve upward
relative to the bowl by an additional $\sim 0.5$--$1.5$\,d at
optical wavelengths, with the largest deviations coming from
configurations that either deepen the shadow lanes (large
$\sigma_\phi$) or extend them radially (large $\sigma_r$). This is
the same direction and amplitude as the \citet{Starkey2023}
rim/ripple correction, and  small pillars ($h_p \lesssim 0.3\,\ld$ or
$\sigma_\phi \lesssim 0.1$\,rad) bring the predicted lag spectrum into
line with the observed values, while larger ones overshoot them. Pillars therefore provide a
non-axisymmetric realization of the same physics: a localized height
structure that hosts hot inward-facing surfaces, driving up the
optical lag, while simultaneously producing the steeper outer-disc
temperature gradient indicated by quasar variability spectra
\citep{Weaver2022}.

In Fig.\,\ref{fig:tau_rel_param_scan} we also overplot the lag
spectrum predicted by the axisymmetric rippled-disc model of
\citet{Starkey2023} (green solid line), evaluated with the same
amplitude ($h_{\rm ripple} = 0.5\,\ld$) and inner radius
($r_{\rm in,ripple} = 2\,\ld$) used for the pillars. The rippled-disc
lag closely matches the central pillar configurations
($h_p \sim 0.5\,\ld$, $\sigma_r \sim 0.5\,\ld$, $\sigma_\phi \sim 0.3\,$rad) at optical wavelengths, while the more
extended pillar configurations bracket it from above. The pillar and
rippled-disc pictures therefore make essentially the same prediction
for the azimuthally-averaged continuum lag spectrum, which is the
quantity probed by inter-band reverberation campaigns. This is an
important consistency check: in the limit of an axisymmetrically
distributed, dense ensemble of pillars at the relevant radii, the
non-axisymmetric pillar geometry reduces to an effectively
axisymmetric height profile, and the continuum lag observable becomes
indistinguishable. The two pictures part company in the
emission-line and time-evolving signatures (Sections\,\ref{sec:vdm}
and\,\ref{sec:results}), where the non-axisymmetric pillar structure
produces 
line-dependent
residuals and barber-pole modulations that
the rippled-disc model cannot.

\begin{figure*}
    \centering
    \includegraphics[width=\textwidth]{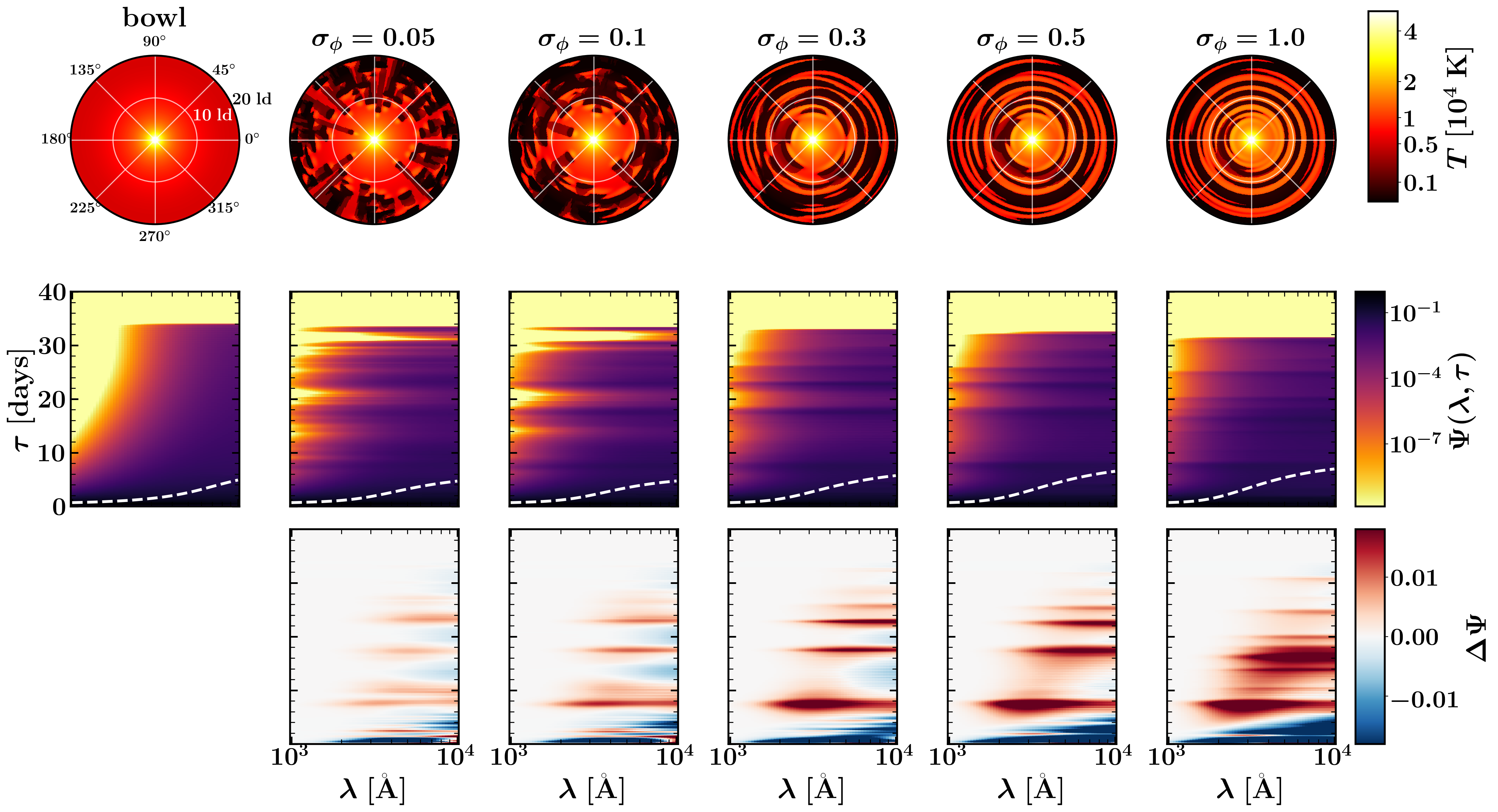}
    \caption{Two-dimensional continuum response function
    $\Psi(\lambda,\tau)$ for the $\sigma_\phi$ scan
    ($h_p = 0.5$, $\sigma_r = 0.5$).
    \textbf{\textit{Top row:}} face-on $T_{\rm eff}(r,\phi)$ map; the
    rightmost colour bar applies to all panels in this row.
    \textbf{\textit{Middle row:}} $\Psi(\lambda,\tau)$ on a shared
    logarithmic colour scale (inverted, so darker indicates higher
    response); its colour bar is shown to the right. The
    dashed white curve traces the  mean-delay relation
    $\langle\tau\rangle(\lambda)$ of
    each model.
    \textbf{\textit{Bottom row:}} difference map
    $\Delta\Psi = \Psi_{\rm pillars} - \Psi_{\rm bowl}$, with its own
    colour bar to the right; red (blue) indicates enhanced (suppressed)
    response. Larger
    $\sigma_\phi$ produces stronger and more coherent residuals,
    explaining the systematic upward shift of the lag curve in
    Fig.\,\ref{fig:tau_rel_param_scan}.}
    \label{fig:psi_vary_sp}
\end{figure*}

To unpack the structure underlying the integrated lag curve, we
computed the full two-dimensional response function
$\Psi(\lambda,\tau)$ (Eq.\,\ref{eq:psi_pillar}) for all four parameter
scans of Section\,\ref{sec:lag_spec}. The four scans are qualitatively
similar, so Fig.\,\ref{fig:psi_vary_sp} shows the $\sigma_\phi$ scan as
representative ($h_p = \sigma_r = 0.5$). The top row is the face-on
$T_{\rm eff}(r,\phi)$ map, where narrow $\sigma_\phi$ gives sharp
fan-shaped shadows and wide $\sigma_\phi$ merges them into
nearly-concentric rings; the middle row is $\Psi(\lambda,\tau)$ with
the bowl mean-delay curve overplotted; the bottom row is the residual
$\Delta\Psi = \Psi_{\rm pillars} - \Psi_{\rm bowl}$.

The bowl response (leftmost column) is smooth and monotonic in $\tau$,
tracing the standard $\langle\tau\rangle\propto\lambda^{4/3}$ relation,
and all configurations share a sharp upper envelope at $\tau \approx
34$\,d set by the maximum geometric delay $r_{\rm out}(1+\sin i)/c$.
Pillars add two features. First, horizontal stripes appear at the
discrete radii where shadow lanes intercept significant emitting area,
broadening and merging as $\sigma_\phi$ grows. Second, the response
centroid shifts to larger $\tau$ at optical wavelengths, leaving a
coherent dipole in $\Delta\Psi$ (a red excess at $\tau \approx
5$--$15$\,d paired with a blue deficit at smaller $\tau$) that reflects
the redistribution of reprocessing radius from the shadowed inner disc
toward the outer pillar belt. The residual amplitude grows with
$\sigma_\phi$, and its wavelength-marginalized projection is exactly
the integrated lag excess of Fig.\,\ref{fig:tau_rel_param_scan}.

\section{Discussion}
\label{sec:discussion}

\subsection{Detectability}
\label{sec:detectability}

The spectral time-evolution residuals reach tens of per cent of the mean line flux density. These amplitudes are well above the
measurement precision of current spectroscopic reverberation campaigns,
$\sim 1$--2 per cent in each spectrum on bright Seyfert\,1 targets
\citep{DeRosa2015,Bentz2009lamp}. The orbital timescale scales as
$T_{\rm orb} = 2\pi (r^3/G M_{\rm BH})^{1/2} \propto r^{3/2}
M_{\rm BH}^{-1/2}$. For NGC\,5548 ($M_{\rm BH} = 7 \times
10^7$\,$\Msun$), the barber-pole feature at $r_p \approx 4\,\ld$ has
$T_{\rm orb} \approx 2$\,yr, while at $r_p \approx 10\,\ld$ it grows to
$\approx$ 9\,yr and at the outer edge $r_p = 20\,\ld$ to $\approx$ 24\,yr.
The shorter periods are already comfortably within reach of existing
year-cadence campaigns such as the Lick AGN Monitoring Project
\citep{Barth2015} and the STORM campaign \citep{DeRosa2015}, both of which
delivered $\sim 100$\,d, daily-cadence light curves on NGC\,5548. Indeed,
\citet{Horne2021} tentatively identified a coherent diagonal feature in
the C\,{\sc iv} residuals of NGC\,5548 with a period consistent with this
estimate, providing the first observational hint that pillar-like
structures may be common in nearby Seyferts.
The pattern is detectable at Type 1 inclinations. Its velocity amplitude
scales with $\sin i$, and its once-per-orbit modulation, which comes
from foreshortening, grows with $i$ (Appendix\,\ref{sec:inclination_appendix}).

Pillars at larger radii ($r \sim 10$--$20\,\ld$) require long-baseline
campaigns spanning $\gtrsim 10$\,yr. Black-hole-mass--scaled BLR sizes
\citep{Kaspi2000,Bentz2013,Peterson2004} indicate that lower-luminosity
AGN ($L_{\rm bol} \sim 10^{43}$--$10^{44}$\,erg\,s$^{-1}$) host BLRs at
$\lesssim 10$\,$\ld$ with correspondingly shorter orbital times, so
moderate-mass Seyferts make natural targets where the entire pillar
population can be sampled within a $\sim 5$\,yr campaign. 
More generally, detecting pillar 
signatures requires (i)~high-cadence spectroscopy
($\Delta t \lesssim 1$\,d to resolve the orbital trajectory in
$(\lambda,t)$), (ii)~long duration ($T_{\rm dur} \gtrsim 3\tau_{\rm max}$
and at least one full $T_{\rm orb}$), and (iii)~high signal-to-noise
(${\rm S/N} \gtrsim 100$ per epoch in the resolved line core), as
outlined by \citet{Horne2004}. Long-baseline photometric monitoring with
LSST/Rubin combined with targeted spectroscopic follow-up of stochastic
flaring episodes \citep{MacLeod2010,Kelly2009} will substantially expand
the sample of AGN in which non-axisymmetric BLR structure can be searched
for.
\subsection{Distinguishing pillars from other structures}
\label{sec:distinguish}

Several mechanisms can produce non-axisymmetric features in velocity--delay
maps. Spiral density waves driven by self-gravity or by embedded perturbers
\citep{goodman2003,chenGI2023,alidib2023} generate extended, logarithmically
curved features in the disc surface and would map onto coherent arcs in
$(v,\tau)$. Disc warps from radiation-driven \citet{Pringle1996} or
Bardeen-Petterson precession produce a single global tilt that modulates
the line profile sinusoidally on the precession timescale, with
equal-amplitude response in all lines because no opacity contrast
is involved. Eccentric or elliptical discs \citep{Eracleous1995} produce
asymmetric, displaced double-peaked profiles that precess on apsidal
timescales rather than orbital ones. Disc winds and outflowing streamlines
\citep{Murray1995,Chiang1996} produce strong asymmetries in the line
profile but their signatures appear at velocities offset from the
Keplerian envelope and modulate primarily the resonance lines (C\,{\sc
iv}, N\,{\sc v}). Magnetically-confined orbiting BLR clouds
\citep{Rees1987,Alexander1997} can also produce localized $(v,\tau)$
features but with no associated line-ratio signature.

Pillars are more likely to tower over massive stars formed in the
marginally (un)stable outer regions (at
$r \gtrsim R_{\rm GI} \sim 10^{2-3}\,R_\bullet$) of AGN accretion discs
 (Section\,\ref{sec:background}). They can be
distinguished from 
intrinsic disc structures by four properties. First, they
produce compact, localized hotspots in the velocity--delay map rather
than extended arcs or global tilts. Second, the differential line
response (Mg\,{\sc ii}/H$\beta$
rising while C\,{\sc iv}/H$\beta$ drops in shadow
regions) is a unique signature of localized opacity sources modifying
the ionization structure rather than redistributing matter or velocity.
Third, the barber-pole period traces $T_{\rm orb} \propto r^{3/2}$,
consistent with Keplerian motion at the inferred pillar radius and
self-consistent with the $(v,\tau)$ position of the hotspot, providing a
joint check on the interpretation. Fourth, the population statistics
predicted by the GI/embedded-star scenario
\citep{cantiello2021ApJ,chenfeedback2024,chenpop2024} predict
$\sim 10^3$--$10^5$ pillars per disc with characteristic radii in the
gravitationally-unstable outer zone $r \gtrsim R_{\rm GI} \sim 10^{2-3} R_\bullet$ 
where Toomre $Q \lesssim 1$ \citep[see also][]{Levin2007,Thompson2005}, leading to a
specific $r$-dependence of the residual amplitude that can be tested
against an ensemble of monitored objects.

\subsection{Connection to continuum lag models}
\label{sec:continuum}

The pillar structures modelled here represent a non-axisymmetric
generalization of the rippled or rim-rising disc surfaces introduced by
\citet{Starkey2023} to explain the larger-than-expected continuum
inter-band lags observed in monitoring campaigns of NGC\,5548 \citep{Fausnaugh2016} and several
other Seyferts \citep[e.g.\,][]{Edelson2019,Lewin2025}. 
Standard \citet{shakura1973}
disc models systematically underpredict the inferred lag-vs-wavelength
slope by factors of $\sim 2$--$3$, motivating geometric models in which
the lamp-post intercepts a non-trivial fraction of the disc surface at
larger radii \citep[see also the BLR-rim
contributions discussed by][]{czerny2015,Sarkar2021}. The bowl-shaped or
rippled surfaces achieve this by raising the outer disc into the
lamp-post pencil beam.

While continuum lags integrate over the entire disc surface and are
therefore sensitive only to the azimuthally-averaged height profile
$\langle H(r) \rangle_\phi$, the emission-line velocity--delay maps are
sensitive to the full two-dimensional structure $H(r,\phi)$ through both
the local emissivity (modulated by the ionizing flux pattern shown in
Section\,\ref{sec:ionflux}) and the local kinematics. This complementarity
makes simultaneous continuum and emission-line reverberation mapping a
powerful probe of disc surface geometry: the continuum constrains the
radial height profile and so the global energetics of disc reprocessing,
while the emission lines reveal the azimuthal structure and allow direct
identification of individual pillars. For the two STORM
targets, NGC\,5548 and Mrk\,817, fitting the continuum and emission-line data
together with a single disc model would over-determine
the geometric model and provide stringent tests of the rippled-disc
hypothesis \citep{Horne2021,Williams2021}.

The same orthogonality applies to the temperature-perturbation
reconstruction of \citet{Neustadt_2022,StoneShen2023}: their axisymmetric
$\delta T(R, t)$ inversion is structurally insensitive to the
non-axisymmetric pillar pattern, because the azimuthal mean of the
shadow distribution is approximately time-stationary even as the
individual pillars orbit. A non-axisymmetric extension of their method
that fits $\delta T(R, \phi, t)$, or the line-based diagnostics
developed in Sections\,\ref{sec:vdm}--\ref{sec:results}, would be the
natural way to test for the pillar mechanism in existing
high-cadence spectroscopic monitoring data.

A further continuum-reverberation signature arises because the
shadowed regions behind opaque pillars are shielded from the variable
irradiation: they still emit their viscous thermal flux but do not
track the driving continuum. The disc light therefore splits into a
blue, variable component from the irradiated surface and a redder,
constant component from the shadows, which is exactly the decomposition
isolated by flux--flux analysis of continuum light curves
\citep{Winkler92}. The constant component is usually assigned to
host-galaxy starlight resolvable with HST; in the pillar picture it
also contains the HST-unresolved emission from the shadowed disc. This
may explain why flux--flux decompositions repeatedly recover a constant
component brighter than the imaged host: \citet{Cai24} find the
flux-variation-gradient host flux exceeds the HST values of
\citet{Bentz09} by factors of $2.2$, $1.7$, and $1.4$ for Mrk\,509,
Mrk\,279, and 3C\,120, respectively. For Mrk\,817, \citet{Montano26}
and \citet{Netzer24} similarly attribute most of the constant flux to
the AGN rather than the host, and the flux--flux host SED of
PG\,0844+349 likewise exceeds its HST image-decomposition estimate
(Huang et al., in preparation). This non-varying,
HST-unresolved emission could be powered by a star embedded in
the disc. Assuming
Eddington accretion,
$L_{\rm Edd}\simeq1.26\times10^{38}\,(M_\star/M_\odot)\,{\rm erg\,s^{-1}}$.
Thermalizing this luminosity over a hemispherical pillar of
radius $R_{\rm p}$ through
$L_{\rm Edd}=2\pi R_{\rm p}^{2}\,\sigma\,(T_{\rm eff}^{\rm pillar})^{4}$ gives
$T_{\rm eff}^{\rm pillar}\simeq7.5\times10^{3}\,{\rm K}\,(M_\star/200\,M_\odot)^{1/4}\,(R_{\rm p}/10\,{\rm AU})^{-1/2}$,
comparable to the blackbody temperature required to reproduce
the constant component probed by the flux--flux analysis.
Modelling this
red component as a multi-temperature
blackbody set by $T_{\rm v}(r)$ and the azimuthal covering fraction of
shadows would constrain the radial distribution of the shadowed
region. The exact extent and surface temperature of the hot spots
produced by such embedded stars will be captured by the
three-dimensional radiation-hydrodynamic simulations now underway.
These simulations will then test whether the hot spots account for the constant flux
observed in excess of the inferred host galaxy light.

\subsection{Future work}
\label{sec:future}

Several extensions of this work are natural and underway. First, we intend
to compute self-consistent pillar heights and azimuthal distributions
from physical disc-instability calculations that solve for the embedded
stellar population, the Toomre-$Q$ profile, and the local thermal balance
between irradiation, viscous heating, and stellar-feedback heating
\citep{cantiello2021ApJ,chenfeedback2024,chenGI2023,owen2025MNRAS}. The
recent population-synthesis models of \citet{chenpop2024} and
\citet{xu2025} predict the radius- and mass-dependent number density
of accreting stars and stellar-mass black holes embedded in AGN discs;
mapping these populations onto thermal feedback bumps following the
prescription of \citet{owocki2012} would replace our parametric pillar
model with a physically motivated $H(r,\phi)$ field. Second, we plan to
perform recovery tests using \memecho{} \citep{Horne1994} and by fitting the pillar model
of this paper to the data to assess whether pillar
signatures can be extracted from realistically noisy simulated data,
including realistic continuum-driving light curves modelled as a damped
random walk \citep{Kelly2009,MacLeod2010}. Third, we aim to apply this
framework to specific well-monitored AGN to obtain quantitative pillar
number densities. The most promising target is NGC\,5548, for which
both continuum lags and emission-line velocity--delay maps are available
from the STORM campaign \citep{Fausnaugh2016,Horne2021}, with the Lick AGN
Monitoring Project sample providing additional constraints
\citep{Bentz2009lamp,Barth2015}. 
Fourth, the framework can be
extended to include partially-transparent pillars  (see Appendix\,\ref{sec:transparent_pillars}), which
will be relevant once realistic optical-depth modelling
replaces our opaque-pillar
approximation. Finally, an obvious avenue is to fold in proper-motion
effects from pillar evolution. Pillar formation, migration, and
destruction occur on viscous and stellar-evolutionary timescales
\citep{huang2023}, and folding these into the model would provide
observational tests on $\sim$decade-long campaigns.

Beyond AGN, the same shadow-mapping framework can be applied to
protoplanetary discs, where embedded planetary or stellar companions
cast analogous radial shadows in scattered-light imaging.
\citet{akansoy2025} use 3D radiative-transfer simulations to
characterize the shadow width and depth as functions of companion mass
and orbital distance and propose this as an indirect probe of
otherwise hard-to-detect companions; extending our velocity--delay and
time-evolving spectral diagnostics to such systems is a natural
follow-up. A key step toward physically motivated pillar transmissions
will be to post-process hydrodynamic radiative-transfer simulations of
embedded stars to obtain self-consistent column densities and emergent
$f_{\rm trans}$ (Appendix\,\ref{sec:transparent_pillars}); the resulting
predictions can then be tested against larger reverberation samples
beyond NGC\,5548, including the SDSS-RM programme
\citep{Stone2025, Fries2024, Shen2024}.


\section{Conclusions}
\label{sec:conclusions}

We have computed velocity--delay maps,  time-evolving emission-line spectra
and continuum lag spectra
for AGN accretion discs with non-axisymmetric pillar structures raised by
massive stars embedded in the disc. Our main
findings are as follows.

\begin{enumerate}
    \item Individual pillars produce localized hotspots in the velocity--delay
    map, with positions determined by their Keplerian orbital velocity and
    the light-travel time delay to the observer.

    \item The Keplerian rotation of pillars produces a barber-pole pattern
    in the time-evolving spectrum $F(\lambda, t)$, with diagonal stripes
    tracing the orbital motion. This pattern is clearly visible for a
    single pillar and persists, though diluted, for $N = 100$ pillars.

    \item Multi-line velocity-resolved reverberation mapping provides
    the most robust diagnostic for detecting pillar structures in AGN
    discs. All lines are suppressed in shadows in absolute terms, but
    when normalized by H$\beta$ (or, equivalently, by the local
    continuum) the ratios move in opposite directions:
    Mg\,{\sc ii}/H$\beta$ rises in shadows because the lower
    ionization parameter favours Mg$^{+}$, while
    C\,{\sc iv}/H$\beta$ drops because C\,{\sc iv} responds more
    steeply than H$\beta$ to the collapse of $\Phi_{\rm H}$.  This line-to-line contrast in the
    response ratios is what distinguishes the pillar mechanism from
    kinematic features that would modulate all lines by similar
    factors.

    \item Pillars also change the disc temperature profile and the
    inter-band continuum lags (Section\,\ref{sec:continuum_lags}). By shadowing
    the outer disc they steepen the azimuthally averaged temperature profile from
    $T \propto r^{-3/4}$ toward $T \propto r^{-7/8}$, and, as a non-axisymmetric
    generalization of the rim and ripples of \citet{Starkey2023}, they lengthen the
    optical continuum lags in the same way. Small pillars bring the predicted lag
    spectrum into line with the AGN STORM lags of NGC\,5548.

    \item These observations are feasible with current monitoring campaigns
    for pillars at small radii around massive SMBHs
    ($T_{\rm orb} \lesssim$\,a few years), and will become accessible at
    larger radii with future long-baseline surveys.
\end{enumerate}


\section*{Acknowledgments}

We thank Y.X. Chen, Z.H. Xu, X.H. Ji, and R.B. Dong for useful discussions.
This work made use of the \cloudy{} photoionization code
\citep{Ferland2017}.


\section*{Data Availability}

The data underlying the findings of this study, together with the analysis code used to produce the results, are available from the corresponding author upon request.


  \bibliographystyle{mnras}
  \bibliography{references} 

@ARTICLE{Alexander1997,
   author = {{Alexander}, T. and {Netzer}, H.},
    title = "{Bloated stars as AGN broad-line clouds: the emission-line profiles}",
  journal = {\mnras},
     year = 1997,
   volume = 284,
    pages = {967-980},
      doi = {10.1093/mnras/284.4.967}
}

@ARTICLE{Barth2015,
   author = {{Barth}, A.~J. and {Pancoast}, A. and {Thorman}, S.~J. and 
	{Bennert}, V.~N. and {Sand}, D.~J. and {Li}, W. and {Malkan}, M.~A. and 
	{Baliber}, N. and {Berlind}, P. and {Calkins}, M.~L. and others},
    title = "{The Lick AGN Monitoring Project 2011: Spectroscopic Campaign and Emission-line Light Curves}",
  journal = {\apjs},
     year = 2015,
   volume = 217,
    pages = {26},
      doi = {10.1088/0067-0049/217/2/26}
}

@ARTICLE{Bentz2013,
       author = {{Bentz}, Misty C. and {Denney}, Kelly D. and {Grier}, Catherine J. and {Barth}, Aaron J. and {Peterson}, Bradley M. and {Vestergaard}, Marianne and {Bennert}, Vardha N. and {Canalizo}, Gabriela and {De Rosa}, Gisella and {Filippenko}, Alexei V. and {Gates}, Elinor L. and {Greene}, Jenny E. and {Li}, Weidong and {Malkan}, Matthew A. and {Pogge}, Richard W. and {Stern}, Daniel and {Treu}, Tommaso and {Woo}, Jong-Hak},
        title = "{The Low-luminosity End of the Radius-Luminosity Relationship for Active Galactic Nuclei}",
      journal = {\apj},
         year = 2013,
        month = apr,
       volume = {767},
       number = {2},
          eid = {149},
        pages = {149},
          doi = {10.1088/0004-637X/767/2/149},
archivePrefix = {arXiv},
       eprint = {1303.1742},
 primaryClass = {astro-ph.CO},
       adsurl = {https://ui.adsabs.harvard.edu/abs/2013ApJ...767..149B}
}

@ARTICLE{Cackett2021,
       author = {{Cackett}, Edward M. and {Bentz}, Misty C. and {Kara}, Erin},
        title = "{Reverberation mapping of active galactic nuclei: from X-ray corona to dusty torus}",
      journal = {iScience},
         year = 2021,
        month = jun,
       volume = {24},
       number = {6},
        pages = {102557},
          doi = {10.1016/j.isci.2021.102557},
archivePrefix = {arXiv},
       eprint = {2105.06926},
 primaryClass = {astro-ph.GA},
       adsurl = {https://ui.adsabs.harvard.edu/abs/2021iSci...24j2557C}
}

@ARTICLE{Bentz2010vdm,
       author = {{Bentz}, Misty C. and {Horne}, Keith and {Barth}, Aaron J. and {Bennert}, Vardha Nicola and {Canalizo}, Gabriela and {Filippenko}, Alexei V. and {Gates}, Elinor L. and {Malkan}, Matthew A. and {Minezaki}, Takeo and {Treu}, Tommaso and {Woo}, Jong-Hak and {Walsh}, Jonelle L.},
        title = "{The Lick AGN Monitoring Project: Velocity-delay Maps from the Maximum-entropy Method for Arp 151}",
      journal = {\apjl},
         year = 2010,
        month = sep,
       volume = {720},
       number = {1},
        pages = {L46-L51},
          doi = {10.1088/2041-8205/720/1/L46},
archivePrefix = {arXiv},
       eprint = {1007.0781},
 primaryClass = {astro-ph.CO},
       adsurl = {https://ui.adsabs.harvard.edu/abs/2010ApJ...720L..46B}
}

@ARTICLE{Bentz2009lamp,
       author = {{Bentz}, Misty C. and {Walsh}, Jonelle L. and {Barth}, Aaron J. and {Baliber}, Nairn and {Bennert}, Vardha Nicola and {Canalizo}, Gabriela and {Filippenko}, Alexei V. and {Ganeshalingam}, Mohan and {Gates}, Elinor L. and {Greene}, Jenny E. and {Hidas}, Marton G. and {Hiner}, Kyle D. and {Lee}, Nicholas and {Li}, Weidong and {Malkan}, Matthew A. and {Minezaki}, Takeo and {Sakata}, Yu and {Serduke}, Frank J.~D. and {Silverman}, Jeffrey M. and {Steele}, Thea N. and {Stern}, Daniel and {Street}, Rachel A. and {Thornton}, Carol E. and {Treu}, Tommaso and {Wang}, Xiaofeng and {Woo}, Jong-Hak and {Yoshii}, Yuzuru},
        title = "{The Lick AGN Monitoring Project: Broad-line Region Radii and Black Hole Masses from Reverberation Mapping of H{\ensuremath{\beta}}}",
      journal = {\apj},
         year = 2009,
        month = nov,
       volume = {705},
       number = {1},
        pages = {199-217},
          doi = {10.1088/0004-637X/705/1/199},
archivePrefix = {arXiv},
       eprint = {0908.0003},
 primaryClass = {astro-ph.CO},
       adsurl = {https://ui.adsabs.harvard.edu/abs/2009ApJ...705..199B}
}

@ARTICLE{Blandford1982,
   author = {{Blandford}, R.~D. and {McKee}, C.~F.},
    title = "{Reverberation mapping of the emission line regions of Seyfert galaxies and quasars}",
  journal = {\apj},
     year = 1982,
   volume = 255,
    pages = {419-439},
      doi = {10.1086/159843}
}

@ARTICLE{Chiang1996,
   author = {{Chiang}, J. and {Murray}, N.},
    title = "{Reverberation Mapping and the Disk-Wind Model of the Broad-Line Region}",
  journal = {\apj},
     year = 1996,
   volume = 466,
    pages = {704},
      doi = {10.1086/177543}
}

@ARTICLE{DeRosa2015,
   author = {{De Rosa}, G. and {Peterson}, B.~M. and {Ely}, J. and others},
    title = "{Space Telescope and Optical Reverberation Mapping Project. I. Ultraviolet Observations of the Seyfert 1 Galaxy NGC 5548 with the Cosmic Origins Spectrograph on Hubble Space Telescope}",
  journal = {\apj},
     year = 2015,
   volume = 806,
    pages = {128},
      doi = {10.1088/0004-637X/806/1/128}
}

@ARTICLE{Eracleous1995,
   author = {{Eracleous}, M. and {Livio}, M. and {Halpern}, J.~P. and 
	{Storchi-Bergmann}, T.},
    title = "{Elliptical Accretion Disks in Active Galactic Nuclei}",
  journal = {\apj},
     year = 1995,
   volume = 438,
    pages = {610},
      doi = {10.1086/175104}
}

@ARTICLE{Ferland2017,
   author = {{Ferland}, G.~J. and {Chatzikos}, M. and {Guzm{\'a}n}, F. and 
	{Lykins}, M.~L. and {van Hoof}, P.~A.~M. and {Williams}, R.~J.~R. and 
	{Abel}, N.~P. and {Badnell}, N.~R. and {Keenan}, F.~P. and 
	{Porter}, R.~L. and {Stancil}, P.~C.},
    title = "{The 2017 Release Cloudy}",
  journal = {Rev. Mexicana Astron. Astrofis.},
     year = 2017,
   volume = 53,
    pages = {385-438}
}

@INPROCEEDINGS{Horne1994,
   author = {{Horne}, K.},
    title = "{Echo mapping problems: Maximum entropy solutions}",
booktitle = {Reverberation Mapping of the Broad-Line Region in Active Galactic Nuclei},
   editor = {{Gondhalekar}, P.~M. and {Horne}, K. and {Peterson}, B.~M.},
   series = {ASP Conference Series},
   volume = 69,
    pages = {23},
     year = 1994
}

@ARTICLE{Horne2004,
   author = {{Horne}, K. and {Peterson}, B.~M. and {Collier}, S.~J. and 
	{Netzer}, H.},
    title = "{Observational Requirements for High-Fidelity Reverberation Mapping}",
  journal = {\pasp},
     year = 2004,
   volume = 116,
    pages = {465-476},
      doi = {10.1086/420755}
}

@ARTICLE{Kaspi2000,
   author = {{Kaspi}, S. and {Smith}, P.~S. and {Netzer}, H. and {Maoz}, D. and 
	{Jannuzi}, B.~T. and {Giveon}, U.},
    title = "{Reverberation Measurements for 17 Quasars and the Size-Mass-Luminosity Relations in Active Galactic Nuclei}",
  journal = {\apj},
     year = 2000,
   volume = 533,
    pages = {631-649},
      doi = {10.1086/308704}
}

@ARTICLE{Kelly2009,
   author = {{Kelly}, B.~C. and {Bechtold}, J. and {Siemiginowska}, A.},
    title = "{Are the Variations in Quasar Optical Flux Driven by Thermal Fluctuations?}",
  journal = {\apj},
     year = 2009,
   volume = 698,
    pages = {895-910},
      doi = {10.1088/0004-637X/698/1/895}
}

@ARTICLE{Krolik1981,
   author = {{Krolik}, J.~H. and {McKee}, C.~F. and {Tarter}, C.~B.},
    title = "{Two-phase models of quasar emission line regions}",
  journal = {\apj},
     year = 1981,
   volume = 249,
    pages = {422-442},
      doi = {10.1086/159303}
}

@ARTICLE{Levin2007,
   author = {{Levin}, Y.},
    title = "{Starbursts near supermassive black holes: young stars in the Galactic Centre, and gravitational waves in LISA band}",
  journal = {\mnras},
     year = 2007,
   volume = 374,
    pages = {515-524},
      doi = {10.1111/j.1365-2966.2006.11155.x}
}

@ARTICLE{MacLeod2010,
   author = {{MacLeod}, C.~L. and {Ivezi{\'c}}, {\v Z}. and {Kochanek}, C.~S. and 
	{Koz{\l}owski}, S. and {Kelly}, B. and {Bulber}, E. and 
	{Kimball}, A. and {Sesar}, B. and {Westman}, D. and {Brooks}, K. and 
	others},
    title = "{Modeling the Time Variability of SDSS Stripe 82 Quasars as a Damped Random Walk}",
  journal = {\apj},
     year = 2010,
   volume = 721,
    pages = {1014-1033},
      doi = {10.1088/0004-637X/721/2/1014}
}

@ARTICLE{Murray1995,
   author = {{Murray}, N. and {Chiang}, J. and {Grossman}, S.~A. and {Voit}, G.~M.},
    title = "{Accretion Disk Winds from Active Galactic Nuclei}",
  journal = {\apj},
     year = 1995,
   volume = 451,
    pages = {498},
      doi = {10.1086/176238}
}

@ARTICLE{Peterson1993,
   author = {{Peterson}, B.~M.},
    title = "{Reverberation mapping of active galactic nuclei}",
  journal = {\pasp},
     year = 1993,
   volume = 105,
    pages = {247-268},
      doi = {10.1086/133140}
}

@ARTICLE{Peterson2004,
       author = {{Peterson}, B.~M. and {Ferrarese}, L. and {Gilbert}, K.~M. and {Kaspi}, S. and {Malkan}, M.~A. and {Maoz}, D. and {Merritt}, D. and {Netzer}, H. and {Onken}, C.~A. and {Pogge}, R.~W. and {Vestergaard}, M. and {Wandel}, A.},
        title = "{Central Masses and Broad-Line Region Sizes of Active Galactic Nuclei. II. A Homogeneous Analysis of a Large Reverberation-Mapping Database}",
      journal = {\apj},
         year = 2004,
        month = oct,
       volume = {613},
       number = {2},
        pages = {682-699},
          doi = {10.1086/423269},
archivePrefix = {arXiv},
       eprint = {astro-ph/0407299},
 primaryClass = {astro-ph},
       adsurl = {https://ui.adsabs.harvard.edu/abs/2004ApJ...613..682P}
}

@ARTICLE{Pringle1996,
   author = {{Pringle}, J.~E.},
    title = "{Self-induced warping of accretion discs}",
  journal = {\mnras},
     year = 1996,
   volume = 281,
    pages = {357-361},
      doi = {10.1093/mnras/281.1.357}
}

@ARTICLE{Rees1987,
   author = {{Rees}, M.~J.},
    title = "{Magnetic confinement of broad-line clouds in active galactic nuclei}",
  journal = {\mnras},
     year = 1987,
   volume = 228,
    pages = {47P-50P},
      doi = {10.1093/mnras/228.1.47P}
}

@ARTICLE{Thompson2005,
   author = {{Thompson}, T.~A. and {Quataert}, E. and {Murray}, N.},
    title = "{Radiation Pressure-supported Starburst Disks and Active Galactic Nucleus Fueling}",
  journal = {\apj},
     year = 2005,
   volume = 630,
    pages = {167-185},
      doi = {10.1086/431923}
}

@ARTICLE{Williams2021,
       author = {{Williams}, Peter R. and {Treu}, Tommaso and {Dahle}, H{\r{a}}kon and {Valenti}, Stefano and {Abramson}, Louis and {Barth}, Aaron J. and {Dyrland}, Karianne and {Gladders}, Michael and {Horne}, Keith and {Sharon}, Keren},
        title = "{The Black Hole Mass of the z = 2.805 Multiply Imaged Quasar SDSS J2222+2745 from Velocity-resolved Time Lags of the C IV Emission Line}",
      journal = {\apj},
         year = 2021,
        month = apr,
       volume = {911},
       number = {1},
          eid = {64},
        pages = {64},
          doi = {10.3847/1538-4357/abe943},
archivePrefix = {arXiv},
       eprint = {2011.02007},
 primaryClass = {astro-ph.GA},
       adsurl = {https://ui.adsabs.harvard.edu/abs/2021ApJ...911...64W}
}

@ARTICLE{shankar2009,
       author = {{Shankar}, Francesco and {Weinberg}, David H. and {Miralda-Escud{\'e}}, Jordi},
        title = "{Self-Consistent Models of the AGN and Black Hole Populations: Duty Cycles, Accretion Rates, and the Mean Radiative Efficiency}",
      journal = {\apj},
         year = 2009,
        month = jan,
       volume = {690},
       number = {1},
        pages = {20-41},
          doi = {10.1088/0004-637X/690/1/20},
archivePrefix = {arXiv},
       eprint = {0710.4488},
 primaryClass = {astro-ph},
       adsurl = {https://ui.adsabs.harvard.edu/abs/2009ApJ...690...20S}
}

@ARTICLE{lyndenbell1969,
       author = {{Lynden-Bell}, D.},
        title = "{Galactic Nuclei as Collapsed Old Quasars}",
      journal = {\nat},
         year = 1969,
        month = aug,
       volume = {223},
       number = {5207},
        pages = {690-694},
          doi = {10.1038/223690a0},
       adsurl = {https://ui.adsabs.harvard.edu/abs/1969Natur.223..690L}
}

@ARTICLE{shakura1973,
       author = {{Shakura}, N.~I. and {Sunyaev}, R.~A.},
        title = "{Black holes in binary systems. Observational appearance.}",
      journal = {\aap},
         year = 1973,
        month = jan,
       volume = {24},
        pages = {337-355},
       adsurl = {https://ui.adsabs.harvard.edu/abs/1973A&A....24..337S}
}

@ARTICLE{goodman2003,
       author = {{Goodman}, Jeremy},
        title = "{Self-gravity and quasi-stellar object discs}",
      journal = {\mnras},
         year = 2003,
        month = mar,
       volume = {339},
       number = {4},
        pages = {937-948},
          doi = {10.1046/j.1365-8711.2003.06241.x},
archivePrefix = {arXiv},
       eprint = {astro-ph/0201001},
 primaryClass = {astro-ph},
       adsurl = {https://ui.adsabs.harvard.edu/abs/2003MNRAS.339..937G}
}

@ARTICLE{chenGI2023,
       author = {{Chen}, Yi-Xian and {Jiang}, Yan-Fei and {Goodman}, Jeremy and {Ostriker}, Eve C.},
        title = "{3D Radiation Hydrodynamic Simulations of Gravitational Instability in AGN Accretion Disks: Effects of Radiation Pressure}",
      journal = {\apj},
         year = 2023,
        month = may,
       volume = {948},
       number = {2},
          eid = {120},
        pages = {120},
          doi = {10.3847/1538-4357/acc023},
archivePrefix = {arXiv},
       eprint = {2302.10868},
 primaryClass = {astro-ph.HE},
       adsurl = {https://ui.adsabs.harvard.edu/abs/2023ApJ...948..120C}
}

@ARTICLE{chenfeedback2024,
       author = {{Chen}, Yi-Xian and {Jiang}, Yan-Fei and {Goodman}, Jeremy and {Lin}, Douglas N.~C.},
        title = "{Radiation Hydrodynamic Simulations of Massive Stars in Gas-rich Environments: Accretion of AGN Stars Suppressed by Thermal Feedback}",
      journal = {\apj},
         year = 2024,
        month = oct,
       volume = {974},
       number = {1},
          eid = {106},
        pages = {106},
          doi = {10.3847/1538-4357/ad6dd4},
archivePrefix = {arXiv},
       eprint = {2408.12017},
 primaryClass = {astro-ph.HE},
       adsurl = {https://ui.adsabs.harvard.edu/abs/2024ApJ...974..106C}
}

@ARTICLE{chenpop2024,
       author = {{Chen}, Yi-Xian and {Lin}, Douglas N.~C.},
        title = "{The Population of Massive Stars in Active Galactic Nuclei Disks}",
      journal = {\apj},
         year = 2024,
        month = jun,
       volume = {967},
       number = {2},
          eid = {88},
        pages = {88},
          doi = {10.3847/1538-4357/ad3c3a},
archivePrefix = {arXiv},
       eprint = {2404.08780},
 primaryClass = {astro-ph.GA},
       adsurl = {https://ui.adsabs.harvard.edu/abs/2024ApJ...967...88C}
}

@ARTICLE{xuetal2026,
       author = {{Xu}, Zheng-Hao and {Chen}, Yi-Xian and {Lin}, Douglas N.~C.},
        title = "{Stellar Evolution with Radiative Feedback in AGN Disks}",
      journal = {\apj},
         year = 2026,
        month = feb,
       volume = {997},
       number = {2},
          eid = {206},
        pages = {206},
          doi = {10.3847/1538-4357/ae2271},
archivePrefix = {arXiv},
       eprint = {2511.03904},
 primaryClass = {astro-ph.GA},
       adsurl = {https://ui.adsabs.harvard.edu/abs/2026ApJ...997..206X}
}

@ARTICLE{cantiello2021ApJ,
       author = {{Cantiello}, Matteo and {Jermyn}, Adam S. and {Lin}, Douglas N.~C.},
        title = "{Stellar Evolution in AGN Disks}",
      journal = {\apj},
         year = 2021,
        month = apr,
       volume = {910},
       number = {2},
          eid = {94},
        pages = {94},
          doi = {10.3847/1538-4357/abdf4f},
archivePrefix = {arXiv},
       eprint = {2009.03936},
 primaryClass = {astro-ph.SR},
       adsurl = {https://ui.adsabs.harvard.edu/abs/2021ApJ...910...94C}
}

@ARTICLE{alidib2023,
       author = {{Ali-Dib}, Mohamad and {Lin}, Douglas N.~C.},
        title = "{The impermanent fate of massive stars in AGN discs}",
      journal = {\mnras},
         year = 2023,
        month = dec,
       volume = {526},
       number = {4},
        pages = {5824-5838},
          doi = {10.1093/mnras/stad2774},
archivePrefix = {arXiv},
       eprint = {2309.04392},
 primaryClass = {astro-ph.GA},
       adsurl = {https://ui.adsabs.harvard.edu/abs/2023MNRAS.526.5824A}
}

@ARTICLE{fryer2025,
       author = {{Fryer}, Chris L. and {Huang}, Jiamu and {Ali-Dib}, Mohamad and {Andrews}, Amaya and {Xu}, Zhenghao and {Lin}, Douglas N.~C.},
        title = "{Stellar population and metal production in AGN discs}",
      journal = {\mnras},
         year = 2025,
        month = feb,
       volume = {537},
       number = {2},
        pages = {1556-1570},
          doi = {10.1093/mnras/staf130},
archivePrefix = {arXiv},
       eprint = {2501.06973},
 primaryClass = {astro-ph.HE},
       adsurl = {https://ui.adsabs.harvard.edu/abs/2025MNRAS.537.1556F}
}

@ARTICLE{huang2023,
       author = {{Huang}, Jiamu and {Lin}, Douglas N.~C. and {Shields}, Gregory},
        title = "{Metal enrichment due to embedded stars in AGN discs}",
      journal = {\mnras},
         year = 2023,
        month = nov,
       volume = {525},
       number = {4},
        pages = {5702-5718},
          doi = {10.1093/mnras/stad2642},
archivePrefix = {arXiv},
       eprint = {2308.15761},
 primaryClass = {astro-ph.GA},
       adsurl = {https://ui.adsabs.harvard.edu/abs/2023MNRAS.525.5702H}
}

@ARTICLE{xu2025,
       author = {{Xu}, Zheng-Hao},
        title = "{The Fate of Stars Embedded in AGN Disks is Determined by an Internal Mixing Threshold}",
      journal = {Research in Astronomy and Astrophysics},
         year = 2025,
        month = nov,
       volume = {25},
       number = {11},
          eid = {115013},
        pages = {115013},
          doi = {10.1088/1674-4527/adfeb9},
       adsurl = {https://ui.adsabs.harvard.edu/abs/2025RAA....25k5013X}
}

@ARTICLE{spruit2002,
       author = {{Spruit}, H.~C.},
        title = "{Dynamo action by differential rotation in a stably stratified stellar interior}",
      journal = {\aap},
         year = 2002,
        month = jan,
       volume = {381},
        pages = {923-932},
          doi = {10.1051/0004-6361:20011465},
archivePrefix = {arXiv},
       eprint = {astro-ph/0108207},
 primaryClass = {astro-ph},
       adsurl = {https://ui.adsabs.harvard.edu/abs/2002A&A...381..923S}
}

@INPROCEEDINGS{owocki2012,
       author = {{Owocki}, S.~P. and {Shaviv}, N.~J.},
        title = "{Instability \& Mass Loss near the Eddington Limit}",
    booktitle = {Eta Carinae and the Supernova Impostors},
         year = 2012,
       editor = {{Davidson}, Kris and {Humphreys}, Roberta M.},
       series = {Astrophysics and Space Science Library},
       volume = {384},
        month = jan,
        pages = {275},
          doi = {10.1007/978-1-4614-2275-4_12},
       adsurl = {https://ui.adsabs.harvard.edu/abs/2012ASSL..384..275O}
}

@ARTICLE{nagao2006,
       author = {{Nagao}, T. and {Marconi}, A. and {Maiolino}, R.},
        title = "{The evolution of the broad-line region among SDSS quasars}",
      journal = {\aap},
         year = 2006,
        month = feb,
       volume = {447},
       number = {1},
        pages = {157-172},
          doi = {10.1051/0004-6361:20054024},
archivePrefix = {arXiv},
       eprint = {astro-ph/0510385},
 primaryClass = {astro-ph},
       adsurl = {https://ui.adsabs.harvard.edu/abs/2006A&A...447..157N}
}

@ARTICLE{lyndenbell1974,
       author = {{Lynden-Bell}, D. and {Pringle}, J.~E.},
        title = "{The evolution of viscous discs and the origin of the nebular variables.}",
      journal = {\mnras},
         year = 1974,
        month = sep,
       volume = {168},
        pages = {603-637},
          doi = {10.1093/mnras/168.3.603},
       adsurl = {https://ui.adsabs.harvard.edu/abs/1974MNRAS.168..603L}
}

@ARTICLE{pringle1981,
       author = {{Pringle}, J.~E.},
        title = "{Accretion discs in astrophysics}",
      journal = {\araa},
         year = 1981,
        month = jan,
       volume = {19},
        pages = {137-162},
          doi = {10.1146/annurev.aa.19.090181.001033},
       adsurl = {https://ui.adsabs.harvard.edu/abs/1981ARA&A..19..137P}
}

@ARTICLE{owen2025MNRAS,
       author = {{Owen}, James E. and {Lin}, Douglas N.~C.},
        title = "{A dust condensation instability in AGN atmospheres: failed winds and the broad-line region}",
      journal = {\mnras},
         year = 2025,
        month = dec,
       volume = {544},
       number = {4},
        pages = {4532-4550},
          doi = {10.1093/mnras/staf1914},
archivePrefix = {arXiv},
       eprint = {2511.02920},
 primaryClass = {astro-ph.GA},
       adsurl = {https://ui.adsabs.harvard.edu/abs/2025MNRAS.544.4532O}
}

@ARTICLE{czerny2015,
       author = {{Czerny}, Bo{\.z}ena and {Modzelewska}, Justyna and {Petrogalli}, Francesco and {Pych}, Wojtek and {Adhikari}, Tek P. and {{\.Z}ycki}, Piotr T. and {Hryniewicz}, Krzysztof and {Krupa}, Magdalena and {{\'S}wie{\c{t}}o{\'n}}, Agnieszka and {Niko{\l}ajuk}, Marek},
        title = "{The dust origin of the Broad Line Region and the model consequences for AGN unification scheme}",
      journal = {Advances in Space Research},
         year = 2015,
        month = apr,
       volume = {55},
       number = {7},
        pages = {1806-1815},
          doi = {10.1016/j.asr.2015.01.004},
archivePrefix = {arXiv},
       eprint = {1409.7312},
 primaryClass = {astro-ph.GA},
       adsurl = {https://ui.adsabs.harvard.edu/abs/2015AdSpR..55.1806C}
}

@ARTICLE{Starkey2023,
       author = {{Starkey}, D.~A. and {Huang}, Jiamu and {Horne}, Keith and {Lin}, Douglas N.~C.},
        title = "{Rimmed and rippled accretion disc models to explain AGN continuum lags}",
      journal = {\mnras},
         year = 2023,
        month = feb,
       volume = {519},
       number = {2},
        pages = {2754-2768},
          doi = {10.1093/mnras/stac3579},
archivePrefix = {arXiv},
       eprint = {2212.01379},
 primaryClass = {astro-ph.GA},
       adsurl = {https://ui.adsabs.harvard.edu/abs/2023MNRAS.519.2754S}
}

@ARTICLE{Horne2021,
       author = {{Horne}, Keith and {De Rosa}, G. and {Peterson}, B.~M. and {Barth}, A.~J. and {Ely}, J. and {Fausnaugh}, M.~M. and {Kriss}, G.~A. and {Pei}, L. and {Bentz}, M.~C. and {Cackett}, E.~M. and {Edelson}, R. and {Eracleous}, M. and {Goad}, M.~R. and {Grier}, C.~J. and {Kaastra}, J. and {Kochanek}, C.~S. and {Krongold}, Y. and {Mathur}, S. and {Netzer}, H. and {Proga}, D. and {Tejos}, N. and {Vestergaard}, M. and {Villforth}, C. and {Adams}, S.~M. and {Anderson}, M.~D. and {Ar{\'e}valo}, P. and {Beatty}, T.~G. and {Bennert}, V.~N. and {Bigley}, A. and {Bisogni}, S. and {Borman}, G.~A. and {Boroson}, T.~A. and {Bottorff}, M.~C. and {Brandt}, W.~N. and {Breeveld}, A.~A. and {Brotherton}, M. and {Brown}, J.~E. and {Brown}, J.~S. and {Canalizo}, G. and {Carini}, M.~T. and {Clubb}, K.~I. and {Comerford}, J.~M. and {Corsini}, E.~M. and {Crenshaw}, D.~M. and {Croft}, S. and {Croxall}, K.~V. and {Dalla Bont{\`a}}, E. and {Deason}, A.~J. and {Dehghanian}, M. and {De Lorenzo-C{\'a}ceres}, A. and {Denney}, K.~D. and {Dietrich}, M. and {Done}, C. and {Efimova}, N.~V. and {Evans}, P.~A. and {Ferland}, G.~J. and {Filippenko}, A.~V. and {Flatland}, K. and {Fox}, O.~D. and {Gardner}, E. and {Gates}, E.~L. and {Gehrels}, N. and {Geier}, S. and {Gelbord}, J.~M. and {Gonzalez}, L. and {Gorjian}, V. and {Greene}, J.~E. and {Grupe}, D. and {Gupta}, A. and {Hall}, P.~B. and {Henderson}, C.~B. and {Hicks}, S. and {Holmbeck}, E. and {Holoien}, T.~W.-S. and {Hutchison}, T. and {Im}, M. and {Jensen}, J.~J. and {Johnson}, C.~A. and {Joner}, M.~D. and {Jones}, J. and {Kaspi}, S. and {Kelly}, P.~L. and {Kennea}, J.~A. and {Kim}, M. and {Kim}, S. and {Kim}, S.~C. and {King}, A. and {Klimanov}, S.~A. and {Korista}, K.~T. and {Lau}, M.~W. and {Lee}, J.~C. and {Leonard}, D.~C. and {Li}, Miao and {Lira}, P. and {Lochhaas}, C. and {Ma}, Zhiyuan and {MacInnis}, F. and {Malkan}, M.~A. and {Manne-Nicholas}, E.~R. and {Mauerhan}, J.~C. and {McGurk}, R. and {McHardy}, I.~M. and {Montuori}, C. and {Morelli}, L. and {Mosquera}, A. and {Mudd}, D. and {M{\"u}ller-S{\'a}nchez}, F. and {Nazarov}, S.~V. and {Norris}, R.~P. and {Nousek}, J.~A. and {Nguyen}, M.~L. and {Ochner}, P. and {Okhmat}, D.~N. and {Pancoast}, A. and {Papadakis}, I. and {Parks}, J.~R. and {Penny}, M.~T. and {Pizzella}, A. and {Pogge}, R.~W. and {Poleski}, R. and {Pott}, J.-U. and {Rafter}, S.~E. and {Rix}, H.-W. and {Runnoe}, J. and {Saylor}, D.~A. and {Schimoia}, J.~S. and {Schn{\"u}lle}, K. and {Scott}, B. and {Sergeev}, S.~G. and {Shappee}, B.~J. and {Shivvers}, I. and {Siegel}, M. and {Simonian}, G.~V. and {Siviero}, A. and {Skielboe}, A. and {Somers}, G. and {Spencer}, M. and {Starkey}, D. and {Stevens}, D.~J. and {Sung}, H.-I. and {Tayar}, J. and {Treu}, T. and {Turner}, C.~S. and {Uttley}, P. and {Van Saders}, J. and {Vican}, L. and {Villanueva}, Jr., S. and {Weiss}, Y. and {Woo}, J.-H. and {Yan}, H. and {Young}, S. and {Yuk}, H. and {Zheng}, W. and {Zhu}, W. and {Zu}, Y.},
        title = "{Space Telescope and Optical Reverberation Mapping Project. IX. Velocity-Delay Maps for Broad Emission Lines in NGC 5548}",
      journal = {\apj},
         year = 2021,
        month = feb,
       volume = {907},
       number = {2},
          eid = {76},
        pages = {76},
          doi = {10.3847/1538-4357/abce60},
archivePrefix = {arXiv},
       eprint = {2003.01448},
 primaryClass = {astro-ph.GA},
       adsurl = {https://ui.adsabs.harvard.edu/abs/2021ApJ...907...76H}
}

@ARTICLE{Sarkar2021,
       author = {{Sarkar}, A. and {Ferland}, G.~J. and {Chatzikos}, M. and {Guzm{\'a}n}, F. and {van Hoof}, P.~A.~M. and {Smyth}, R.~T. and {Ramsbottom}, C.~A. and {Keenan}, F.~P. and {Ballance}, C.~P.},
        title = "{Improved Fe II Emission-line Models for AGNs Using New Atomic Data Sets}",
      journal = {\apj},
         year = 2021,
        month = jan,
       volume = {907},
       number = {1},
          eid = {12},
        pages = {12},
          doi = {10.3847/1538-4357/abcaa6},
archivePrefix = {arXiv},
       eprint = {2011.09007},
 primaryClass = {astro-ph.GA},
       adsurl = {https://ui.adsabs.harvard.edu/abs/2021ApJ...907...12S}
}

@ARTICLE{FerlandPersson1989,
       author = {{Ferland}, G.~J. and {Persson}, S.~E.},
        title = "{Implications of Ca II emission for physical conditions in the broad-line region of active galactic nuclei}",
      journal = {\apj},
         year = 1989,
        month = dec,
       volume = {347},
        pages = {656},
          doi = {10.1086/168156},
       adsurl = {https://ui.adsabs.harvard.edu/abs/1989ApJ...347..656F}
}

@ARTICLE{Panda2020,
       author = {{Panda}, S. and {Mart{\'i}nez-Aldama}, M.~L. and {Marinello}, M. and {Czerny}, B. and {Marziani}, P. and {Dultzin}, D.},
        title = "{The CaFe Project: Optical Fe II and Near-infrared Ca II Triplet Emission in Active Galaxies. I. Photoionization Modeling}",
      journal = {\apj},
         year = 2020,
        month = oct,
       volume = {902},
       number = {1},
          eid = {76},
        pages = {76},
          doi = {10.3847/1538-4357/abb5b8},
archivePrefix = {arXiv},
       eprint = {2004.05201},
 primaryClass = {astro-ph.GA},
       adsurl = {https://ui.adsabs.harvard.edu/abs/2020ApJ...902...76P}
}

@ARTICLE{Gunasekera2023,
       author = {{Gunasekera}, Chamani M. and {van Hoof}, Peter A.~M. and {Chatzikos}, Marios and {Ferland}, Gary J.},
        title = "{The 23.01 Release of Cloudy}",
      journal = {Research Notes of the American Astronomical Society},
         year = 2023,
        month = nov,
       volume = {7},
       number = {11},
          eid = {246},
        pages = {246},
          doi = {10.3847/2515-5172/ad0e75},
archivePrefix = {arXiv},
       eprint = {2311.10163},
 primaryClass = {astro-ph.GA},
       adsurl = {https://ui.adsabs.harvard.edu/abs/2023RNAAS...7..246G}
}

@ARTICLE{Weaver2022,
       author = {{Weaver}, John R. and {Horne}, Keith},
        title = "{Dust and the intrinsic spectral index of quasar variations: hints of finite stress at the innermost stable circular orbit}",
      journal = {\mnras},
         year = 2022,
        month = may,
       volume = {512},
       number = {1},
        pages = {899-916},
          doi = {10.1093/mnras/stac248},
archivePrefix = {arXiv},
       eprint = {2201.11134},
 primaryClass = {astro-ph.GA},
       adsurl = {https://ui.adsabs.harvard.edu/abs/2022MNRAS.512..899W}
}

@ARTICLE{StoneShen2023,
       author = {{Stone}, Zachary and {Shen}, Yue},
        title = "{Temperature fluctuations in quasar accretion discs from spectroscopic monitoring data}",
      journal = {\mnras},
         year = 2023,
        month = sep,
       volume = {524},
       number = {3},
        pages = {4521-4542},
          doi = {10.1093/mnras/stad2034},
archivePrefix = {arXiv},
       eprint = {2306.13062},
 primaryClass = {astro-ph.GA},
       adsurl = {https://ui.adsabs.harvard.edu/abs/2023MNRAS.524.4521S}
}

@ARTICLE{Baldwin1995,
       author = {{Baldwin}, J. and {Ferland}, G. and {Korista}, K. and {Verner}, D.},
        title = "{Locally Optimally Emitting Clouds and the Origin of Quasar Emission Lines}",
      journal = {\apjl},
         year = 1995,
       volume = {455},
        pages = {L119},
          doi = {10.1086/309827},
       adsurl = {https://ui.adsabs.harvard.edu/abs/1995ApJ...455L.119B}
}

@ARTICLE{KoristaGoad2000,
       author = {{Korista}, Kirk T. and {Goad}, Michael R.},
        title = "{Locally Optimally Emitting Clouds and the Variable Broad Emission Line Spectrum of NGC 5548}",
      journal = {\apj},
         year = 2000,
       volume = {536},
        pages = {284},
          doi = {10.1086/308930},
       adsurl = {https://ui.adsabs.harvard.edu/abs/2000ApJ...536..284K}
}

@ARTICLE{KoristaGoad2001,
       author = {{Korista}, Kirk T. and {Goad}, Michael R.},
        title = "{The Variable Diffuse Continuum Emission of Broad-Line Clouds}",
      journal = {\apj},
         year = 2001,
       volume = {553},
        pages = {695},
          doi = {10.1086/320964},
       adsurl = {https://ui.adsabs.harvard.edu/abs/2001ApJ...553..695K}
}

@ARTICLE{KoristaGoad2004,
       author = {{Korista}, Kirk T. and {Goad}, Michael R.},
        title = "{What the Optical Recombination Lines Can Tell Us about the Broad-Line Regions of Active Galactic Nuclei}",
      journal = {\apj},
         year = 2004,
       volume = {606},
        pages = {749},
          doi = {10.1086/383193},
       adsurl = {https://ui.adsabs.harvard.edu/abs/2004ApJ...606..749K}
}

@ARTICLE{Shen2015,
       author = {{Shen}, Yue and {Brandt}, W.~N. and {Dawson}, K.~S. and {Hall}, P.~B. and {McGreer}, I.~D. and {Anderson}, S.~F. and {Chen}, Y. and {Denney}, K.~D. and {Eftekharzadeh}, S. and {Fan}, X. and others},
        title = "{The Sloan Digital Sky Survey Reverberation Mapping Project: Technical Overview}",
      journal = {\apjs},
         year = 2015,
       volume = {216},
        pages = {4},
          doi = {10.1088/0067-0049/216/1/4},
       adsurl = {https://ui.adsabs.harvard.edu/abs/2015ApJS..216....4S}
}

@ARTICLE{Stone2025,
       author = {{Stone}, Zachary and {Shen}, Yue and {Anderson}, Scott F. and {Bauer}, Franz and {Brandt}, W.~N. and {Chakraborty}, Priyanka and {Davis}, Megan C. and {Fries}, Logan B. and {Grier}, Catherine J. and {Hall}, P.~B. and {Horne}, Keith and {Koekemoer}, Anton M. and {Mart{\'\i}nez-Aldama}, Mary Loli and {Long}, Knox and {Morrison}, Sean and {Ricci}, Claudio and {Schneider}, Donald P. and {Temple}, Matthew J. and {Trump}, Jonathan R.},
        title = "{The SDSS-V Black Hole Mapper Reverberation Mapping Project: Multiline Dynamical Modeling of a Highly Variable Active Galactic Nucleus with Decade-long Light Curves}",
      journal = {\apj},
         year = 2025,
        month = oct,
       volume = {991},
       number = {2},
          eid = {218},
        pages = {218},
          doi = {10.3847/1538-4357/adfd4c},
archivePrefix = {arXiv},
       eprint = {2408.04789},
 primaryClass = {astro-ph.GA},
       adsurl = {https://ui.adsabs.harvard.edu/abs/2025ApJ...991..218S}
}

@ARTICLE{Fries2024,
       author = {{Fries}, Logan B. and {Trump}, Jonathan R. and {Horne}, Keith and {Davis}, Megan C. and {Grier}, Catherine J. and {Shen}, Yue and {Anderson}, Scott F. and {Dwelly}, Tom and {Homayouni}, Y. and {Morrison}, Sean and {Runnoe}, Jessie C. and {Trakhtenbrot}, Benny and {Assef}, Roberto J. and {Bizyaev}, Dmitry and {Brandt}, W.~N. and {Breiding}, Peter and {Brownstein}, Joel and {Chakraborty}, Priyanka and {Hall}, P.~B. and {Koekemoer}, Anton M. and {Ibarra-Medel}, H{\'e}ctor J. and {Mart{\'\i}nez-Aldama}, Mary Loli and {Negrete}, C. Alenka and {Pan}, Kaike and {Ricci}, Claudio and {Schneider}, Donald P. and {Sharp}, Hugh W. and {Smith}, Theodore B. and {Stone}, Zachary and {Temple}, Matthew J.},
        title = "{The SDSS-V Black Hole Mapper Reverberation Mapping Project: A Kinematically Variable Broad-line Region and Consequences for the Masses of Luminous Quasars}",
      journal = {\apj},
         year = 2024,
        month = nov,
       volume = {975},
       number = {2},
          eid = {239},
        pages = {239},
          doi = {10.3847/1538-4357/ad7c42},
archivePrefix = {arXiv},
       eprint = {2409.12229},
 primaryClass = {astro-ph.GA},
       adsurl = {https://ui.adsabs.harvard.edu/abs/2024ApJ...975..239F}
}

@ARTICLE{Almeida2023,
       author = {{Almeida}, Andr{\'e}s and {Anderson}, Scott F. and {Argudo-Fern{\'a}ndez}, Maria and {Badenes}, Carles and {Barger}, Kat and {Barrera-Ballesteros}, Jorge K. and {Bender}, Chad F. and {Benitez}, Erika and {Besser}, Felipe and {Bird}, Jonathan C. and {Bizyaev}, Dmitry and {Blanton}, Michael R. and {Bochanski}, John and {Bovy}, Jo and {Brandt}, William Nielsen and {Brownstein}, Joel R. and {Buchner}, Johannes and {Bulbul}, Esra and {Burchett}, Joseph N. and {Cano D{\'\i}az}, Mariana and {Carlberg}, Joleen K. and {Casey}, Andrew R. and {Chandra}, Vedant and {Cherinka}, Brian and {Chiappini}, Cristina and {Coker}, Abigail A. and {Comparat}, Johan and {Conroy}, Charlie and {Contardo}, Gabriella and {Cortes}, Arlin and {Covey}, Kevin and {Crane}, Jeffrey D. and {Cunha}, Katia and {Dabbieri}, Collin and {Davidson}, James W. and {Davis}, Megan C. and {de Andrade Queiroz}, Anna Barbara and {De Lee}, Nathan and {M{\'e}ndez Delgado}, Jos{\'e} Eduardo and {Demasi}, Sebastian and {Di Mille}, Francesco and {Donor}, John and {Dow}, Peter and {Dwelly}, Tom and {Eracleous}, Mike and {Eriksen}, Jamey and {Fan}, Xiaohui and {Farr}, Emily and {Frederick}, Sara and {Fries}, Logan and {Frinchaboy}, Peter and {G{\"a}nsicke}, Boris T. and {Ge}, Junqiang and {Gonz{\'a}lez {\'A}vila}, Consuelo and {Grabowski}, Katie and {Grier}, Catherine and {Guiglion}, Guillaume and {Gupta}, Pramod and {Hall}, Patrick and {Hawkins}, Keith and {Hayes}, Christian R. and {Hermes}, J.~J. and {Hern{\'a}ndez-Garc{\'\i}a}, Lorena and {Hogg}, David W. and {Holtzman}, Jon A. and {Ibarra-Medel}, Hector Javier and {Ji}, Alexander and {Jofre}, Paula and {Johnson}, Jennifer A. and {Jones}, Amy M. and {Kinemuchi}, Karen and {Kluge}, Matthias and {Koekemoer}, Anton and {Kollmeier}, Juna A. and {Kounkel}, Marina and {Krishnarao}, Dhanesh and {Krumpe}, Mirko and {Lacerna}, Ivan and {Lago}, Paulo Jakson Assuncao and {Laporte}, Chervin and {Liu}, Chao and {Liu}, Ang and {Liu}, Xin and {Lopes}, Alexandre Roman and {Macktoobian}, Matin and {Majewski}, Steven R. and {Malanushenko}, Viktor and {Maoz}, Dan and {Masseron}, Thomas and {Masters}, Karen L. and {Matijevic}, Gal and {McBride}, Aidan and {Medan}, Ilija and {Merloni}, Andrea and {Morrison}, Sean and {Myers}, Natalie and {M{\'e}sz{\'a}ros}, Szabolcs and {Negrete}, C. Alenka and {Nidever}, David L. and {Nitschelm}, Christian and {Oravetz}, Daniel and {Oravetz}, Audrey and {Pan}, Kaike and {Peng}, Yingjie and {Pinsonneault}, Marc H. and {Pogge}, Rick and {Qiu}, Dan and {Ramirez}, Solange V. and {Rix}, Hans-Walter and {Fern{\'a}ndez Rosso}, Daniela and {Runnoe}, Jessie and {Salvato}, Mara and {Sanchez}, Sebastian F. and {Santana}, Felipe A. and {Saydjari}, Andrew and {Sayres}, Conor and {Schlaufman}, Kevin C. and {Schneider}, Donald P. and {Schwope}, Axel and {Serna}, Javier and {Shen}, Yue and {Sobeck}, Jennifer and {Song}, Ying-Yi and {Souto}, Diogo and {Spoo}, Taylor and {Stassun}, Keivan G. and {Steinmetz}, Matthias and {Straumit}, Ilya and {Stringfellow}, Guy and {S{\'a}nchez-Gallego}, Jos{\'e} and {Taghizadeh-Popp}, Manuchehr and {Tayar}, Jamie and {Thakar}, Ani and {Tissera}, Patricia B. and {Tkachenko}, Andrew and {Hernandez Toledo}, Hector and {Trakhtenbrot}, Benny and {Fern{\'a}ndez-Trincado}, Jos{\'e} G. and {Troup}, Nicholas and {Trump}, Jonathan R. and {Tuttle}, Sarah and {Ulloa}, Natalie and {Vazquez-Mata}, Jose Antonio and {Vera Alfaro}, Pablo and {Villanova}, Sandro and {Wachter}, Stefanie and {Weijmans}, Anne-Marie and {Wheeler}, Adam and {Wilson}, John and {Wojno}, Leigh and {Wolf}, Julien and {Xue}, Xiang-Xiang and {Ybarra}, Jason E. and {Zari}, Eleonora and {Zasowski}, Gail},
        title = "{The Eighteenth Data Release of the Sloan Digital Sky Surveys: Targeting and First Spectra from SDSS-V}",
      journal = {\apjs},
         year = 2023,
        month = aug,
       volume = {267},
       number = {2},
          eid = {44},
        pages = {44},
          doi = {10.3847/1538-4365/acda98},
archivePrefix = {arXiv},
       eprint = {2301.07688},
 primaryClass = {astro-ph.GA},
       adsurl = {https://ui.adsabs.harvard.edu/abs/2023ApJS..267...44A}
}

@ARTICLE{Shen2024,
       author = {{Shen}, Yue and {Grier}, Catherine J. and {Horne}, Keith and {Stone}, Zachary and {Li}, Jennifer I. and {Yang}, Qian and {Homayouni}, Yasaman and {Trump}, Jonathan R. and {Anderson}, Scott F. and {Brandt}, W.~N. and {Hall}, Patrick B. and {Ho}, Luis C. and {Jiang}, Linhua and {Petitjean}, Patrick and {Schneider}, Donald P. and {Tao}, Charling and {Donnan}, Fergus. R. and {AlSayyad}, Yusra and {Bershady}, Matthew A. and {Blanton}, Michael R. and {Bizyaev}, Dmitry and {Bundy}, Kevin and {Chen}, Yuguang and {Davis}, Megan C. and {Dawson}, Kyle and {Fan}, Xiaohui and {Greene}, Jenny E. and {Gr{\"o}ller}, Hannes and {Guo}, Yucheng and {Ibarra-Medel}, H{\'e}ctor and {Jiang}, Yuanzhe and {Keenan}, Ryan P. and {Kollmeier}, Juna A. and {Lejoly}, Cassandra and {Li}, Zefeng and {de la Macorra}, Axel and {Moe}, Maxwell and {Nie}, Jundan and {Rossi}, Graziano and {Smith}, Paul S. and {Tee}, Wei Leong and {Weijmans}, Anne-Marie and {Xu}, Jiachuan and {Yue}, Minghao and {Zhou}, Xu and {Zhou}, Zhimin and {Zou}, Hu},
        title = "{The Sloan Digital Sky Survey Reverberation Mapping Project: Key Results}",
      journal = {\apjs},
         year = 2024,
        month = jun,
       volume = {272},
       number = {2},
          eid = {26},
        pages = {26},
          doi = {10.3847/1538-4365/ad3936},
archivePrefix = {arXiv},
       eprint = {2305.01014},
 primaryClass = {astro-ph.GA},
       adsurl = {https://ui.adsabs.harvard.edu/abs/2024ApJS..272...26S}
}

@ARTICLE{Ogborn2026,
       author = {{Ogborn}, Mary and {Eracleous}, Michael and {Runnoe}, Jessie C. and {Dotti}, Massimo and {Mohammed}, Niana N. and {Dabbieri}, Collin M.},
        title = "{Simulating Disky Broad-line Region Reverberation}",
      journal = {\apjs},
         year = 2026,
        month = apr,
       volume = {283},
       number = {2},
          eid = {50},
        pages = {50},
          doi = {10.3847/1538-4365/ae48e7},
archivePrefix = {arXiv},
       eprint = {2511.17334},
 primaryClass = {astro-ph.GA},
       adsurl = {https://ui.adsabs.harvard.edu/abs/2026ApJS..283...50O}
}

@ARTICLE{Pancoast2011,
       author = {{Pancoast}, A. and {Brewer}, B.~J. and {Treu}, T.},
        title = "{Geometric and Dynamical Models of Reverberation Mapping Data}",
      journal = {\apj},
         year = 2011,
       volume = {730},
        pages = {139},
          doi = {10.1088/0004-637X/730/2/139},
       adsurl = {https://ui.adsabs.harvard.edu/abs/2011ApJ...730..139P}
}

@ARTICLE{Li2013,
       author = {{Li}, Yan-Rong and {Wang}, Jian-Min and {Ho}, Luis C. and others},
        title = "{A Bayesian Approach to Estimate the Parameters and Uncertainties of Reverberation Mapped Active Galactic Nuclei}",
      journal = {\apj},
         year = 2013,
       volume = {779},
        pages = {110},
          doi = {10.1088/0004-637X/779/2/110},
       adsurl = {https://ui.adsabs.harvard.edu/abs/2013ApJ...779..110L}
}

@ARTICLE{GRAVITY2020,
       author = {{GRAVITY Collaboration}},
        title = "{The size of the broad-line emission region of 3C 273}",
      journal = {\aap},
         year = 2020,
       volume = {643},
        pages = {A154},
          doi = {10.1051/0004-6361/202039067},
       adsurl = {https://ui.adsabs.harvard.edu/abs/2020A&A...643A.154G}
}

@ARTICLE{BaskinLaor2018,
       author = {{Baskin}, Alexei and {Laor}, Ari},
        title = "{Dust inflated accretion disc as the origin of the broad-line region in active galactic nuclei}",
      journal = {\mnras},
         year = 2018,
       volume = {474},
        pages = {1970},
          doi = {10.1093/mnras/stx2850},
       adsurl = {https://ui.adsabs.harvard.edu/abs/2018MNRAS.474.1970B}
}

@ARTICLE{NaddafCzerny2022,
       author = {{Naddaf}, M.~H. and {Czerny}, B.},
        title = "{Radiation pressure on dust explaining the low ionized broad emission lines in active galactic nuclei. Dust as an important driver of line shape}",
      journal = {\aap},
         year = 2022,
        month = jul,
       volume = {663},
          eid = {A77},
        pages = {A77},
          doi = {10.1051/0004-6361/202142806},
archivePrefix = {arXiv},
       eprint = {2111.14963},
 primaryClass = {astro-ph.GA},
       adsurl = {https://ui.adsabs.harvard.edu/abs/2022A&A...663A..77N}
}

@ARTICLE{Goad2014,
       author = {{Goad}, M.~R. and {Korista}, K.~T.},
        title = "{Interpreting broad emission-line variations - II. Constraints on the structure of the broad-line region in NGC 5548}",
      journal = {\mnras},
         year = 2014,
       volume = {444},
        pages = {43},
          doi = {10.1093/mnras/stu1456},
       adsurl = {https://ui.adsabs.harvard.edu/abs/2014MNRAS.444...43G}
}

@ARTICLE{Park2012,
       author = {{Park}, Daeseong and {Kelly}, Brandon C. and {Woo}, Jong-Hak and {Treu}, Tommaso},
        title = "{Recalibration of the Virial Factor and {M}$_{BH}$-{$\sigma$} Relation for Local Active Galaxies}",
      journal = {\apjs},
         year = 2012,
       volume = {203},
        pages = {6},
          doi = {10.1088/0067-0049/203/1/6},
       adsurl = {https://ui.adsabs.harvard.edu/abs/2012ApJS..203....6P}
}

@ARTICLE{CackettHorne2006,
       author = {{Cackett}, Edward M. and {Horne}, Keith},
        title = "{Photoionized H{\ensuremath{\beta}} emission in NGC 5548: it breathes!}",
      journal = {\mnras},
         year = 2006,
        month = feb,
       volume = {365},
       number = {4},
        pages = {1180-1190},
          doi = {10.1111/j.1365-2966.2005.09795.x},
archivePrefix = {arXiv},
       eprint = {astro-ph/0510800},
 primaryClass = {astro-ph},
       adsurl = {https://ui.adsabs.harvard.edu/abs/2006MNRAS.365.1180C}
}

@ARTICLE{GuoGu2014,
       author = {{Guo}, Hengxiao and {Gu}, Minfeng},
        title = "{The Optical Variability of SDSS Quasars from Multi-epoch Spectroscopy. I. Results from 60 Quasars with >= Six-epoch Spectra}",
      journal = {\apj},
         year = 2014,
        month = sep,
       volume = {792},
       number = {1},
          eid = {33},
        pages = {33},
          doi = {10.1088/0004-637X/792/1/33},
archivePrefix = {arXiv},
       eprint = {1407.3025},
 primaryClass = {astro-ph.HE},
       adsurl = {https://ui.adsabs.harvard.edu/abs/2014ApJ...792...33G}
}

@ARTICLE{Wang2020breathing,
       author = {{Wang}, Shu and {Shen}, Yue and {Jiang}, Linhua and {Grier}, C.~J. and {Horne}, Keith and {Homayouni}, Y. and {Peterson}, B.~M. and {Trump}, Jonathan R. and {Brandt}, W.~N. and {Hall}, P.~B. and {Ho}, Luis C. and {Li}, Jennifer I.-Hsiu and {Hernandez Santisteban}, J.~V. and {Kinemuchi}, K. and {McGreer}, Ian D. and {Schneider}, D.~P.},
        title = "{The Sloan Digital Sky Survey Reverberation Mapping Project: How Broad Emission Line Widths Change When Luminosity Changes}",
      journal = {\apj},
         year = 2020,
        month = nov,
       volume = {903},
       number = {1},
          eid = {51},
        pages = {51},
          doi = {10.3847/1538-4357/abb36d},
archivePrefix = {arXiv},
       eprint = {2006.06178},
 primaryClass = {astro-ph.GA},
       adsurl = {https://ui.adsabs.harvard.edu/abs/2020ApJ...903...51W}
}

@ARTICLE{Wang2026dyn,
       author = {{Wang}, Shu and {Woo}, Jong-Hak and {Villafa{\~n}a}, Lizvette and {Treu}, Tommaso and {Gallo}, Elena},
        title = "{Dynamical Modeling of the Broad-line Region with High-mass Active Galactic Nuclei and Constraints on the Virial Factor}",
      journal = {\apj},
         year = 2026,
        month = apr,
       volume = {1000},
       number = {2},
          eid = {180},
        pages = {180},
          doi = {10.3847/1538-4357/ae4699},
archivePrefix = {arXiv},
       eprint = {2604.24721},
 primaryClass = {astro-ph.GA},
       adsurl = {https://ui.adsabs.harvard.edu/abs/2026ApJ..1000..180W}
}

@INPROCEEDINGS{scott2004,
       author = {{Scott}, J. and {Kriss}, G. and {Brotherton}, M. and {Green}, R. and {Hutchings}, J. and {Shull}, J. and {Zheng}, W.},
        title = "{A Composite Extreme Ultraviolet QSO Spectrum from FUSE}",
    booktitle = {AGN Physics with the Sloan Digital Sky Survey},
         year = 2004,
       editor = {{Richards}, Gordon T. and {Hall}, Patrick B.},
       series = {Astronomical Society of the Pacific Conference Series},
       volume = {311},
        pages = {31},
       eprint = {astro-ph/0403662}
}

@ARTICLE{akansoy2025,
       author = {{Akansoy}, Deniz and {Petrou}, Helen and {Ballabio}, Giulia and {Penzlin}, Anna},
        title = "{Modelling shadows in scattered light observations as signals from companions in protoplanetary discs}",
      journal = {\mnras},
         year = 2025,
       month = jul,
       volume = {540},
       number = {4},
        pages = {3186-3203},
          doi = {10.1093/mnras/staf925},
       eprint = {2506.04415}
}

@ARTICLE{kingpringle2007,
       author = {{King}, A.~R. and {Pringle}, J.~E.},
        title = "{Fuelling active galactic nuclei}",
      journal = {\mnras},
         year = 2007,
       volume = {377},
        pages = {L25-L28},
       eprint = {astro-ph/0701679}
}

@ARTICLE{Cai24,
       author = {{Cai}, Minxuan and {Wan}, Zhen and {Cai}, Zhenyi and {Fan}, Lulu and {Wang}, Junxian},
        title = "{The Host Galaxy Fluxes of Active Galaxy Nuclei Are Generally Overestimated by the Flux Variation Gradient Method}",
      journal = {Universe},
         year = 2024,
        month = jun,
       volume = {10},
       number = {7},
          eid = {282},
        pages = {282},
          doi = {10.3390/universe10070282},
archivePrefix = {arXiv},
       eprint = {2407.03597},
 primaryClass = {astro-ph.GA},
       adsurl = {https://ui.adsabs.harvard.edu/abs/2024Univ...10..282C}
}

@ARTICLE{Winkler92,
       author = {{Winkler}, H. and {Glass}, I.~S. and {van Wyk}, F. and {Marang}, F. and {Jones}, J.~H.~S. and {Buckley}, D.~A.~H. and {Sekiguchi}, K.},
        title = "{Variability studies of seyfert galaxies - I. Broad-band optical photometry .}",
      journal = {\mnras},
         year = 1992,
        month = aug,
       volume = {257},
        pages = {659-676},
          doi = {10.1093/mnras/257.4.659},
       adsurl = {https://ui.adsabs.harvard.edu/abs/1992MNRAS.257..659W}
}

@ARTICLE{Bentz09,
       author = {{Bentz}, Misty C. and {Peterson}, Bradley M. and {Netzer}, Hagai and {Pogge}, Richard W. and {Vestergaard}, Marianne},
        title = "{The Radius-Luminosity Relationship for Active Galactic Nuclei: The Effect of Host-Galaxy Starlight on Luminosity Measurements. II. The Full Sample of Reverberation-Mapped AGNs}",
      journal = {\apj},
         year = 2009,
        month = may,
       volume = {697},
       number = {1},
        pages = {160-181},
          doi = {10.1088/0004-637X/697/1/160},
archivePrefix = {arXiv},
       eprint = {0812.2283},
 primaryClass = {astro-ph},
       adsurl = {https://ui.adsabs.harvard.edu/abs/2009ApJ...697..160B}
}

@ARTICLE{Montano26,
       author = {{Montano}, John W. and {Barth}, Aaron J. and {Horne}, Keith and {Cackett}, Edward M. and {De Rosa}, Gisella and {Homayouni}, Y. and {Kara}, Erin A. and {Kriss}, Gerard A. and {Landt}, Hermine and {Apolonio}, Gilvan G. and {Arav}, Nahum and {Boizelle}, Benjamin D. and {Dalla Bont{\`a}}, Elena and {Chelouche}, Doron and {Dehghanian}, Maryam and {Edelson}, Rick and {Ferland}, Gary J. and {Fian}, Carina and {Sobrino Figaredo}, Catalina and {Goad}, Michael R. and {Gonzalez-Buitrago}, Diego H. and {Guo}, Wei-Jian and {Hu}, Chen and {Ili{\'c}}, Dragana and {Joner}, Michael D. and {Kaspi}, Shai and {Kochanek}, Christopher S. and {Kova{\v{c}}evi{\'c}}, Andjelka B. and {Lewin}, Collin and {Li}, Sha-Sha and {Li}, Yan-Rong and {Liu}, Jun-Rong and {Miller}, Jake A. and {Neustadt}, Jack M.~M. and {Netzer}, Hagai and {Ochner}, Paolo and {Partington}, Ethan R. and {Pizzella}, Alessandro and {Plesha}, Rachel and {Popovi{\'c}}, Luka {\v{C}}. and {Sanmartim}, David and {Hern{\'a}ndez Santisteban}, Juan V. and {Vestergaard}, Marianne and {Wooley}, Jack H.~F. and {Yang}, Sen and {Yao}, Zhu-Heng and {Zaidouni}, Fatima},
        title = "{AGN STORM 2. XII. Ground-based Optical Photometry and Lag Measurements of Mrk 817}",
      journal = {\apj},
         year = 2026,
        month = jun,
       volume = {1003},
       number = {2},
          eid = {147},
        pages = {147},
          doi = {10.3847/1538-4357/ae6103},
archivePrefix = {arXiv},
       eprint = {2605.02875},
 primaryClass = {astro-ph.GA},
       adsurl = {https://ui.adsabs.harvard.edu/abs/2026ApJ..1003..147M}
}

@ARTICLE{Netzer24,
       author = {{Netzer}, Hagai and {Goad}, Michael R. and {Barth}, Aaron J. and {Cackett}, Edward M. and {Horne}, Keith and {Hu}, Chen and {Kara}, Erin and {Korista}, Kirk T. and {Kriss}, Gerard A. and {Lewin}, Collin and {Montano}, John and {Arav}, Nahum and {Behar}, Ehud and {Brotherton}, Michael S. and {Chelouche}, Doron and {De Rosa}, Gisella and {Dalla Bont{\`a}}, Elena and {Dehghanian}, Maryam and {Ferland}, Gary J. and {Fian}, Carina and {Homayouni}, Yasaman and {Ili{\'c}}, Dragana and {Kaspi}, Shai and {Kova{\v{c}}evi{\'c}}, Andjelka B. and {Landt}, Hermine and {{\v{C}}. Popovi{\'c}}, Luka and {Storchi-Bergmann}, Thaisa and {Wang}, Jian-Min and {Zaidouni}, Fatima},
        title = "{AGN STORM 2. X. The Origin of the Interband Continuum Delays in Mrk 817}",
      journal = {\apj},
         year = 2024,
        month = nov,
       volume = {976},
       number = {1},
          eid = {59},
        pages = {59},
          doi = {10.3847/1538-4357/ad8160},
archivePrefix = {arXiv},
       eprint = {2410.02652},
 primaryClass = {astro-ph.GA},
       adsurl = {https://ui.adsabs.harvard.edu/abs/2024ApJ...976...59N}
}

@ARTICLE{Welsh1991,
       author = {{Welsh}, William F. and {Horne}, Keith},
        title = "{Echo Images of Broad-Line Regions in Active Galactic Nuclei}",
      journal = {\apj},
         year = 1991,
        month = oct,
       volume = {379},
        pages = {586},
          doi = {10.1086/170530},
       adsurl = {https://ui.adsabs.harvard.edu/abs/1991ApJ...379..586W}
}

@ARTICLE{Fausnaugh2016,
       author = {{Fausnaugh}, M.~M. and {Denney}, K.~D. and {Barth}, A.~J. and {Bentz}, M.~C. and {Bottorff}, M.~C. and {Carini}, M.~T. and {Croxall}, K.~V. and {De Rosa}, G. and {Goad}, M.~R. and {Horne}, Keith and {Joner}, M.~D. and {Kaspi}, S. and {Kim}, M. and {Klimanov}, S.~A. and {Kochanek}, C.~S. and {Leonard}, D.~C. and {Netzer}, H. and {Peterson}, B.~M. and {Schn{\"u}lle}, K. and {Sergeev}, S.~G. and {Vestergaard}, M. and {Zheng}, W.-K. and {Zu}, Y. and {Anderson}, M.~D. and {Ar{\'e}valo}, P. and {Bazhaw}, C. and {Borman}, G.~A. and {Boroson}, T.~A. and {Brandt}, W.~N. and {Breeveld}, A.~A. and {Brewer}, B.~J. and {Cackett}, E.~M. and {Crenshaw}, D.~M. and {Dalla Bont{\`a}}, E. and {De Lorenzo-C{\'a}ceres}, A. and {Dietrich}, M. and {Edelson}, R. and {Efimova}, N.~V. and {Ely}, J. and {Evans}, P.~A. and {Filippenko}, A.~V. and {Flatland}, K. and {Gehrels}, N. and {Geier}, S. and {Gelbord}, J.~M. and {Gonzalez}, L. and {Gorjian}, V. and {Grier}, C.~J. and {Grupe}, D. and {Hall}, P.~B. and {Hicks}, S. and {Horenstein}, D. and {Hutchison}, T. and {Im}, M. and {Jensen}, J.~J. and {Jones}, J. and {Kaastra}, J. and {Kelly}, B.~C. and {Kennea}, J.~A. and {Kim}, S.~C. and {Korista}, K.~T. and {Kriss}, G.~A. and {Lee}, J.~C. and {Lira}, P. and {MacInnis}, F. and {Manne-Nicholas}, E.~R. and {Mathur}, S. and {McHardy}, I.~M. and {Montouri}, C. and {Musso}, R. and {Nazarov}, S.~V. and {Norris}, R.~P. and {Nousek}, J.~A. and {Okhmat}, D.~N. and {Pancoast}, A. and {Papadakis}, I. and {Parks}, J.~R. and {Pei}, L. and {Pogge}, R.~W. and {Pott}, J.-U. and {Rafter}, S.~E. and {Rix}, H.-W. and {Saylor}, D.~A. and {Schimoia}, J.~S. and {Siegel}, M. and {Spencer}, M. and {Starkey}, D. and {Sung}, H.-I. and {Teems}, K.~G. and {Treu}, T. and {Turner}, C.~S. and {Uttley}, P. and {Villforth}, C. and {Weiss}, Y. and {Woo}, J.-H. and {Yan}, H. and {Young}, S.},
        title = "{Space Telescope and Optical Reverberation Mapping Project. III. Optical Continuum Emission and Broadband Time Delays in NGC 5548}",
      journal = {\apj},
         year = 2016,
        month = apr,
       volume = {821},
       number = {1},
          eid = {56},
        pages = {56},
          doi = {10.3847/0004-637X/821/1/56},
archivePrefix = {arXiv},
       eprint = {1510.05648},
 primaryClass = {astro-ph.GA},
       adsurl = {https://ui.adsabs.harvard.edu/abs/2016ApJ...821...56F}
}

@ARTICLE{Edelson2019,
       author = {{Edelson}, R. and {Gelbord}, J. and {Cackett}, E. and {Peterson}, B.~M. and {Horne}, K. and {Barth}, A.~J. and {Starkey}, D.~A. and {Bentz}, M. and {Brandt}, W.~N. and {Goad}, M. and {Joner}, M. and {Korista}, K. and {Netzer}, H. and {Page}, K. and {Uttley}, P. and {Vaughan}, S. and {Breeveld}, A. and {Cenko}, S.~B. and {Done}, C. and {Evans}, P. and {Fausnaugh}, M. and {Ferland}, G. and {Gonzalez-Buitrago}, D. and {Gropp}, J. and {Grupe}, D. and {Kaastra}, J. and {Kennea}, J. and {Kriss}, G. and {Mathur}, S. and {Mehdipour}, M. and {Mudd}, D. and {Nousek}, J. and {Schmidt}, T. and {Vestergaard}, M. and {Villforth}, C.},
        title = "{The First Swift Intensive AGN Accretion Disk Reverberation Mapping Survey}",
      journal = {\apj},
         year = 2019,
        month = jan,
       volume = {870},
       number = {2},
          eid = {123},
        pages = {123},
          doi = {10.3847/1538-4357/aaf3b4},
archivePrefix = {arXiv},
       eprint = {1811.07956},
 primaryClass = {astro-ph.HE},
       adsurl = {https://ui.adsabs.harvard.edu/abs/2019ApJ...870..123E}
}

@ARTICLE{Lewin2025,
       author = {{Lewin}, Collin and {Kara}, Erin and {Panagiotou}, Christos and {Cackett}, Edward M. and {Gelbord}, Jonathan and {Hern{\'a}ndez Santisteban}, Juan V. and {Horne}, Keith and {Kriss}, Gerard A.},
        title = "{The Accretion Disk Size Problem in AGN Disk Reverberation Mapping Is an Obscuration Effect: A Uniform AGN Sample Study with Swift}",
      journal = {\apj},
         year = 2025,
        month = nov,
       volume = {993},
       number = {2},
          eid = {245},
        pages = {245},
          doi = {10.3847/1538-4357/ae0b6c},
archivePrefix = {arXiv},
       eprint = {2509.25315},
 primaryClass = {astro-ph.HE},
       adsurl = {https://ui.adsabs.harvard.edu/abs/2025ApJ...993..245L}
}

@ARTICLE{Gerling-Dunsmore2026,
       author = {{Gerling-Dunsmore}, Hannalore J. and {Begelman}, Mitchell C. and {Simon}, Jacob B. and {Armitage}, Philip J.},
        title = "{Magnetic Pressure Dominance Stabilizes AGN Disks against Gravitational Instability}",
      journal = {\apj},
         year = 2026,
        month = mar,
       volume = {1000},
       number = {1},
          eid = {10},
        pages = {10},
          doi = {10.3847/1538-4357/ae42bd},
archivePrefix = {arXiv},
       eprint = {2508.16842},
 primaryClass = {astro-ph.HE},
       adsurl = {https://ui.adsabs.harvard.edu/abs/2026ApJ..1000...10G}
}

@ARTICLE{wu2024,
       author = {{Wu}, Yinhao and {Chen}, Yi-Xian and {Lin}, Douglas N.~C.},
        title = "{Chaotic Type I migration in turbulent discs}",
      journal = {\mnras},
         year = 2024,
        month = feb,
       volume = {528},
       number = {1},
        pages = {L127-L132},
          doi = {10.1093/mnrasl/slad183},
archivePrefix = {arXiv},
       eprint = {2311.15747},
 primaryClass = {astro-ph.EP},
       adsurl = {https://ui.adsabs.harvard.edu/abs/2024MNRAS.528L.127W}
}

@ARTICLE{Neustadt_2022,
       author = {{Neustadt}, J.~M.~M. and {Kochanek}, C.~S.},
        title = "{Using AGN light curves to map accretion disc temperature fluctuations}",
      journal = {\mnras},
         year = 2022,
        month = jun,
       volume = {513},
       number = {1},
        pages = {1046-1062},
          doi = {10.1093/mnras/stac888},
archivePrefix = {arXiv},
       eprint = {2201.10565},
 primaryClass = {astro-ph.GA},
       adsurl = {https://ui.adsabs.harvard.edu/abs/2022MNRAS.513.1046N}
}

@ARTICLE{Hopkins2024,
       author = {{Hopkins}, Philip F. and {Grudic}, Michael Y. and {Su}, Kung-Yi and {Wellons}, Sarah and {Angles-Alcazar}, Daniel and {Steinwandel}, Ulrich P. and {Guszejnov}, David and {Murray}, Norman and {Faucher-Giguere}, Claude-Andre and {Quataert}, Eliot and {Keres}, Dusan},
        title = "{FORGE'd in FIRE: Resolving the End of Star Formation and Structure of AGN Accretion Disks from Cosmological Initial Conditions}",
      journal = {The Open Journal of Astrophysics},
         year = 2024,
        month = mar,
       volume = {7},
          eid = {18},
        pages = {18},
          doi = {10.21105/astro.2309.13115},
archivePrefix = {arXiv},
       eprint = {2309.13115},
 primaryClass = {astro-ph.GA},
       adsurl = {https://ui.adsabs.harvard.edu/abs/2024OJAp....7E..18H}
}

@ARTICLE{Newman1997,
       author = {{Newman}, Jeffrey A. and {Eracleous}, Michael and {Filippenko}, Alexei V. and {Halpern}, Jules P.},
        title = "{Measurement of an Active Galactic Nucleus Central Mass on Centiparsec Scales: Results of Long-Term Optical Monitoring of Arp 102B}",
      journal = {\apj},
         year = 1997,
        month = aug,
       volume = {485},
       number = {2},
        pages = {570-580},
          doi = {10.1086/304460},
       adsurl = {https://ui.adsabs.harvard.edu/abs/1997ApJ...485..570N},
}

@ARTICLE{FlohicEracleous2008,
       author = {{Flohic}, H{\'e}l{\`e}ne M.~L.~G. and {Eracleous}, Michael},
        title = "{Interpreting the Variability of Double-Peaked Emission Lines in Active Galactic Nuclei with Stochastically Perturbed Accretion Disk Models}",
      journal = {\apj},
         year = 2008,
        month = oct,
       volume = {686},
       number = {1},
        pages = {138-147},
          doi = {10.1086/590778},
       adsurl = {https://ui.adsabs.harvard.edu/abs/2008ApJ...686..138F},
}

@ARTICLE{Jovanovic2010,
       author = {{Jovanovi{\'c}}, P. and {Popovi{\'c}}, L.~{\v{C}}. and {Stalevski}, M. and {Shapovalova}, A.~I.},
        title = "{Variability of the H$\beta$ Line Profiles as an Indicator of Orbiting Bright Spots in Accretion Disks of Quasars: A Case Study of 3C 390.3}",
      journal = {\apj},
         year = 2010,
        month = jul,
       volume = {718},
       number = {1},
        pages = {168-176},
          doi = {10.1088/0004-637X/718/1/168},
       adsurl = {https://ui.adsabs.harvard.edu/abs/2010ApJ...718..168J},
}

@ARTICLE{DexterAgol2011,
       author = {{Dexter}, Jason and {Agol}, Eric},
        title = "{Quasar Accretion Disks are Strongly Inhomogeneous}",
      journal = {\apjl},
         year = 2011,
        month = jan,
       volume = {727},
       number = {1},
        pages = {L24},
          doi = {10.1088/2041-8205/727/1/L24},
       adsurl = {https://ui.adsabs.harvard.edu/abs/2011ApJ...727L..24D},
}

@ARTICLE{MarshHorne1988,
       author = {{Marsh}, T.~R. and {Horne}, K.},
        title = "{Images of accretion discs - II. Doppler tomography}",
      journal = {\mnras},
         year = 1988,
        month = nov,
       volume = {235},
        pages = {269-286},
          doi = {10.1093/mnras/235.1.269},
       adsurl = {https://ui.adsabs.harvard.edu/abs/1988MNRAS.235..269M},
}

@ARTICLE{Manser2019,
       author = {{Manser}, Christopher J. and {G{\"a}nsicke}, Boris T. and {Eggl}, Siegfried and {Hollands}, Mark and {Izquierdo}, Paula and {Koester}, Detlev and {Landstreet}, John D. and {Lyra}, Wladimir and {Marsh}, Thomas R. and {Meru}, Farzana and {Mustill}, Alexander J. and {Rodr{\'\i}guez-Gil}, Pablo and {Toloza}, Odette and {Veras}, Dimitri and {Wilson}, David J. and {Burleigh}, Matthew R. and {Davies}, Melvyn B. and {Farihi}, Jay and {Gentile Fusillo}, Nicola and {de Martino}, Domitilla and {Parsons}, Steven G. and {Quirrenbach}, Andreas and {Raddi}, Roberto and {Reffert}, Sabine and {Del Santo}, Melania and {Schreiber}, Matthias R. and {Silvotti}, Roberto and {Toonen}, Silvia and {Villaver}, Eva and {Wyatt}, Mark and {Xu}, Siyi and {Portegies Zwart}, Simon},
        title = "{A planetesimal orbiting within the debris disc around a white dwarf star}",
      journal = {Science},
         year = 2019,
        month = apr,
       volume = {364},
       number = {6435},
        pages = {66-69},
          doi = {10.1126/science.aat5330},
       adsurl = {https://ui.adsabs.harvard.edu/abs/2019Sci...364...66M},
}


\appendix

\section{Inclination and Pillar-Height Dependence of the ``Barber Pole'' Effect}
\label{sec:inclination_appendix}

The half-period brightness asymmetry between near-side and far-side
pillar phases is the result of two distinct geometric effects, both of
which strengthen with the inclination $i$ and with the pillar height
$h_p$, but with different functional dependences. This appendix
isolates each of the two parameters with a single-pillar sweep,
holding everything else fixed at the fiducial single-pillar
configuration of Section\,\ref{sec:pillar_dist} ($r_p = 4$\,\ld,
$\phi_p = 3\pi/4$, $\sigma_r = 0.5$\,\ld, $\sigma_\phi = 0.2$\,rad,
$H_{\rm LP} = 0.05$\,\ld). Figures\,\ref{fig:inclination_civ}
and\,\ref{fig:height_civ} use a common colour scale,
$0 \le F/\langle F\rangle \le 3$, so the modulation amplitude can be
read off directly across both panels and across both sweeps.

\subsection{Inclination sweep}
\label{sec:inclination_sweep}

\textbf{Foreshortening.}
The first effect is the smooth modulation of the foreshortening
factor $\hat{\bf n}\cdot\hat{\bf o}$, where $\hat{\bf n}$ is the local
disc surface normal and $\hat{\bf o} = (\sin i, 0, \cos i)$ is the
observer direction. The actual normal used in the code is computed from
both gradients $dh/dr$ and $dh/d\phi$ (Appendix\,\ref{sec:appendix_flux}); to
expose the inclination scaling, consider the simplified case in which
the azimuthal slope is neglected so the normal at a cell on the
pillar's lamp-facing radial inner slope reduces to
$\hat{\bf n} = (-\sin\theta_t \cos\phi_p, -\sin\theta_t \sin\phi_p,
\cos\theta_t)$, where $\theta_t = \arctan(|dh/dr|)$ is the local
radial tilt. Under this approximation,
\begin{equation}
\hat{\bf n}\cdot\hat{\bf o}
   = \cos i\,\cos\theta_t - \sin i\,\sin\theta_t \cos\phi_p .
\label{eq:foreshort}
\end{equation}
The first term is azimuth-independent and dominates at low
inclination; the second term carries the once-per-orbit modulation
through $\cos\phi_p$, with amplitude proportional to $\sin i$. At
face-on ($i = 0^\circ$), $\sin i = 0$ and the foreshortening
modulation vanishes entirely. As $i$ grows, the modulation amplitude
scales linearly in $\sin i$.

\textbf{Geometric self-occultation.}
At sufficiently high
inclination, the pillar's own body blocks the observer's line of sight
to its lamp-facing inner slope when the pillar crosses the line of
sight on the near side ($\cos\phi_p > 0$). For a vertical wall of
height $h_p$ at radius $r_p$, the radial occultation depth is
$\Delta r_{\rm occ} \approx h_p\,\tan i\,|\cos\phi_p|$, set by the
projection of the pillar wall onto the observer's line of sight (this
is what the \texttt{observer\_occult} block in the worker function
applies). Unlike foreshortening, occultation switches on relatively
sharply with $i$: for small $i$ the line of sight passes
above the pillar top, while for $\tan i$ comparable to
$h_p / \sigma_r$ the wall starts to occult its own inner face on the
near side. For our fiducial pillar
($h_p = 0.15\,\ld$, $\sigma_r = 0.5\,\ld$) this transition lies near
$i \sim 70^\circ$.

\begin{figure}
    \centering
    \includegraphics[width=\columnwidth]{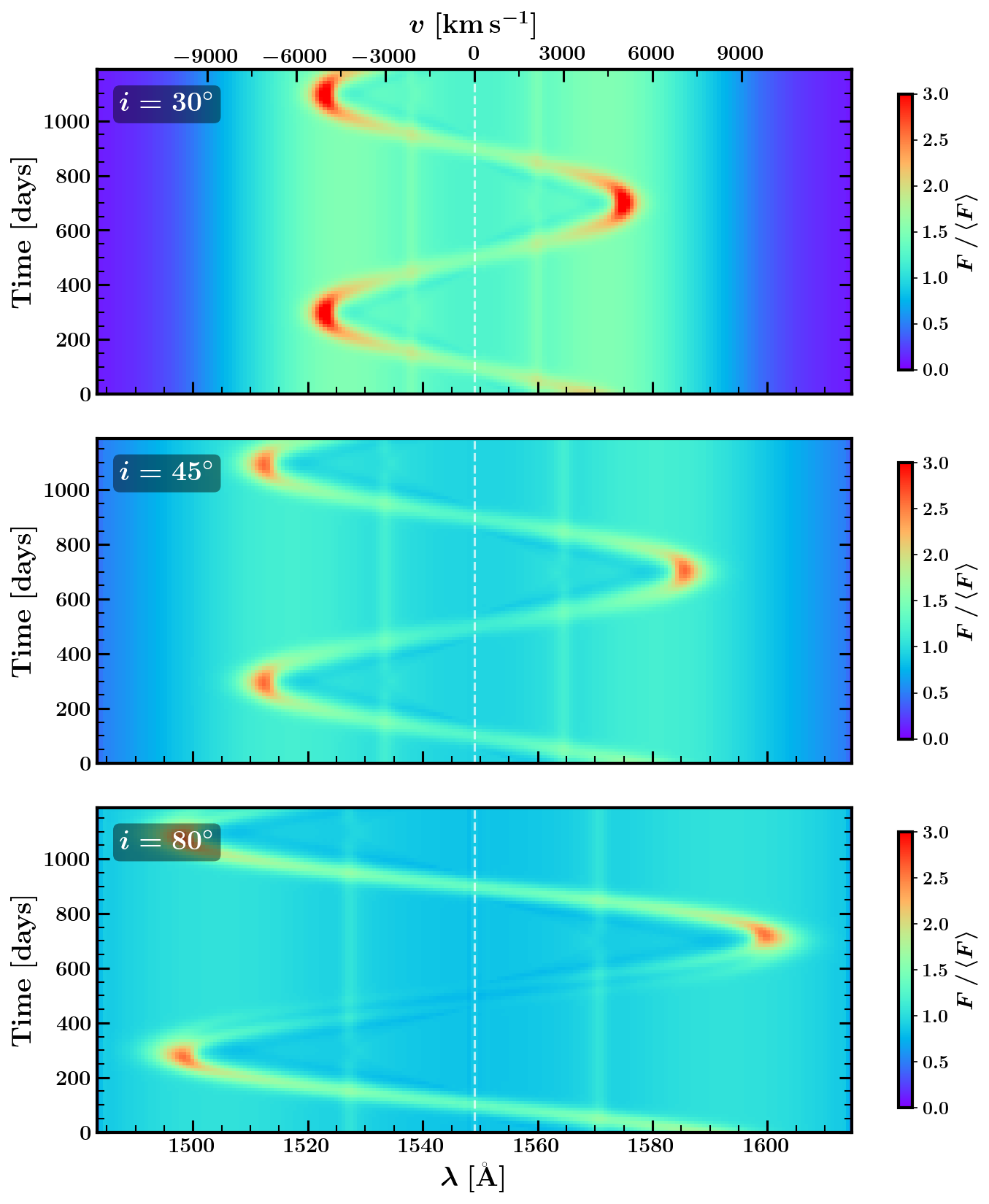}
    \caption{C\,{\sc iv} velocity-time map $F(\lambda, t)/\langle F\rangle$
    for the single-pillar configuration at three inclination angles
    (top to bottom: $i = 30^\circ, 45^\circ, 80^\circ$). All three panels
    use the same disc, pillar, and lamp parameters; only the inclination
    differs. The smooth once-per-orbit modulation grows linearly in
    $\sin i$ from foreshortening (Eq.\,\ref{eq:foreshort}); the
    additional sharp on/off contrast at $i = 80^\circ$ is geometric
    self-occultation of the pillar's lamp-facing inner slope by its
    own body when the pillar crosses the line of sight on the near
    side.}
    \label{fig:inclination_civ}
\end{figure}

The two effects combine to produce the inclination scaling shown in
Fig.\,\ref{fig:inclination_civ}: at $i = 30^\circ$ both terms are weak
and the trail is nearly uniform along the orbit; at $i = 45^\circ$
the foreshortening (Eq.\,\ref{eq:foreshort}) is the dominant
modulator, giving a moderate near/far contrast; at $i = 80^\circ$
self-occultation kicks in and the contrast becomes near-binary.

Figure\,\ref{fig:inclination_civ} illustrates this scaling for
C\,{\sc iv} (the line with the steepest emissivity response to
$\Phi_{\rm H}$, hence the cleanest tracer of the modulation): the
single-pillar velocity-time map shows a faint sinusoidal trail with
modest brightness modulation at $i = 30^\circ$, more visible
modulation at $i = 45^\circ$, and pronounced near/far asymmetry at
$i = 80^\circ$, where the inner illuminated face is occulted on the
near side and visible on the far side.

\subsection{Pillar-height sweep}
\label{sec:height_sweep}

Both effects in Eq.\,(\ref{eq:foreshort}) and the occultation depth
$\Delta r_{\rm occ} \approx h_p\tan i\,|\cos\phi_p|$ depend explicitly
on the pillar height $h_p$, so the modulation can also be enhanced by
making the pillar taller while holding the inclination fixed. The
radial tilt of the lamp-facing wall scales as
$\tan\theta_t \sim h_p/\sigma_r$, so the foreshortening term in
Eq.\,(\ref{eq:foreshort}) grows roughly linearly with $h_p$ at fixed
$\sigma_r$. The occultation depth $\Delta r_{\rm occ}$ is also linear
in $h_p$. The pillar's own shadow, which removes ionizing flux from a
strip of disc material behind it, lengthens in proportion to $h_p$
through Eq.\,(\ref{eq:shadow_extent}), enlarging the spatial region
where the line emissivity is suppressed.

\begin{figure}
    \centering
    \includegraphics[width=\columnwidth]{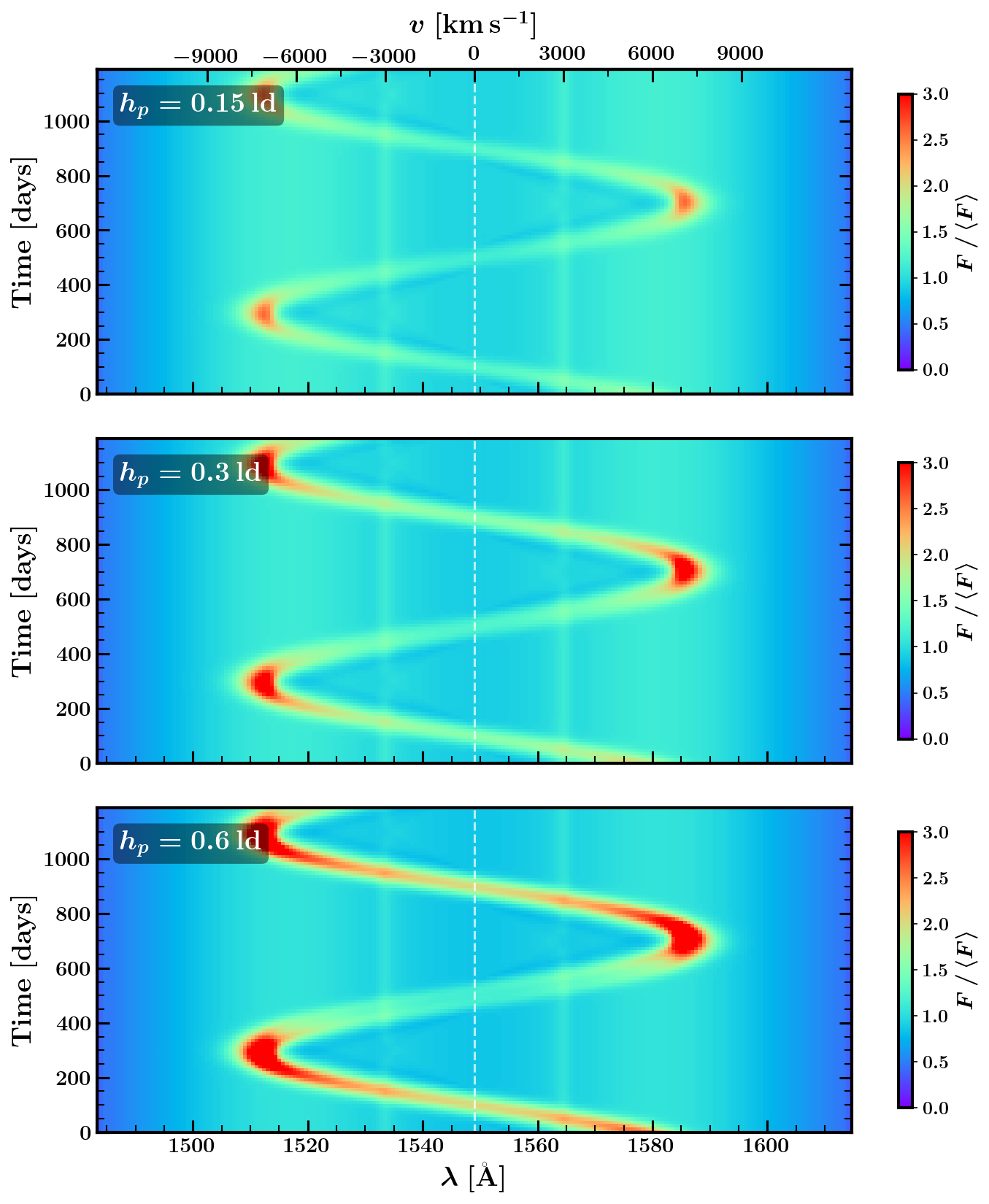}
    \caption{C\,{\sc iv} velocity-time map $F(\lambda, t)/\langle F\rangle$
    for the single-pillar configuration at three pillar heights
    (top to bottom: $h_p = 0.15, 0.3, 0.6$\,\ld), with the inclination
    held fixed at $i = 45^\circ$. The colour scale is identical to
    Fig.\,\ref{fig:inclination_civ}, so the two figures can be read on
    a common amplitude scale. The modulation strengthens monotonically
    with $h_p$: at $h_p = 0.15$\,\ld\ the trail is moderate, at
    $0.3$\,\ld\ it brightens and sharpens, and at $0.6$\,\ld\ it
    becomes near-binary. The $h_p = 0.6$\,\ld\ panel produces a
    similar near/far contrast to the high-inclination ($i = 80^\circ$)
    panel of Fig.\,\ref{fig:inclination_civ}, showing that pillar
    height and inclination act as complementary amplifiers of the
    same geometric foreshortening + self-occultation effect.}
    \label{fig:height_civ}
\end{figure}

Figure\,\ref{fig:height_civ} shows the C\,{\sc iv} velocity-time map
at fixed $i = 45^\circ$ for $h_p = 0.15, 0.3, 0.6$\,\ld. At the
fiducial $h_p = 0.15$\,\ld\ the modulation is moderate and matches
the $i = 45^\circ$ panel of Fig.\,\ref{fig:inclination_civ}. Doubling
the height to $h_p = 0.3$\,\ld\ noticeably brightens the trail and
sharpens the contrast between near-side and far-side phases.
Quadrupling to $h_p = 0.6$\,\ld\ produces a near-binary on/off
signature comparable to the $i = 80^\circ$ panel of
Fig.\,\ref{fig:inclination_civ}: very tall pillars at moderate
inclination reproduce the same observational signature as
short pillars at near-edge-on inclination, because both routes
enhance the foreshortening + occultation modulation amplitude.

\section{Computational Notes on the Ionizing Flux}
\label{sec:appendix_flux}

In Section\,\ref{sec:ionflux} we described how the ionizing flux map
$\Phi_{\rm H}(r, \phi)$ is built from the lamp-post irradiation, the $Q$
calibration and the shadow mask. Here we give further computational details
of that calculation.

\subsection{Viscous Wien-tail contribution}
\label{sec:visc_appendix}

The disc's own thermal emission can in principle contribute hydrogen
ionizing photons.  For a local blackbody at viscous temperature
$T_{\rm v}(r)$, the Wien-tail approximation gives
\begin{equation}
  \Phi_{\rm visc}(r) \approx \frac{2\pi\,\nu_0^2}{c^2}
    \left(\frac{k_{\rm B} T_{\rm v}}{h}\right)
    \exp\!\left(-\frac{h\nu_0}{k_{\rm B}\,T_{\rm v}}\right),
  \label{eq:phi_visc}
  \end{equation}
where $\nu_0$ is the hydrogen ionization frequency. The prefactor
$2\pi/c^2$ follows from weighting the upward hemispheric blackbody
field by the foreshortening factor $\cos\theta$, so that
$\Phi_{\rm visc}$ is the photon flux through the disc surface,
consistent with the $\cos\theta_{\rm inc}$ foreshortening adopted for
the lamp flux in Eq.\,\eqref{eq:phi_lamp}. Weighting all directions
in the hemisphere equally instead, which matches the photoionization
rate per atom at the surface, would double $\Phi_{\rm visc}$; this
0.3\,dex ambiguity does not affect the conclusions below.

\begin{figure}
    \centering
    \includegraphics[width=\columnwidth]{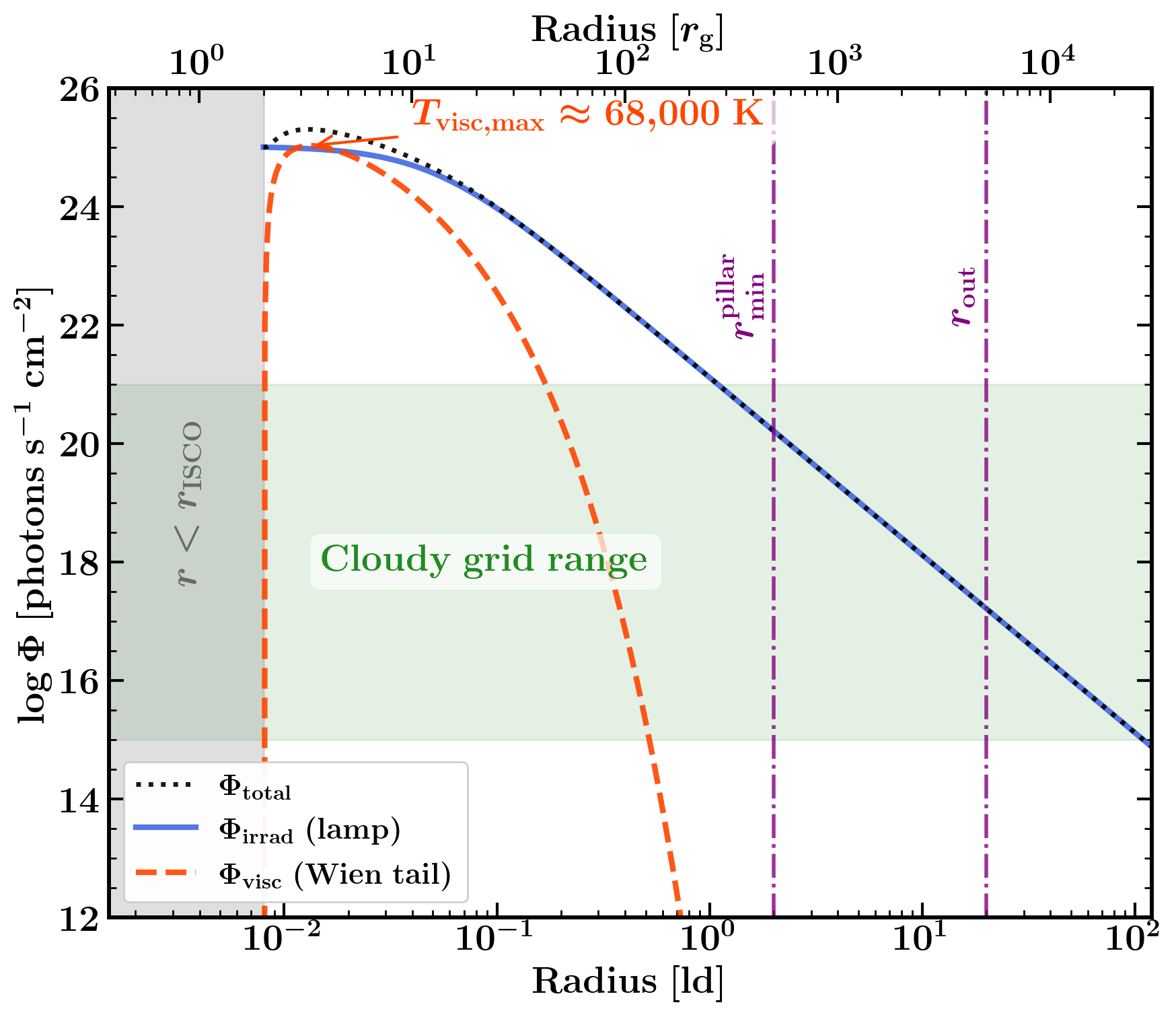}
    \caption{Radial profile of the ionizing photon flux for a disc
    with no pillars.  The blue solid line shows the lamp-post
    irradiation flux $\Phi_{\rm irrad}$ (Eq.\,\ref{eq:phi_lamp}),
    the orange dashed line shows the viscous Wien-tail contribution
    $\Phi_{\rm visc}$ (Eq.\,\ref{eq:phi_visc}), and the black
    dotted line is the total ionizing flux $\Phi_{\rm total}$.
    The green shaded band marks the \cloudy{} photoionization grid
    range ($15 \le \log\Phi \le 21$).  The grey shaded region marks
    $r < r_{\rm ISCO} = 2\,r_{\rm g}$ (Kerr SMBH with
    $M_{\rm BH} = 7 \times 10^7\,\Msun$); the purple dot-dashed
    vertical lines mark the inner cutoff of the pillar radial
    distribution, $r_{\rm min}^{\rm pillar} = 2$\,ld, and the disc
    outer edge $r_{\rm out} = 20$\,\ld.  The top axis shows radius
    in gravitational radii.  Although the viscous contribution
    peaks at $\log\Phi \approx 25$ near $r \approx 0.013$\,\ld{} where
    $T_{\rm v,max} \approx 68{,}000$\,K, it falls below the
    \cloudy{} floor by $r \sim 0.5$\,ld.  Throughout the region
    where pillars are placed ($r \ge r_{\rm min}^{\rm pillar}$),
    $\Phi_{\rm visc}$ is negligible compared with $\Phi_{\rm irrad}$,
    so the lamp-post irradiation sets the ionization state of every
    pillar in our model.}
    \label{fig:ionflux_profile}
\end{figure}

For our fiducial parameters, the viscous temperature follows
$T_{\rm v}(r) \propto r^{-3/4}$ and peaks at
$T_{\rm v,max} \approx 68{,}000$\,K near
$r \approx$ 1.7\,$r_{\rm in}$.  At this maximum the Wien-tail
ionizing flux reaches $\log\Phi_{\rm visc} \approx 25$, comparable
to the lamp at the innermost radii.  Because of the
$\exp(-h\nu_0 / k T_{\rm v})$ suppression, however, $\Phi_{\rm visc}$
drops steeply with radius: by $r \sim 0.5$\,\ld{} it has already
fallen below the \cloudy{} grid floor ($\log\Phi = 15$), and
throughout the region where pillars are placed
($r \ge r_{\rm min}^{\rm pillar} = 2$\,ld) it is many orders of
magnitude below the lamp contribution
(Fig.\,\ref{fig:ionflux_profile}).  The viscous contribution is
therefore negligible for the ionization budget of every pillar in
our model, and in particular it does not provide a meaningful
``floor'' to $\Phi_{\rm H}$ inside opaque shadows: in shadows
where $S = 0$ the total ionizing flux drops effectively to zero.
Values of $\log\Phi_{\rm H}$ from Eq.\,(\ref{eq:phi_total}) are
clipped to the \cloudy{} grid range $[15, 21]$ before
interpolation.

\subsection{Linear temperature-rescaling}
\label{sec:trescale_appendix}

Earlier versions of the code computed $\log\Phi_{\rm H}$ via a linear
rescaling of the disc effective temperature rather than from the
direct lamp-post integral of Section\,\ref{sec:ionflux}. We retain a
description here for reference; the rescaling is enabled in the code
by setting \texttt{flux\_method = "scale\_temp"} and was used to
generate some of the earliest versions of the line-emission maps in
this work.

In this approach the effective temperature is first computed as
\begin{equation}
T_{\rm e}^4(r, \phi) = T_{\rm v}^4(r) + S^2(r, \phi)\,T_{\rm irr}^4(r),
\label{eq:teff_rescale}
\end{equation}
where the viscous and irradiation profiles share the
$T \propto r^{-3/4}\sqrt{(r-r_{\rm in})/r}$ scaling but with a free
ratio $(T_{\rm irr}/T_{\rm v})_{r_0}$ at the reference radius (set in
the configuration; here $T_{\rm irr}/T_{\rm v} = 10$ at $r_0 = r_{\rm
out}$). $T_{\rm e}$ is then mapped onto the \cloudy{} ionizing-flux
grid by a linear rescaling of the temperature ratio:
\begin{equation}
\log\Phi_{\rm H} = 15.0 + \frac{T_{\rm e}/T_{\rm ref} - 0.1}{10.0 - 0.1}
\,(21.5 - 15.0),
\label{eq:logphi_rescale}
\end{equation}
where $T_{\rm ref} = T_{\rm v,0} \times (T_{\rm irr}/T_{\rm v})_{r_0}$
is the reference irradiation temperature at $r_0$, and the temperature
ratio is clipped to the range $[0.1, 10]$ before rescaling. By
construction, the coolest shadow regions are mapped to
$\log\Phi \approx 15$ (the \cloudy{} grid floor) and the hottest
illuminated regions to $\log\Phi \approx 21.5$.

This rescaling is computationally trivial but has no physical
justification: the disc thermal continuum does not emit ionizing
photons at the level needed to populate the \cloudy{} grid (see
Appendix\,\ref{sec:visc_appendix}), and the linear remap from $T_{\rm e}$ to
$\log\Phi_{\rm H}$ has no underlying radiative-transfer interpretation.
The direct lamp-post computation in Section\,\ref{sec:ionflux} replaces
it as our default method, and all results in this paper are produced
with \texttt{flux\_method = "direct"}.

\subsection{Periodic Azimuthal Gaussian for Pillar Height Profile}
\label{sec:periodic_gaussian}

Each pillar is modelled as a 2D Gaussian bump added to the bowl
baseline $H_0(r)$ of Eq.\,(\ref{eq:h_profile}). The pillar centred at
$(r_p, \phi_p)$ contributes a height perturbation

\begin{equation}
    \Delta H(r, \phi) = h_p \, G_r(r) \, G_\phi(\phi),
\end{equation}

\noindent where $h_p$ is the peak pillar height (in light-days), $G_r$ is the radial Gaussian, and $G_\phi$ is the azimuthal Gaussian.

The radial factor is a standard Gaussian:

\begin{equation}
    G_r(r) = \exp\!\left[-\frac{1}{2}\left(\frac{r - r_p}{\sigma_r}\right)^2\right],
\end{equation}

\noindent with radial width $\sigma_r$ (in light-days).

\subsubsection{Non-periodic azimuthal Gaussian}

The simplest approach wraps the azimuthal difference $\Delta\phi = \phi - \phi_p$ into the interval $(-\pi, \pi]$ and evaluates a single Gaussian:

\begin{equation}
    G_\phi^{\rm (single)}(\phi) = \exp\!\left[-\frac{1}{2}\left(\frac{\Delta\phi_{\rm w}}{\sigma_\phi}\right)^2\right],
    \label{eq:gauss_single}
\end{equation}

\noindent where $\Delta\phi_{\rm w} = [(\phi - \phi_p + \pi) \bmod 2\pi] - \pi$. This correctly identifies the shortest angular path to the pillar centre, but the Gaussian is not periodic on the circle: the profile and its derivative are discontinuous at $\Delta\phi_{\rm w} = \pm\pi$ (the antipodal point). For narrow pillars ($\sigma_\phi \lesssim 1$\,rad) the Gaussian decays to negligible values before reaching $\pm\pi$, so the discontinuity is invisible. For broad pillars ($\sigma_\phi \gtrsim 1$\,rad), however, a cusp appears at the antipodal azimuth.

\subsubsection{Periodic azimuthal Gaussian}

To enforce smooth periodicity on the circle $\phi \in [0, 2\pi)$, we sum over periodic images of the Gaussian, shifted by integer multiples of $2\pi$:

\begin{equation}
    G_\phi^{\rm (periodic)}(\phi) = \sum_{n=-\infty}^{+\infty} \exp\!\left[-\frac{1}{2}\left(\frac{\Delta\phi_{\rm w} - 2\pi n}{\sigma_\phi}\right)^2\right].
    \label{eq:gauss_periodic}
\end{equation}

\noindent In practice, only the $n = -1, 0, +1$ terms contribute significantly, giving

\begin{align}
    G_\phi^{\rm (periodic)}(\phi) \approx\;
    & \exp\!\left[-\frac{1}{2}\left(\frac{\Delta\phi_{\rm w}}{\sigma_\phi}\right)^2\right]
    + \exp\!\left[-\frac{1}{2}\left(\frac{\Delta\phi_{\rm w} - 2\pi}{\sigma_\phi}\right)^2\right] \nonumber \\
    +\; & \exp\!\left[-\frac{1}{2}\left(\frac{\Delta\phi_{\rm w} + 2\pi}{\sigma_\phi}\right)^2\right].
    \label{eq:gauss_3image}
\end{align}

\noindent The $n=0$ term is the original Gaussian centred at $\phi_p$, while the $n=\pm1$ terms are copies of the same Gaussian shifted by one full revolution in each direction. At any point on the disc, the contributions from all three copies add smoothly, eliminating the cusp that arises in the single-wrap approach at $\Delta\phi = \pm\pi$. Physically, this construction ensures that the pillar height varies continuously around the full $2\pi$ circle with no artificial features at any azimuth. The total pillar height profile is then

\begin{equation}
    H(r, \phi) = H_0(r) + \sum_{k} h_{p,k} \, G_{r,k}(r) \, G_{\phi,k}^{\rm (periodic)}(\phi),
\end{equation}

\noindent where the sum runs over all pillars $k$.

Note that for $\sigma_\phi \gtrsim 2$\,rad, the periodic Gaussian approaches a nearly uniform ring: $G_\phi^{\rm (periodic)} \to 1$ for all $\phi$, and the peak value at $\phi = \phi_p$ slightly exceeds unity due to the overlapping images. For the pillar parameters adopted in this work ($\sigma_\phi = 0.05$\,rad), the periodic correction is negligible and equations~(\ref{eq:gauss_single}) and~(\ref{eq:gauss_periodic}) are indistinguishable.

\begin{figure}
    \centering
    \includegraphics[width=\columnwidth]{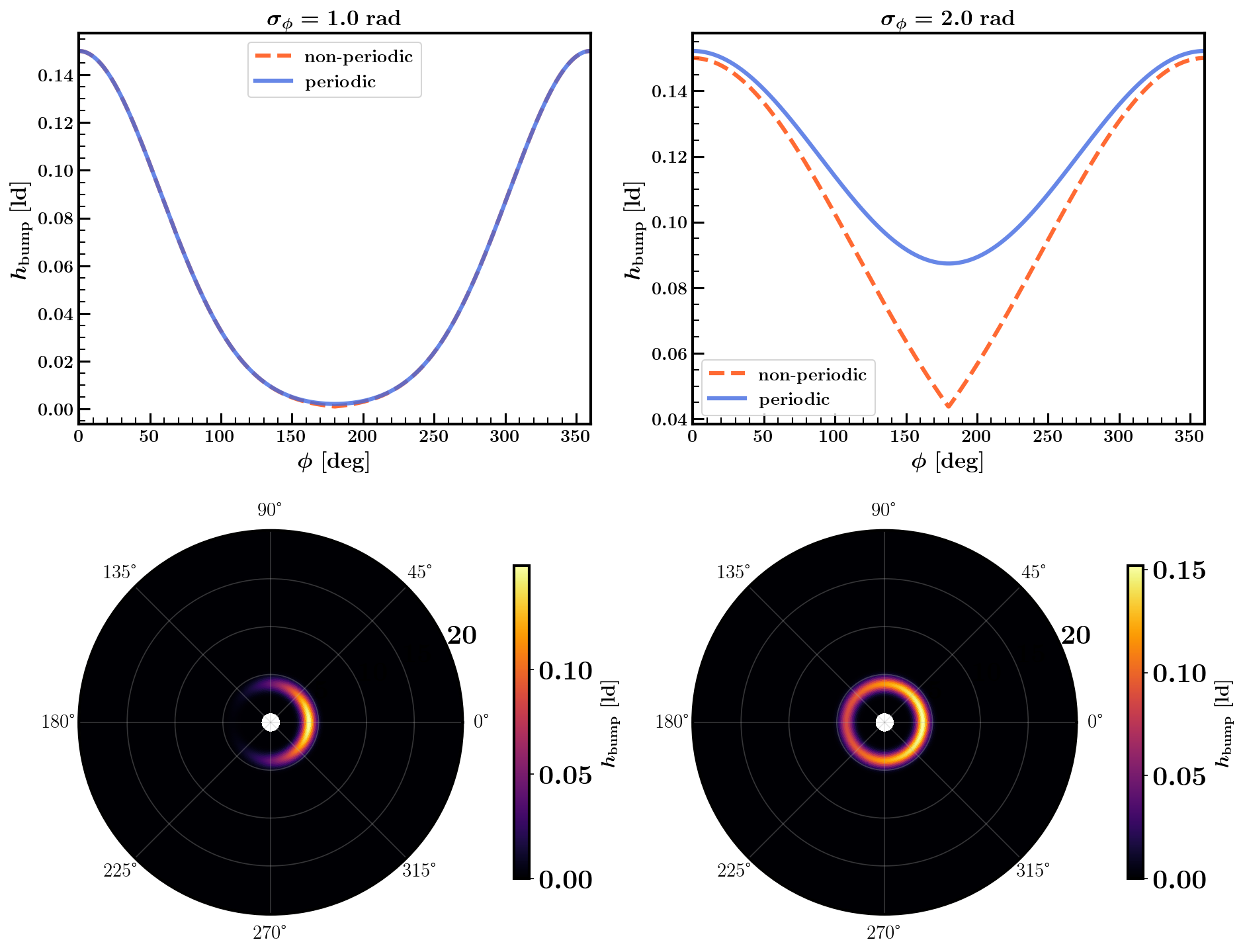}
    \caption{Comparison of the non-periodic (single-wrap; orange dashed) and periodic (three-image sum; blue solid) azimuthal Gaussian profiles for a pillar at $\phi_p = 0$ with peak height $h_p = 0.15$\,\ld{} and $\sigma_r = 0.5$\,\ld.
    \textbf{\textit{Top left:}}~pillar bump height $h_{\rm bump}(\phi)$ at the pillar radius $r = r_p$ for $\sigma_\phi = 1$\,rad; the two methods are nearly identical.
    \textbf{\textit{Top right:}}~same for $\sigma_\phi = 2$\,rad; the non-periodic profile develops a cusp at the antipodal azimuth ($\phi = 180^\circ$), while the periodic profile remains smooth.
    \textbf{\textit{Bottom left:}}~face-on polar map of $h_{\rm bump}(r, \phi)$ from the periodic formulation (Eq.\,\ref{eq:gauss_3image}) for $\sigma_\phi = 1$\,rad, producing a localized bump.
    \textbf{\textit{Bottom right:}}~same for $\sigma_\phi = 2$\,rad, where the bump spreads into a nearly complete ring around the pillar radius.}
    \label{fig:periodic_gaussian}
\end{figure}

\subsection{Non-opaque Pillars}
\label{sec:transparent_pillars}

The fiducial model adopts a partially-transparent pillar with $f_{\rm trans} = 0.3$, and $f_{\rm trans}$ is allowed to take any value in $[0, 1]$: $f_{\rm trans} = 0$ corresponds to fully opaque pillars (a hard shadow on the disc behind the pillar) and $f_{\rm trans} = 1$ to fully transparent ones (no shadow).

Physically, the transmission of a pillar depends on its column density $N_{\rm H}$ and the photon wavelength. For the \cloudy{} models used throughout this work we adopt $N_{\rm H} = 10^{25}$\,cm$^{-2}$ (Section\,\ref{sec:cloudy_grid}), at which the pillar interior is optically thick to ionizing photons and the shadow on the disc should be effectively complete. Even so, some ionizing continuum can leak through the pillar edges, and recombination and resonance line photons formed within the pillar itself can leak out to reach the disc surface behind it. A self-consistent treatment of these effects requires the pillar column density and the emergent line transmission from a hydrodynamic radiative-transfer simulation of an embedded star, which we defer to a future paper. Here we represent both effects with a single effective transmission $f_{\rm trans} \in [0, 1]$ applied identically to the ionizing continuum and the line emission.

\subsubsection{Shadow transmission}

For an opaque pillar, the shadow mask $S(r, \phi)$ is binary: $S = 1$ for illuminated regions and $S = 0$ in the geometric shadow behind the pillar. With partial transparency, a fraction $f_{\rm trans}$ of the lamp flux penetrates through the pillar body and reaches the shadowed disc. The shadow mask becomes

\begin{equation}
    S(r, \phi) = 1 - s(r, \phi)\,(1 - f_{\rm trans}),
    \label{eq:shadow_trans}
\end{equation}

\noindent where $s(r, \phi) \in [0, 1]$ is the geometric shadow strength (determined by ray-tracing from the lamp through the pillar profile, as described in Section\,\ref{sec:shadows}). For $f_{\rm trans} = 0$ this reduces to the opaque case $S = 1 - s$, while for $f_{\rm trans} > 0$ the shadow is partially filled in, with a floor of $S = f_{\rm trans}$ in the deepest part of the shadow.

\subsubsection{Effect on velocity-delay maps}

\begin{figure}
    \centering
    \includegraphics[width=\columnwidth]{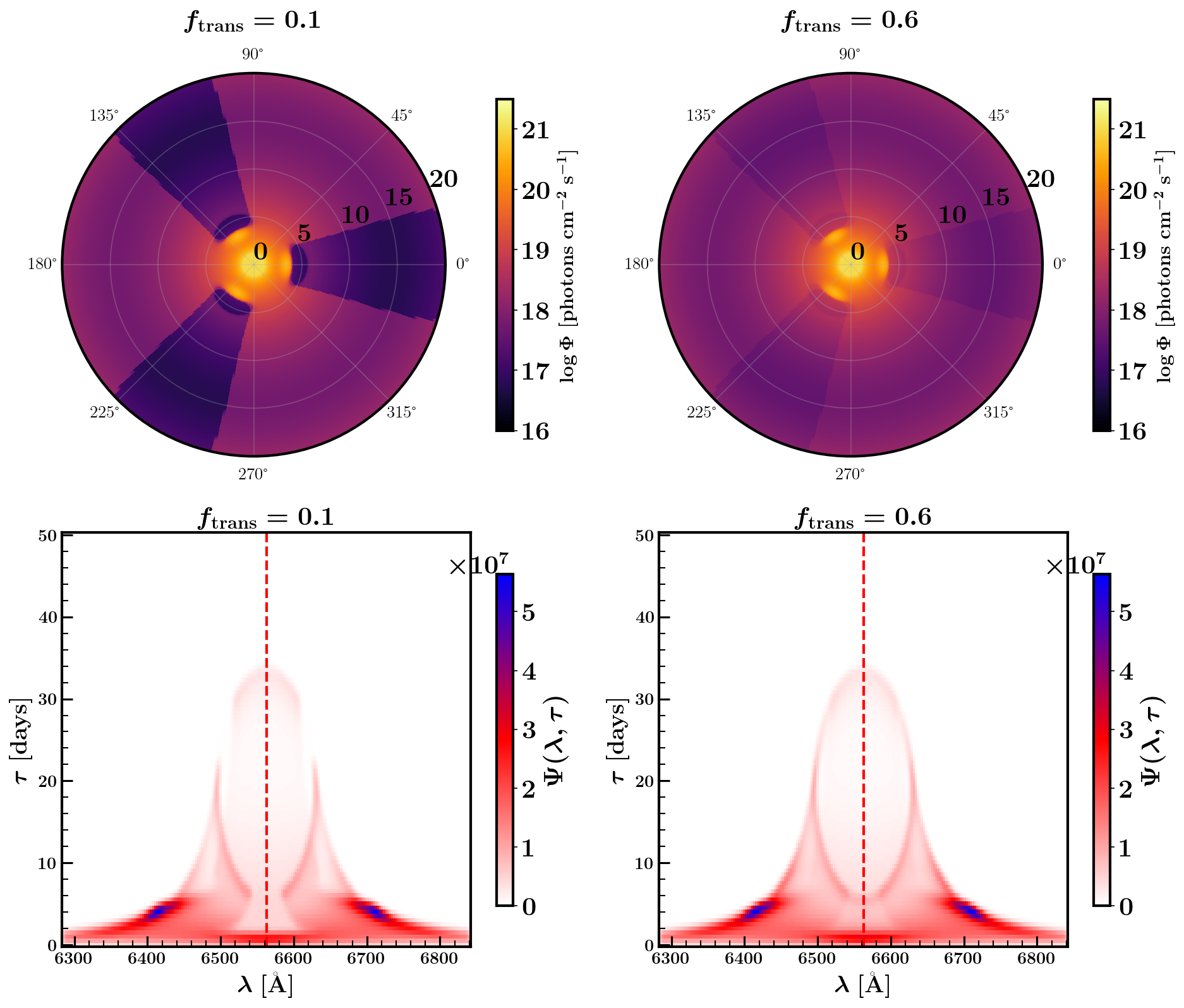}
    \caption{Effect of pillar transparency on ionizing flux and H$\alpha$ velocity-delay maps for three pillars spaced $120^\circ$ apart at $r_p = 4$\,ld.
    \textbf{\textit{Top left:}}~face-on ionizing flux map $\log\Phi_{\rm H}(r, \phi)$ for $f_{\rm trans} = 0.1$, on a fixed colour scale of $\log\Phi = 15$--$21.5$.
    \textbf{\textit{Top right:}}~same for $f_{\rm trans} = 0.6$. The shadow tails are visibly deeper for the lower $f_{\rm trans}$.
    \textbf{\textit{Bottom left:}}~corresponding H$\alpha$ velocity-delay map $\Psi(\lambda, \tau)$ for $f_{\rm trans} = 0.1$.
    \textbf{\textit{Bottom right:}}~same for $f_{\rm trans} = 0.6$. With increasing transparency, emission from the previously shadowed disc regions partially fills in, subtly modifying the velocity-delay structure at intermediate delays.}
    \label{fig:ftrans_vdm}
\end{figure}

Figure\,\ref{fig:ftrans_vdm} illustrates the effect of pillar transparency on both the ionizing flux distribution and the H$\alpha$ velocity-delay map for a configuration with three pillars spaced $120^\circ$ apart at $r_p = 4$\,ld. The top row shows face-on maps of the ionizing flux $\log\Phi_{\rm H}$, while the bottom row shows the corresponding velocity-delay maps $\Psi(\lambda, \tau)$.

For nearly opaque pillars ($f_{\rm trans} = 0.1$, left column), deep shadow tails extend radially outward from each pillar, suppressing the ionizing flux and hence the line emission behind them. For more transparent pillars ($f_{\rm trans} = 0.6$, right column), the shadows are partially filled in, allowing moderate line emission from the previously shadowed regions. The velocity-delay maps reflect this change: with increasing $f_{\rm trans}$, emission from the shadowed disc fills in, modifying the response function at delays corresponding to the shadow regions.

\section{Emissivity- versus Responsivity-Weighted Velocity--Delay Maps}
\label{sec:appendix_emiss_weighting}

In Eq.\,\ref{eq:psi_sum} (Section\,\ref{sec:vdm}) we adopt the
responsivity weighting
$F_{\rm line}\,\eta_{\rm line}
 = \partial F_{\rm line}/\partial\log\Phi_{\rm H}$
per disc element. The observed continuum scales linearly with the
lamp luminosity, $F_c \propto Q$, and the local ionizing flux at cell
$ij$ is linear in $Q$ with a $Q$-independent geometric prefactor (set
by $S$, $\cos\theta_{\rm inc}$, and $d$ in Eq.\,\ref{eq:phi_lamp}).
Taking logarithms gives $\log\Phi_{ij} - \log F_c = c_{ij}$, a
cell-dependent constant independent of $Q$. Hence $d\log\Phi_{ij} = d\log F_c$ at every cell,
and by the chain rule
\begin{equation}
\frac{\partial F_{\rm line}}{\partial \log F_c}
 = \frac{\partial F_{\rm line}}{\partial \log\Phi_{ij}}
 = F_{\rm line}\,
   \frac{\partial \log F_{\rm line}}{\partial \log\Phi_{ij}}
 \equiv F_{\rm line}\,\eta_{\rm line}.
\label{eq:log_deriv_equality}
\end{equation}
The summed $\Psi(\lambda, \tau)$ in Eq.\,\ref{eq:psi_sum} is therefore
the line-response function with the continuum variation expressed in
log/fractional units. The alternative weighting,
$w_{ij} \to F_{\rm line}(\log\Phi_{ij})$ with no responsivity factor,
produces instead the steady-state line surface brightness distribution.

The two weightings differ by a multiplicative factor of
$\eta_{\rm line}(\log\Phi_{ij})$ per cell. Over the
$\log\Phi_{\rm H}$ range that the disc actually populates, the LOC
responsivity is everywhere positive and varies modestly:
$\eta_{\rm line} \sim 0.5\text{--}2$ for the three lines we consider
(Fig.\,\ref{fig:loc_emissivity}, bottom panel). 

\begin{figure*}
    \centering
    \includegraphics[width=\textwidth]{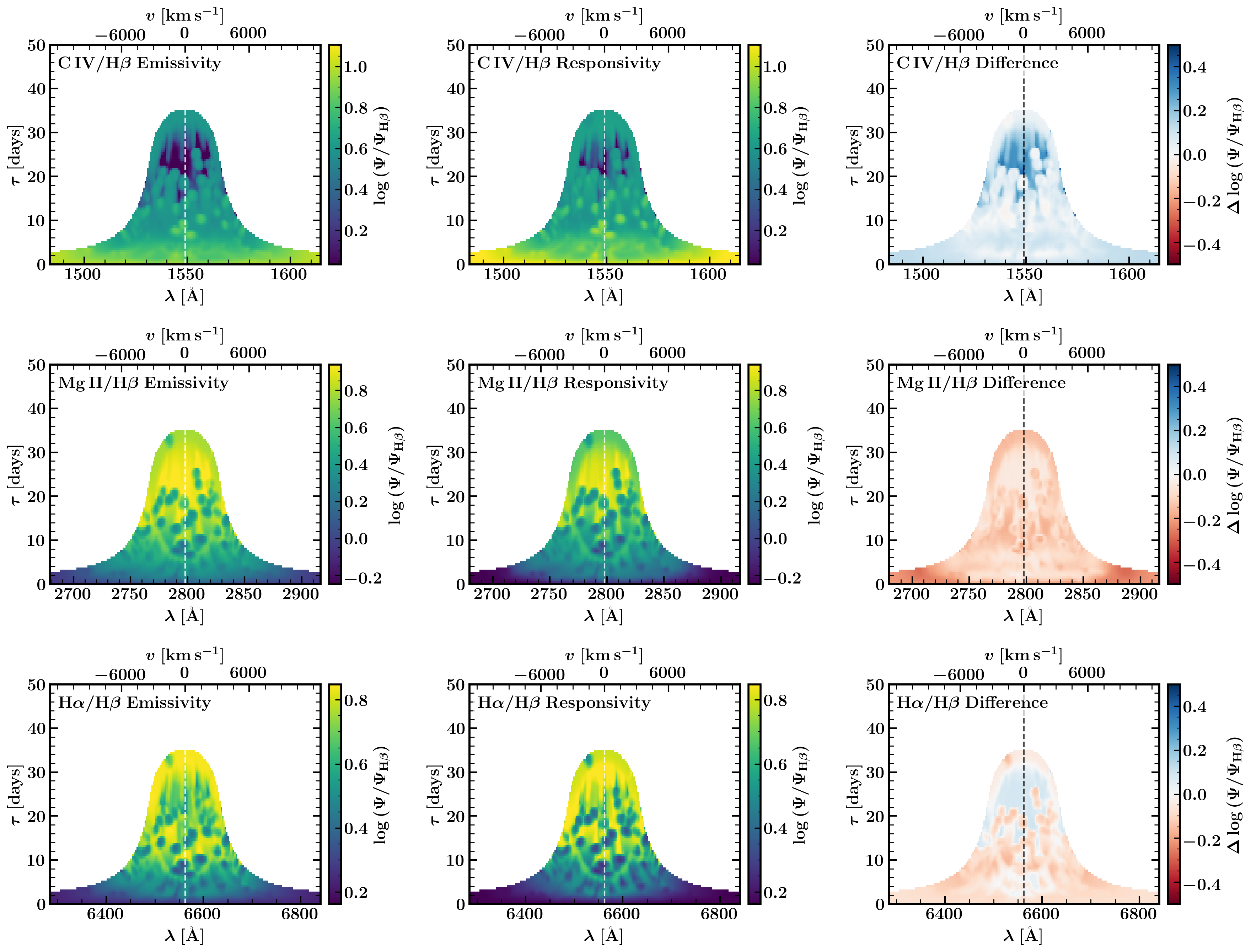}
    \caption{Comparison of the emissivity- and responsivity-weighted
    velocity--delay maps for the 100-pillar disc, each shown relative to the H$\beta$ map
    with the same weighting, $\log(\Psi/\Psi_{\rm H\beta})$. Rows correspond to
    C\,\textsc{iv}, Mg\,\textsc{ii}, and H$\alpha$.
    \textbf{\textit{Left column:}}~emissivity weighting,
    $w \propto F_{\rm line}$, equivalent to the steady-state
    line surface brightness $\bar{F}_l(v, \tau)$ of
    \citet[][Eq.\,6]{Horne2004}.
    \textbf{\textit{Middle column:}}~responsivity weighting,
    $w \propto F_{\rm line}\,\eta_{\rm line}
    = \partial F_{\rm line}/\partial\log\Phi_{\rm H}$ (the
    linear-response function of \citealt{Horne2004} Eq.\,3), as used
    throughout this paper.
    \textbf{\textit{Right column:}}~difference of the two
    ratio maps, responsivity minus emissivity, which
    equals $\log(\eta_{\rm line}/\eta_{\rm H\beta})$. It is negative where the line
    responds less strongly than H$\beta$ in proportion to its emission
    (Mg\,\textsc{ii} everywhere), and positive where it responds more strongly
    (C\,\textsc{iv}).}
    \label{fig:vdm_weighting_compare}
\end{figure*}

Multiplying $F_{\rm line}$ by this factor preserves the overall morphology
of $\Psi(\lambda, \tau)$, so the two weightings give maps that look alike.
Figure\,\ref{fig:vdm_weighting_compare} therefore shows each map relative to the
H$\beta$ map with the same weighting, which removes the common disc geometry.
The difference between the two ratio maps (third column) is
$\log(\eta_{\rm line}/\eta_{\rm H\beta})$, a property of the photoionization
response alone, and is typically 0.1\,dex. Mg\,\textsc{ii} responds less
strongly than H$\beta$ at every ionizing flux, C\,\textsc{iv} more strongly,
especially in the faint outer disc, and H$\alpha$ less strongly in the inner
disc and slightly more strongly further out. Line ratios measured from
reverberation responses therefore differ from the steady-state line ratios by
up to 0.3\,dex.

\label{lastpage}
\end{document}